\documentclass[a4paper,11pt]{article}
\pdfoutput=1
\usepackage{jcappub}

\usepackage{amsfonts}
\usepackage{slashed,subfigure}
\usepackage{latexsym}
\usepackage{graphicx}
\usepackage{bm}
\usepackage{cancel}
\usepackage{multirow}
\compress

\newcommand{\hepth}[1]{\href{https://arxiv.org/abs/hep-th/#1}{\tt hep-th/#1}}
\newcommand{\hepph}[1]{\href{https://arxiv.org/abs/hep-ph/#1}{\tt hep-ph/#1}}

\newcommand{\grqc}[1]{\href{https://arxiv.org/abs/gr-qc/#1}{\tt gr-qc/#1}}

\newcommand{\astroph}[1]{\href{https://arxiv.org/abs/astro-ph/#1}{\tt astro-ph/#1}}

\newcommand{\arXivid}[1]{\href{https://arxiv.org/abs/#1}{\tt arXiv:#1}}
\newcommand{\Math}[2]{\if!#1!\href{https://arxiv.org/abs/math/#2}{\tt math/#2}\else\href{https://arxiv.org/abs/math.#1/#2}{\tt math.#1/#2}\fi}
\newcommand\inspire[1]{\href{http://inspirehep.net/search?p=find+#1}{{\tiny IN}{\footnotesize SPIRE}}}
\newcommand\erratum[4][ibid.\ ]{\emph{Erratum #1}{\bf #2} (#3) #4}
\newcommand{\jhep}[3] {\ifnum#2>2009\href{http://dx.doi.org/10.1007/JHEP#1(#2)#3}{\emph{JHEP} {\bf #1} (#2) #3}\else\href{http://dx.doi.org/10.1088/1126-6708/#2/#1/#3}{\emph{JHEP} {\bf #1} (#2) #3}\fi}
\newcommand{\jcap}[3] {\href{http://dx.doi.org/10.1088/1475-7516/#2/#1/#3}{\emph{JCAP} {\bf #1} (#2) #3}}

\def\be{\begin{equation}}
\def\ee{\end{equation}}
\def\ba{\begin{eqnarray}}
\def\ea{\end{eqnarray}}
\def\epsone{\epsilon_1}
\def\epstwo{\epsilon_2}
\def\Mpl{M_p}
\def\ns{n_s}
\def\Mpc{{\rm Mpc}}

\newcommand{\msun}{M_\odot}

\newcommand{\la}{\lesssim}

\title{Science with the space-based interferometer LISA. IV: probing inflation with gravitational waves}

\author[a,b,c]{Nicola Bartolo,}
\author[d]{Chiara Caprini,}
\author[d]{Valerie Domcke,}
\author[e,1]{Daniel~G.~Figueroa,\note{Group coordinators.}}
\author[f]{Juan Garcia-Bellido,}
\author[a,b]{Maria~Chiara~Guzzetti,}
\author[a,b,c]{Michele Liguori,}
\author[a,b,c,g]{Sabino~Matarrese,}
\author[h]{Marco~Peloso,}
\author[d]{Antoine Petiteau,}
\author[i,1]{Angelo~Ricciardone,}
\author[l]{Mairi~Sakellariadou,}
\author[m]{Lorenzo Sorbo}
\author[n]{and~Gianmassimo~Tasinato}

\emailAdd{nicola.bartolo@pd.infn.it}
\emailAdd{caprini@apc.in2p3.fr}
\emailAdd{valerie.domcke@apc.univ-paris7.fr}
\emailAdd{daniel.figueroa@cern.ch}
\emailAdd{juan.garciabellido@uam.es}
\emailAdd{mariachiara.guzzetti@pd.infn.it}
\emailAdd{michele.liguori@pd.infn.it}
\emailAdd{sabino.matarrese@pd.infn.it}
\emailAdd{peloso@physics.umn.edu}
\emailAdd{antoine.petiteau@apc.univ-paris7.fr}
\emailAdd{angelo.ricciardone@uis.no}
\emailAdd{mairi.sakellariadou@kcl.ac.uk}
\emailAdd{sorbo@physics.umass.edu}
\emailAdd{g.tasinato@swansea.ac.uk}

\affiliation[a]{Dipartimento di Fisica e Astronomia ``G. Galilei'', Universit\`a degli Studi di Padova,\\
via Marzolo 8, I-35131, Padova, Italy}
\affiliation[b]{INFN, Sezione di Padova,\\
via Marzolo 8, I-35131, Padova, Italy}
\affiliation[c]{INAF-Osservatorio Astronomico di Padova,\\
Vicolo dell'Osservatorio 5, I-35122 Padova, Italy}
\affiliation[d]{APC, Universit\'e Paris Diderot, CNRS UMR 7164, Observatoire de Paris,\\
Sorbonne Paris Cit\'e, 10 rue Alice Domon et L\'eonie Duquet,\\
75205 Paris Cedex 13, France}
\affiliation[e]{Theoretical Physics Department, CERN,\\
Geneva, Switzerland}
\affiliation[f]{Instituto de F\'isica Te\'orica UAM-CSIC, Universidad Auton\'oma de Madrid,\\
Cantoblanco, 28049 Madrid, Spain}
\affiliation[g]{Gran Sasso Science Institute, INFN,\\
Viale F. Crispi 7, I-67100 L'Aquila, Italy}
\affiliation[h]{School of Physics and Astronomy, and Minnesota Institute for Astrophysics,
\\
University of Minnesota, Minneapolis, 55455, U.S.A.}
\affiliation[i]{Faculty of Science and Technology, University of Stavanger,\\
4036, Stavanger, Norway}
\affiliation[l]{Theoretical Particle Physics and Cosmology Group, Department of Physics,\\
King's College London, University of London, Strand, London WC2R 2LS, U.K.}
\affiliation[m]{Amherst Center for Fundamental Interactions, Department of Physics,\\
University of Massachusetts, Amherst, MA 01003, U.S.A.}
\affiliation[n]{Department of Physics, Swansea University,\\
Swansea, SA2 8PP, U.K.}

\abstract{We investigate the potential for the LISA space-based interferometer to detect the
stochastic gravitational wave background produced from different mechanisms during inflation. Focusing on well-motivated scenarios, we study the resulting contributions from particle production during inflation, inflationary spectator fields with varying speed of sound, effective field theories of inflation with specific patterns of symmetry breaking and models leading to the formation of primordial black holes. The projected sensitivities of LISA are used in a model-independent way for various detector designs and configurations. We demonstrate that LISA is able to probe these well-motivated inflationary scenarios beyond the irreducible vacuum tensor modes expected from any inflationary background.}

\begin{document}
\maketitle\flushbottom

\section{Introduction}
\label{sec:intro}

Gravitational waves (GWs) are ripples of the space-time metric, corresponding to a tensor perturbation $h_{ij}$ of the Friedmann-Lemaitre-Robertson-Walker (FLRW) \hbox{line~element,}
\begin{equation}\label{eq:FRW}
ds^2\,=\,-d t^2+a^2(t)\,\left(\delta_{ij}+h_{ij}\right)\,d x^i d x^j\,,
\end{equation}
which is transverse ($\partial_{i}h^{i}_{\,j} = 0$) and traceless ($h^{i}_{i}=0$). Here {\emph{t}} denotes the physical time and $a(t)$ represents the scale factor. The transverse and traceless conditions leave only two independent and physical degrees of freedom, the two polarizations of the GWs.

The recent direct detection of GWs~\cite{Abbott-ml-2016blz,Abbott-ml-2016nmj} by \emph{Advanced LIGO} (Laser Interferometer Gravitational-Wave Observatory)~\cite{Harry-ml-2010zz} represents a milestone in astronomy. This detection has opened a new window for exploring both the late and early stages of the Universe. In the coming years, many astrophysical sources are expected to be detected by LIGO and other planned detectors, like Advanced VIRGO~\cite{Acernese-ml-2015gua}, KAGRA~\cite{Somiya-ml-2011np}, and eventually LIGO-India~\cite{LIGO-India} and Einstein Telescope (ET)~\cite{Sathyaprakash-ml-2012jk}. The European Space Agency (ESA) has recently approved the first GW observer in space, and the Laser Interferometer Space Antenna (LISA) project~\cite{AmaroSeoane-ml-2012km} is the main candidate for this mission. LISA will have the potential to detect, not only astrophysical sources, but also cosmological sources, or at least to constrain early Universe scenarios. Gravitational waves are in fact the most promising cosmic relic to probe the unknown aspects of the early Universe. Sufficiently energetic processes in the early Universe imprinted characteristic signatures in relic GW backgrounds. It is important therefore to characterize all possible GW signals in order to achieve a better understanding of \hbox{a future~detection.}

A main goal of modern cosmology is to detect GWs produced in the early Universe. As GWs decouple immediately upon production, they travel freely through space, carrying information about the source that produced them. From non-equilibrium phenomena in the early Universe, we expect a strong production of GWs from e.g.\ (p)reheating~\cite{Khlebnikov-ml-1997di,Easther-ml-2006gt,Easther-ml-2006vd,GarciaBellido-ml-2007dg,GarciaBellido-ml-2007af,Dufaux-ml-2007pt,Dufaux-ml-2008dn,Figueroa-ml-2011ye,Bethke-ml-2013aba,Bethke-ml-2013vca,Enqvist-ml-2012im,Figueroa-ml-2013vif,Figueroa-ml-2014aya,Figueroa-ml-2016ojl,Antusch-ml-2016con}, phase transitions~\cite{Kosowsky-ml-1991ua,Kosowsky-ml-1992rz,Kosowsky-ml-1992vn,Kamionkowski-ml-1993fg,Grojean-ml-2006bp,Caprini-ml-2007xq,Huber-ml-2008hg,Kahniashvili-ml-2008pf,Kahniashvili-ml-2009mf,Caprini-ml-2009fx,Child-ml-2012qg,Hindmarsh-ml-2013xza,Giblin-ml-2014qia,Hindmarsh-ml-2015qta,Kisslinger-ml-2015hua,Caprini-ml-2015zlo,Weir-ml-2016tov}, or cosmic defects~\cite{Vachaspati-ml-1984gt,Caldwell-ml-1996en,Damour-ml-2000wa,Damour-ml-2001bk,Damour-ml-2004kw,Siemens-ml-2006yp,Siemens-ml-2006vk,JonesSmith-ml-2007ne,Fenu-ml-2009qf,Figueroa-ml-2012kw,Olmez-ml-2010bi,Dufaux-ml-2010cf,Binetruy-ml-2012ze,Sanidas-ml-2012ee}. Gravitational waves with sufficiently large amplitude from preheating are naturally peaked at very high frequencies, and hence out of the reach of LISA or other planned detectors. Gravitational waves from phase transitions are however peaked at frequencies depending on the energy scale of the phase transition, hence both high and low frequency peaked backgrounds with sufficiently large amplitude can be expected. In particular, the GW background from the electroweak phase transition lies precisely in the LISA frequency window of $f \sim (10^{-5}-0.1)$ Hz. The GW background(s) from cosmic defects span many decades in frequency, and are therefore expected to cross through the frequency window of all planned detectors. Whether the GW signal from cosmic defects can be detected, depends on the scenario, mostly on the energy scale of the phase transition that created the defects in the first place. The detection of any of these GW backgrounds from the early Universe, will allow us to access into physics beyond the reach of high-energy particle colliders, like the Large Hadron Collider (LHC).

In this paper we rather focus on the GWs expected from cosmic inflation. In the absence of any source, GWs are always generated quantum mechanically during inflation~\cite{Starobinsky-ml-1979ty}. Moreover, depending of the modeling of the inflationary sector, active sources can also be present during inflation, giving rise to a further contribution to the GWs signal, besides that generated by quantum fluctuations, see e.g.~\cite{Guzzetti-ml-2016mkm} for a recent review. The features of the GWs produced by quantum fluctuations of the gravitational field, reflect the properties of the theory of gravity which underlines the inflationary model, while the GWs contribution induced by the presence of a source term, reflects the presence of further fields besides the inflaton. At the end, from the inflationary stage we expect the universe to be filled in, at the present time, by a GW spectral-energy density given by two contributions: one due to quantum fluctuations of the gravitational field, and in some cases by a second contribution due to the presence of a source term. In general, modifying the gravity theory which underlines the inflationary physics, and/or assuming the presence of active sources during inflation, gives rise to the production of GWs with a large amplitude and tilt. A detection of any of these primordial GW signals will provide information about the energy scale and other relevant parameters of inflation, opening a window into the inflationary physics beyond the reach of (and complementary to) the Cosmic Microwave Background (CMB). It will also help to discriminate inflationary models from each other, ruling out entire classes of models.

\paragraph{The irreducible background of gravitational waves from inflation.} During inflation, GWs are always expected to be generated by the amplification of vacuum metric fluctuations. This background represents an irreducible contribution from any inflationary scenario. Its amplitude encodes direct information about the energy scale of inflation, or more precisely, about the Hubble parameter during inflation. In the standard inflation scenarios, where the accelerated expansion is driven by a scalar field slowly rolling down along its flat potential, tensor fluctuations are characterized by an almost scale invariant spectrum, slightly red tilted. Denoting by $\Omega_{\rm GW}$ today's GW fractional energy density per logarithmic wave-number interval, the amplitude of this irreducible background, at the frequencies corresponding to the CMB scales $f_{\rm CMB} \sim 10^{-18}-10^{-17}$ Hz, is
\begin{equation}\label{eq:omegainfl}
h^2\Omega_{\rm GW}^{\rm CMB}\equiv h^2\Omega_{\rm{GW}}\left(f_{\rm CMB}\right)\approx 5 \cdot 10^{-16}
\left({H\over H_{\rm max}}\right)^2\,,
\end{equation}
where $H$ is the inflationary Hubble rate (evaluated at the CMB scales), and $H_{\rm max} \simeq 8.8\times 10^{13}$\,GeV is the current upper bound on $H$~\cite{Ade-ml-2015lrj}. If we parametrize the GW energy-density spectrum at different frequencies by a power law around a pivot scale at the CMB frequencies, we can write
\begin{equation}
\Omega_{\rm GW}(f) = \Omega_{\rm GW}^{\rm CMB}\left(\frac{f}{f_{\rm CMB}}\right)^{n_T} \,,
\label{eq:GWen}
\end{equation}
with $n_T$ a spectral index. In the case of standard single-field slow-roll inflation models, it must be satisfied the consistency relation~\cite{Liddle-ml-1992wi}
\begin{equation}\label{eq:CR}
n_{T} = -r/8\,,
\end{equation}
where $r \equiv A_{T}/A_{S}$ is the so-called tensor-to-scalar perturbation ratio, with $A_{T}$ and $A_{S}$ the amplitude of the primordial tensor and scalar power spectra. In standard inflation models we expect therefore, a slightly red tilted spectrum, i.e.~$n_T < 0$ with $|n_T| \ll 1$, as the current bounds from the CMB indicate $r \lesssim 0.1$, see discussion at the end of section~\ref{sec:LISAcurves}.

A detection of this background will provide extremely useful information about the early Universe. It will help to differentiate inflationary models, ruling out entire model families. It will also probe some aspects of the quantum nature of fields and gravity.
This irreducible background leaves a precise imprint on the CMB, resulting in a specific polarization pattern of B-modes, which is the primary probe for its detection~\cite{Kosowsky-ml-1994cy, Hu-ml-1997hv}. A large number of experiments are presently active or they are proposed for searching such a signal through indirect effects on the CMB. However, given the current strong bounds from the CMB~\cite{Ade-ml-2015lrj} on the amplitude of the spectrum, eq.~(\ref{eq:omegainfl}), and the fact that it is predicted to be red tilded, eq.~(\ref{eq:CR}), this signal cannot be detected by LISA or any of the ground-based planned detectors. Even in a best-case scenario, assuming an almost scale invariant spectrum, the amplitude $\Omega_{\rm{gw}}\left(f\right) \sim 10^{-15}$ is simply too small. This tiny amplitude remains therefore only potentially interesting for some next-to-next-generation of space-based observatories, like Big Bang Observatory (BBO)~\cite{Corbin-ml-2005ny} and maybe Deci-hertz Interferometer Gravitational wave Observatory (DECIGO)~\cite{Kawamura-ml-2011zz}.

\paragraph{Beyond the irreducible background of gravitational waves.} We demonstrate in this work that the details of the GWs produced during inflation, and hence the perspective of detecting such primordial GW backgrounds, change completely if:
\begin{enumerate}
\item[$i)$] additional degrees of freedom, besides the inflaton, are present during inflation
\item[$ii)$] new symmetry patterns are considered in the inflationary sector
\item[$iii)$] large peaks in the inflationary scalar spectrum collapse into primordial black holes after horizon re-entry.
\end{enumerate}

In all these circumstances, the spectrum of GWs associated to these new ingredients can be rather large and blue-tilted, or exhibit a large-amplitude bump at specific scales. In the case of additional degrees of freedom, these provide a source term in the GW evolution equation, that in Fourier space reads
\begin{equation}\label{eq:GWeq}
\ddot{h}_{ij}\left(\bold{k},t\right)+3H\,\dot{h}_{ij}\left(\bold{k},t\right)+k^2\, h_{ij}\left(\bold{k},t\right)=\frac{2}{M_{{\rm Pl}}^2} \Pi_{ij}^{TT}\left(\bold{k},t\right)\,,
 \end{equation}
where a dot denotes derivative with respect to {\emph{t}}, {\emph{H}} is the Hubble rate, ${{M_{\rm Pl}}} \simeq 2.44\cdot 10^{18}$\,GeV is the reduced Planck mass, {\emph{k}} is the physical momentum, and $\Pi_{ij}^{TT}$ is the source of the GWs, corresponding to the transverse-traceless part of the anisotropic stress $\Pi_{ij}$. The latter is given by $a^2 \Pi_{ij}\,=\,T_{ij}-p a^2\left(\delta_{ij}+h_{ij}\right)$, where $T_{ij}$ denotes the spatial components of the energy-momentum tensor of the additional sources and $p$ the background value of the pressure.
The amplitude of the GW background predicted whenever either of the circumstances $i), ii)$ or $iii)$ are met during inflation, can significantly overtake the irreducible GW signal~(\ref{eq:omegainfl}) due to quantum fluctuations. The latter are characterized by the same equation~\eqref{eq:GWeq} but with negligible anisotropic stress, $\Pi_{ij}^{TT}=0$ (in this case, tensor perturbations are generated by the fast accelerated expansion of the~Universe).

The possibility of detecting these inflation-related backgrounds with GW interferometers, is therefore very compelling. These scenarios represent a new source of GWs, with an amplitude much larger than the standard irreducible inflationary background,\footnote{Note that there are also alternative scenarios that may produce a large background of GWs, possibly accessible to LISA, see e.g.~\cite{Gasperini-ml-2016gre}. In this paper, however, we only focus on the inflation-related scenarios listed in page 5.} providing an attractive target for the upcoming first space-based GW observer, LISA, which will have the ability to probe a significant fraction of their parameter space.

In order to design the best configuration for the LISA mission, it becomes important to determine what information can be extracted from a detection (or an absence of it) of signals at the frequencies probed by LISA, underlining the importance of the complementarity with the CMB scales. In this paper we address, specifically for the LISA mission, the scientific goal of extracting information from the inflationary era, studying the parameter space compatible with a detection/non-detection of a GW signal with LISA. We have combined our results for LISA with independent constraints coming from other probes at different scales. From our analysis we will argue that measurements of a GW signal on the small scales accessible to LISA, will become of fundamental importance in order to provide constraints on tensor perturbations complementary to the CMB. Spanning 16 orders of magnitudes in frequency, from the CMB to the LISA frequencies, this represents a unique opportunity to test the latest stage of the inflationary period, to probe the couplings of the inflaton to the latter, the presence of extra fields besides the inflaton, and to probe the degree of violation of the inflationary consistency relation. Concretely, we focus on four well-motivated scenarios:
\begin{itemize}
\item {\emph{Particle production during inflation}}: in a broad class of well-motivated models of inflation the inflaton $\phi$ sources gauge fields via the coupling $\phi\,F^{\mu\nu}\,\tilde{F}_{\mu\nu}$. In its turn, the gauge field sources a population of GWs that generally have a blue spectrum and can therefore rise to an observable level at LISA scales. Contrary to astrophysical backgrounds, this population has a net chirality and is highly non-Gaussian.
\item {\emph{Spectator field(s) during inflation}}: if, besides the inflaton, some spectator field(s) are present during inflation, a \emph{classical} production of GWs can take place. The amplitude and spectral index of such GW background, turn out to be specified by the sound speed of the spectator field(s), as well as by  the time variation of the latter. Interestingly, this GW background is expected to be blue-tilted.

\item {\emph{Effective Field Theory (EFT) of space-reparametrization}}: when space reparameterization invariance is broken during inflation, the graviton can acquire a mass. Then the tensor spectrum can be blue and get enhanced at small scales, not because of interactions between the inflaton and other auxiliary fields, but due to the specific symmetry breaking pattern induced by the fields driving inflation.

\item {\emph{Primordial Black Holes (PBHs)}}: certain models of inflation can produce large peaks in the matter power spectrum, that later collapse forming primordial black holes upon horizon reentry, during the radiation-dominated era. These PBHs are clustered and merge within the age of the Universe, generating a stochastic background of GWs that could be detected by LISA. 

\end{itemize}

In this paper we will quantify the ability of LISA to probe inflation with gravitational waves. We will focus on the four well motivated scenarios cited above. The paper is structured as follows. In section~\ref{sec:LISAcurves} we discuss the LISA sensitivity to a stochastic background. In section~\ref{sec:ParticleProduction} we study the GW signal from particle production during inflation, in section~\ref{sec:Spectators} the GW signal from inflationary spectator fields, in section~\ref{sec:EFT} the GW production in the context of the effective field theory of inflation new symmetry patterns, and in section~\ref{sec:PBH} the GW production from merging of primordial black holes. In section~\ref{sec:conclusions} we summarize our results.

\section{LISA sensitivity to a stochastic background}
\label{sec:LISAcurves}

In 2013 the European Space Agency (ESA) approved a GW observer in space as the L3 mission. The main candidate for this mission is a space-borne interferometer based on the long-standing, ESA-NASA joint project LISA (Laser Interferometer Space Antenna). The goal of the LISA mission is to detect GWs in the frequency range $(10^{-5}-0.1)$ Hz with high sensitivity, see e.g.~ref.~\cite{Seoane-ml-2013qna} and references therein. This frequency band is unexplored so far and very rich with both astrophysical and cosmological sources: the main target is the GW signal from massive black hole binaries (MBHB) (masses in the range $10^4-10^7\,M_\odot$) with high signal-to-noise ratio (SNR) and up to high redshift, see e.g.~ref.~\cite{Klein-ml-2015hvg} and references therein. However, low-mass black hole binaries, as those detected by LIGO in the range of few tens of solar masses, will also be visible far from merging~\cite{Sesana-ml-2016ljz,Barausse-ml-2016eii}, together with galactic binaries~\cite{Nelemans-ml-2013yg}, extreme mass ratio inspirals (EMRIs)~\cite{Gair-ml-2012vi}, and possibly a stochastic background from the early Universe~\cite{Caprini-ml-2015zlo}.

\begin{table}
\centering\renewcommand\arraystretch{1.2}
\begin{tabular}{|l|c|c|c|c|c|c|}
\hline
Name & A5M5 & A5M2 & A2M5 &  A2M2 & A1M5 & A1M2 \\ \hline
Arm length [$10^6$\,Km] & 5 & 5 & 2 & 2 & 1 & 1 \\ \hline
Duration [years] & 5 & 2 & 5 & 2 & 5 & 2 \\ \hline
\end{tabular}
\caption{\label{tab:configurations}The six representative
  LISA configurations chosen for the analysis (number of links fixed to six and noise level to N2 (for a definition, c.f~\cite{Klein-ml-2015hvg})), where in the notation $AiMj$, $i$ refers to the length of the arms in millions of Km and $j$ to the duration of the mission.}
\end{table}%

In 2015, in preparation for the L3 mission, ESA appointed the ``Gravitational Observatory Advisory Team'' (GOAT) to provide advice on the science return of a range of  possible configurations for the eLISA (evolved LISA) detector. Several analyses were then conducted on the scientific performance of different (e)LISA designs to specify the  science case: the present work is part of this series of papers. The first paper of this series dealt with the GW signal from massive black hole binaries~\cite{Klein-ml-2015hvg}, the second paper with the stochastic background from first order phase transitions occurring in the early Universe~\cite{Caprini-ml-2015zlo}, and the third one with the use of massive black hole binaries as standard sirens to probe the expansion of the Universe~\cite{Tamanini-ml-2016zlh}. A paper on the GW signal from EMRIs is in preparation, and other studies dealing with the scientific performances of (e)LISA have also been completed outside the series, see for example~\cite{Sesana-ml-2016ljz,Barausse-ml-2016eii,Bonvin-ml-2016qxr}. Here, we address specifically the potential of several LISA configurations to detect a stochastic background of GWs coming from inflation.

The variable characteristics of the (e)LISA configuration analysed in the aforementioned papers were the low-frequency noise level (N1 and N2, see~\cite{Klein-ml-2015hvg}), the number of laser links (4 or 6), the length of the interferometer arm (1, 2 or 5 million km), and the duration of the mission (2 or 5 years). Since then, a major achievement has been reached: the LISA Pathfinder satellite has flown and demonstrated that the expected instrumental noise in (e)LISA can be reduced six times below the original requirement~\cite{Armano-ml-2016bkm}. The noise that we adopt in this analysis is therefore the so-called N2 noise level~\cite{Klein-ml-2015hvg}: this has been tested by the pathfinder at frequencies $f > 1$ mHz, but the forecast is that it will be finally achieved over the whole frequency spectrum. Moreover, the outcome of the GOAT study accompanied by the renewed international interest in the (e)LISA mission, in particular from NASA, following both the first GW direct detection by the LIGO and Virgo collaborations and the successful flight of the Pathfinder, prompted the community to anticipate that the number of laser links of the future GW Observer can be six. Correspondingly, the name goes back to LISA. Therefore, in this work we consider six LISA configurations: having fixed the number of laser links to six (L6) and the best low-frequency noise level (N2), we let vary the length of the arms (A1, A2, and A5 for respectively 1, 2, and 5 million km) and the mission duration (M2 and M5, for respectively 2 and 5 years). Table~\ref{tab:configurations} summarizes the characteristics of these configurations.

\begin{figure}
\centerline{\includegraphics[width=12cm]{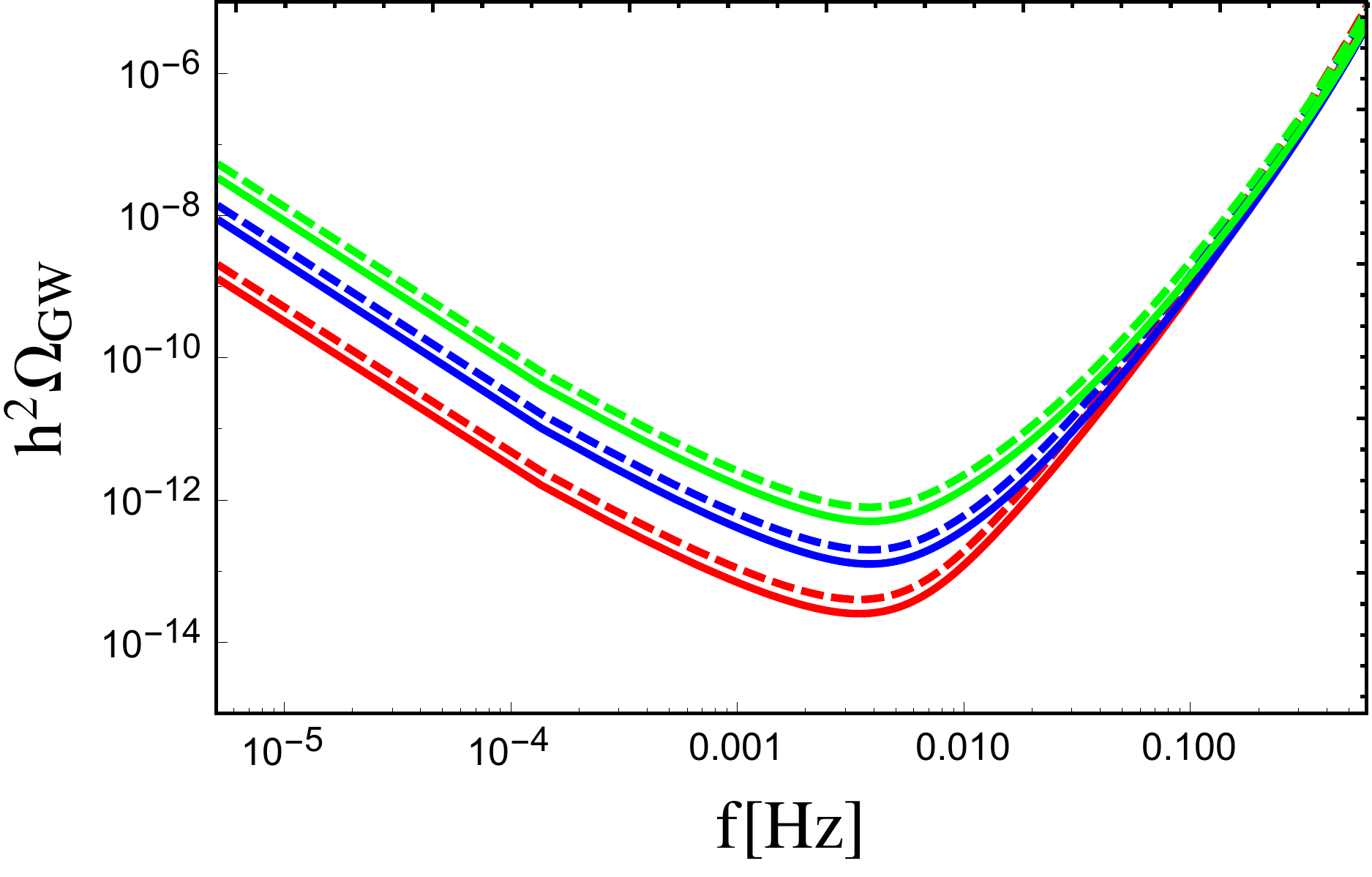}\qquad}
\caption{Power law sensitivity curves for the six LISA configurations considered in this work: red A5M5, red dashed A5M2, blue A2M5, blue dashed A2M2, green A1M5, green dashed A1M2. \label{FigPLSC}}
\end{figure}

\begin{figure}
\vspace{-2mm}
\centerline{
\includegraphics[width=7.2cm]{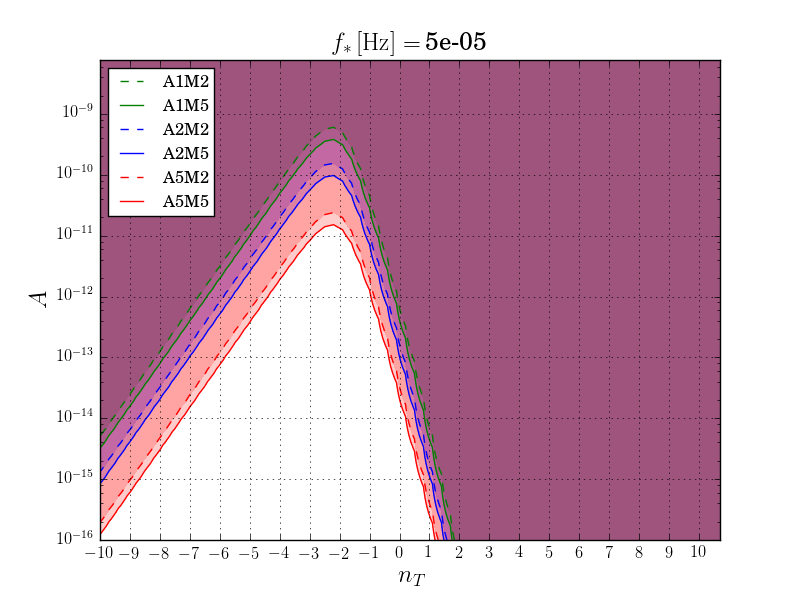}
\includegraphics[width=7.2cm]{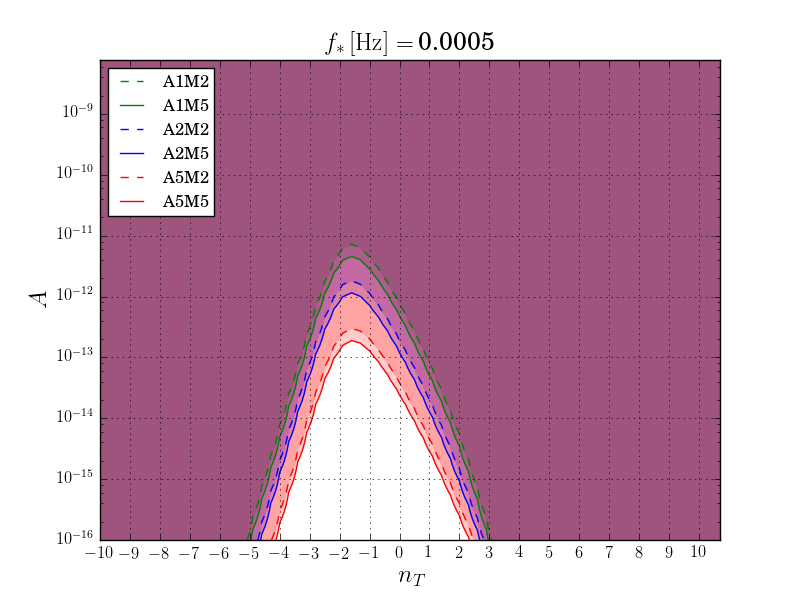}
}
\centerline{
\includegraphics[width=7.2cm]{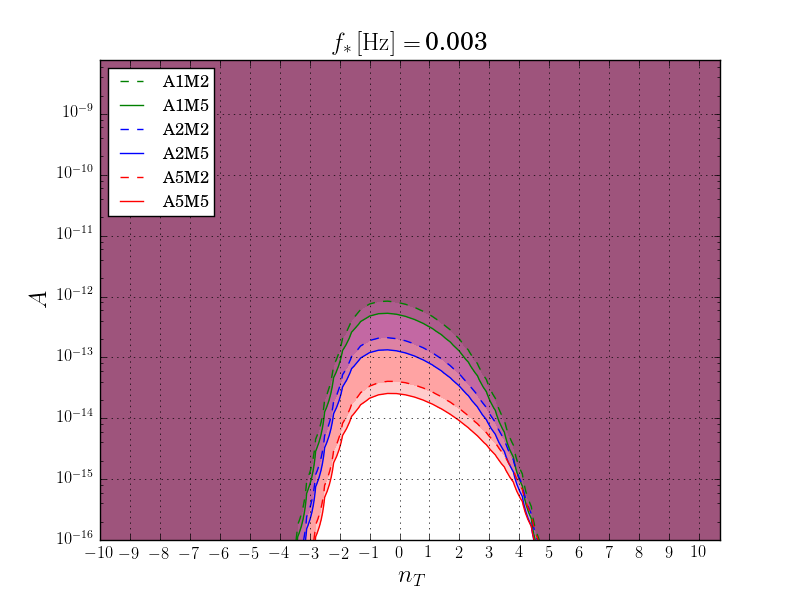}
\includegraphics[width=7.2cm]{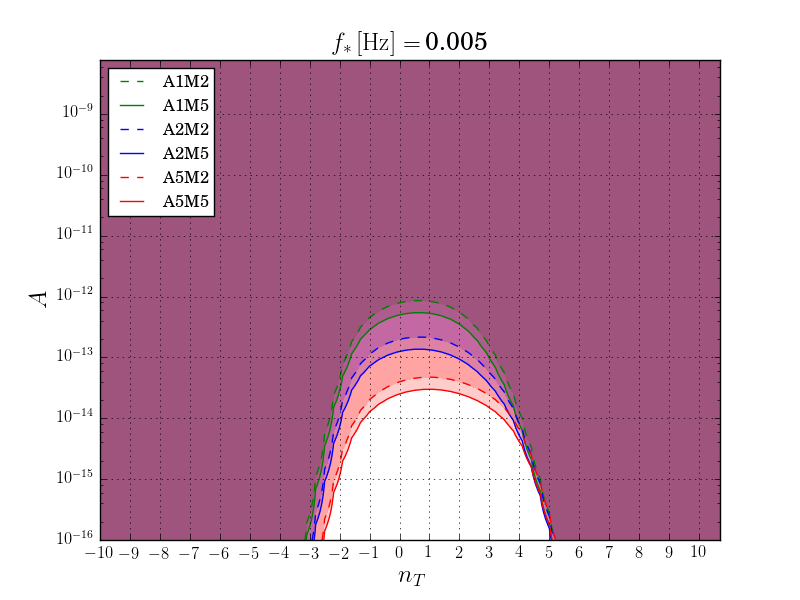}
}
\centerline{
\includegraphics[width=7.2cm]{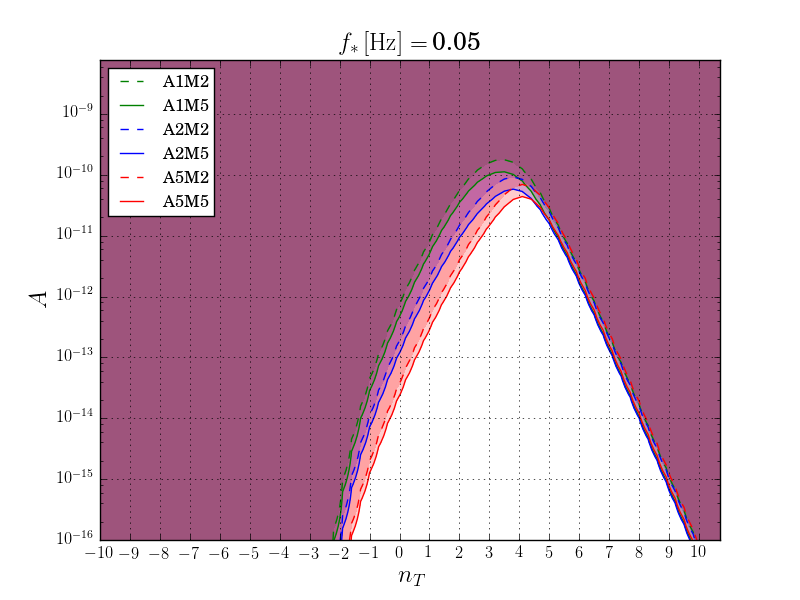}
\includegraphics[width=7.2cm]{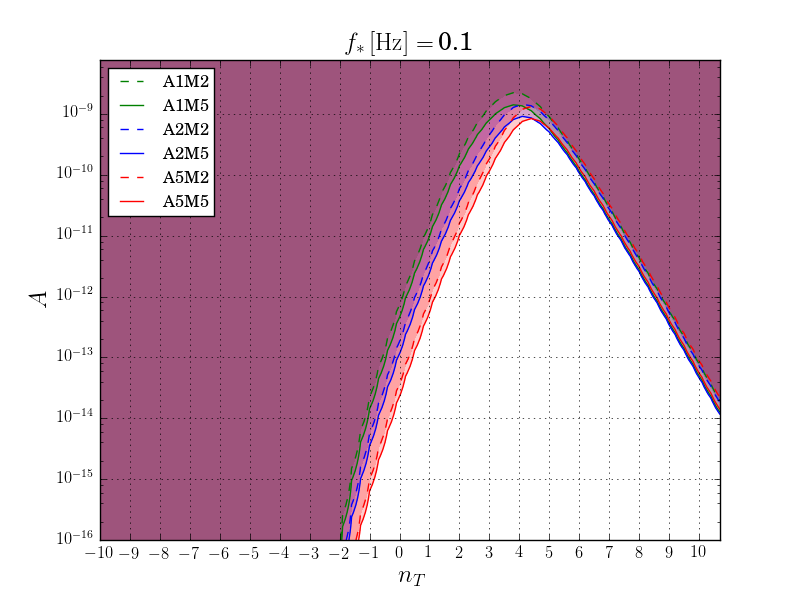}
}
\caption{For a power-law stochastic background of the form $\Omega_{\rm gw}=A(f/f_*)^{n_T}$, the shaded regions represent the detectable regions in the $(n_T\,,\,A)$ parameter space visible by the six LISA configurations under analysis: red A5M5, red dashed A5M2, blue A2M5, blue dashed A2M2, green A1M5, green dashed A1M2. We have chosen six representative pivot frequencies, \hbox{$f_*=0.05\,,\,0.5\,,\,3\,,\,5\,,\,50\,,\,100$~mHz.}}
\label{fig:FigAnT}
\end{figure}

The sensitivity curves to a stochastic background of GW have been discussed in~\cite{Caprini-ml-2015zlo} for four representative LISA configurations: two with four links and two with six links (for all configurations, a paper is in preparation~\cite{antoine}). We briefly revise the strategy adopted there to assess the detectability of a generic GW background, and present the new sensitivity curves of the six configurations under analysis here. Applying a Bayesian method, refs.~\cite{Adams-ml-2010vc,Adams-ml-2013qma} found that, over one year, the best 6-link configuration (with N2 noise level and 5 million km arms) can detect a white noise background at the level of $h^2\Omega_{\rm gw}=10^{-13}$. One can use this result and convert it into a threshold SNR above which the signal is visible. In order to do so, we compute for every LISA configuration the power law sensitivity curve defined in~\cite{Thrane-ml-2013oya}. With respect to the power law sensitivity curve, the SNR corresponding to a white noise spectrum with $h^2\Omega_{\rm gw}=10^{-13}$ is ${\rm SNR}=10$; we therefore classify every signal with  ${\rm SNR}>10$ as visible by a six-link LISA configuration. The power law sensitivity curves for the six configurations considered in this work are shown in figure~\ref{FigPLSC}.

In figure~\ref{fig:FigAnT} we present the detectability, by the six LISA configurations, of a generic GW background parametrised by a single power law, $\Omega_{\rm gw}=A(f/f_*)^{n_T}$. The regions in parameter space $(n_T\,,\,A)$, for several values of the pivot frequency $f_*$, have been derived applying the strategy described above, in particular they represent values of the parameters for which the signal is visible with ${\rm SNR}>10$. We have chosen representative values of the pivot frequency $f_*$, ranging from far smaller to far larger than the frequency of maximal sensitivity of the instrument configurations. Values of the spectral index close to zero are only visible for high enough amplitudes.

The parametrization of the GW energy-density spectrum by a power law opens the possibility to constrain cosmological parameters which are strictly connected with the inflationary period.
We expect the related GW background to cover a wide range of frequencies, from CMB scales up to the scales where laser interferometers are sensitive. Since current CMB measurements provide an upper bound on the inflationary GWs amplitude, it is useful to take into account such a constraint.
 In particular, we can constrain the GW spectral index $n_{T}$ and tensor-to-scalar ratio $r\equiv A_{T}/A_{S}$. Let us assume a power law spectrum as in eq.~(\ref{eq:GWen}), $\Omega_{\rm GW}(f) = \Omega_{\rm GW}^{\rm CMB}(f/f_{\rm CMB})^{n_T}$, but with $n_T$ not constrained to follow the consistency relation eq.~(\ref{eq:CR}) between $n_T$ and $r$. We can then re-express $\Omega_{\rm GW}^{\rm CMB}$ in terms of \emph{r} and the amplitude of the primordial scalar power spectrum at CMB scales, estimated by Planck~\cite{Ade-ml-2015lrj}. In this way we can combine constraints on \emph{r} and $n_{T}$ from the CMB scales with constraints to $\Omega_{\rm GW}(f)$ and $n_{T}$ from direct detection experiments, in particular obtained by current constraints from aLIGO, and with those expected by LISA. Up to now a constraint $n_{T}=0.06^{+0.63}_{-0.89}$ at $95\%$ C.L.~\cite{Meerburg-ml-2015zua} has been found\footnote{This constraint is determined  assuming a hypothetical detection of a tensor-to-scalar ratio at $0.01\,{\rm{Mpc^{-1}}}$ of $r_{0.01}=0.02$.} combining BICEP2/Keck Array and Planck (BKP), Planck $2013$, WMAP low $\ell$ polarization, HST data, Barion Acoustic Oscillations (BAO) measurements from SDSS and the upper limit on the energy density of stochastic GW background from LIGO. The most recent constraint on the
tensor-to-scalar ratio provided by BKP and other data gives $r_{0.05}<0.07$ ($95\%$ C.L.), at $0.05$ ${\rm{Mpc^{-1}}}$~\cite{Array-ml-2015xqh}, assuming the consistency relation of eq.~(\ref{eq:CR}) [$r=-8n_{T}$] of single-field slow-roll models of inflation.

\begin{figure}
\vspace{-2mm}
\centerline{\includegraphics[width=9.cm]{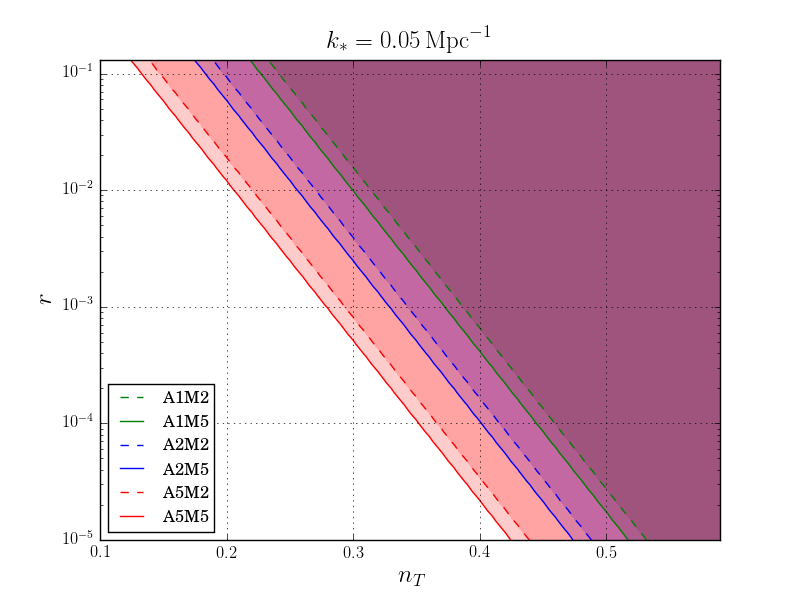}}
\vspace{-1mm}
\caption{Limits on the tensor spectral tilt $n_T$ and the tensor-to-scalar ratio $r$ for the six LISA configurations listed in table 1, assuming a power-law spectrum~\eqref{eq:GWen} (with arbitrary $n_{T}$, not~\eqref{eq:CR}) with reference scale $k_{*}=0.05\, \rm{Mpc}^{-1}$. This highlights the ability of LISA to test the $r - n_T$ relation.
\label{fig:rvsnt}}
\end{figure}

Recently, it has been shown how CMB experiments alone are not able to put strong constraints on the spectral tilt, finding $n_{T}\lesssim5$ at $95\%$ C.L. for $r_{0.01}=0.02$~\cite{Lasky-ml-2015lej}, even in the case of a detection of B-modes. CMB experiments focus on a narrow range of frequency around $10^{-17}${\rm{Hz}}; so, it becomes clear the importance of the combination of several experiments that cover different range of scales.
In~\cite{Lasky-ml-2015lej, Cabass-ml-2015jwe} it has been pointed out how the combination of GW experiments on a large range of frequencies, including ground and space-based interferometers, indirect measurements from CMB, BAO, and Big Bang Nucleosynthesis (BBN), puts stronger constraints on the tensor spectral index. In light of this we forecast the ability of LISA to  obtain constraints on the spectral tilt $n_T$, considering $r$ consistent with the current CMB upper bound. We repeat the analysis performed to obtain figure~\ref{fig:FigAnT}, relating the amplitude $A$ with the tensor-to-scalar ratio $r$, and $f_*=f_{\rm CMB}=7.7 \cdot 10^{-17}$ Hz (corresponding to the reference scale $k_{*}=0.05\, \rm{Mpc}^{-1}$), for the six LISA configurations under analysis. The results are shown in figure~\ref{fig:rvsnt}, where we can see that assuming the best LISA configuration with 6 links, 5 million km arm length and 5 year mission, we can constrain the spectral index up to $n_{T}\lesssim 0.35$. This number can be compared with the bound coming from initial LIGO O1, $n_{T}<0.54$ at $95\%$C.L. at a reference value $r_{0.05}=0.11$~\cite{Lasky-ml-2015lej}.
It becomes clear that this ability of LISA to constrain the tensor spectral tilt, becomes also a test for (strong) deviations of the consistency relation $n_T=-r/8<0$. In fact, as we know that all single-field slow-roll models of inflation follow the consistency relation, any evidence of a blue tilt $n_T > 0$ (necessary for a detection at the LISA frequencies), would be an indication of a deviation from single-field slow-roll models. As we will see in the next sections, this ability of LISA, is extremely flexible so it can be used for the different scenarios taken into account to put constraints on the related parameter space.


\section{Particle production during inflation}
\label{sec:ParticleProduction}

Requirements of \emph{radiative stability} play a crucial role in the construction of models of inflation, as they help discriminating between technically natural potentials, for which the properties in the full quantum theory are close, in a controllable way, to those of the classical theory, and models that require some form of fine tuning of the parameters. The most popular way of ensuring radiative stability of a scalar potential such as that of the inflaton $\phi$ is to assume a (softly broken) \emph{shift symmetry}, i.e.~an invariance under the transformation $\phi\to\phi+\phi_0$ with $\phi_0$ an arbitrary constant. One of the very few operators of low dimension that are allowed by an exact shift symmetry is the axionic coupling of the inflaton to a ${\rm U}(1)$ gauge field,
\begin{align}\label{eq:phiFF}
\Delta{\cal L}=-\frac{1}{4\,f}\,\phi\,F_{\mu\nu}\,\tilde{F}^{\mu\nu}\,,
\end{align}
where $1/f$ is a coupling constant with the dimension of a length. The operator~(\ref{eq:phiFF}) is thus expected to be generated in a large of class of technically natural models of inflation. Moreover, it leads to a very rich phenomenology. In fact, the equation of motion of the $\pm 1$-helicity modes  $A_\pm({\bf k},\,\tau)$ of the gauge field, in the presence of a time-dependent inflaton $\phi(t)$, reads~\cite{Anber-ml-2006xt}
\begin{align}\label{eq:apm_eulerlagrange}
\frac{d^2A_\pm({\bf k},\,\tau)}{d\tau^2}+\left[k^2\pm 2\,\xi\,\frac{k}{\tau}\right]\,A_\pm({\bf k},\,\tau)=0\,,
\end{align}
where we have defined
\begin{align}\label{eq:XiParam}
\xi\equiv\frac{\dot\phi}{2\,f\,H}\,,
\end{align}
and where $\tau$ denotes conformal time, whereas $\dot\phi$ is the derivative of the inflaton expectation value with respect to cosmic time $t$, and $H$ is the Hubble parameter. The presence of the $\pm$ sign in eq.~(\ref{eq:apm_eulerlagrange}) implies that for long wavelengths $-k\,\tau<2\,\xi$ one of the two helicity modes is exponentially amplified.  More specifically, an exact solution~\cite{Anber-ml-2006xt} of eq.~(\ref{eq:apm_eulerlagrange}) for constant $\xi$ shows that only the mode $A_+$ is amplified by a factor $\sim e^{\pi\,\xi}$ for\footnote{It is worth noting that in slow-roll inflation with negligible back reaction of the gauge field,  $\xi=\frac{M_{\rm Pl}}{f}\,\sqrt{\frac{\epsilon_\phi}{2}}$, where $\epsilon_{\phi}\equiv\frac{\dot{\phi}^2}{2H^2 M_{\rm Pl}^2}$, see eq.~(\ref{eq:f_epsilon}) below. Therefore, unless inflation happens at a very low scale (so that COBE normalization implies that $\epsilon\simeq \epsilon_\phi$ is many orders of magnitude below unity), a value of $f$ within an order of magnitude or two from $M_{\rm Pl}$ will lead to a value of $\xi\gtrsim {\cal O}(1)$.} $\xi\gtrsim {\cal O}(1)$. Here we assume (without loss of generality) that $\xi>0$, so that only positive helicity photons are amplified. The fact that only one of the photon helicities is amplified is reminiscent of the parity-violating nature of the operator~(\ref{eq:phiFF}) in the presence of $\dot\phi\neq 0$.

The exponentially large mode functions of one of the helicities of the gauge field act in their turn as a source of scalar perturbations (related to the perturbations in the inflaton, $\delta\phi$) and of gravitational waves $\delta g$, through processes  $A+A\to \delta\phi$ and $A+A\to \delta g$. Since the photon modes result from the amplification of Gaussian vacuum modes, and since  $\delta\phi$ and $\delta g$ are sourced by $2\to 1$ processes, the scalar and tensor perturbations sourced by the photons satisfy fully non-Gaussian statistics (i.e., their three point function is given, up to ${\cal O}(1)$ factors, by their two point function to the power $3/2$). The bispectrum of the sourced scalar perturbations has an approximate equilateral shape, with~\cite{Barnaby-ml-2010vf}\footnote{Loosely speaking, the non-linear parameter is defined by noting that the primordial curvature perturbations $\zeta$ are extremely close to Gaussian, and by parametrizing the departure from non-Gaussianity as $\zeta = \zeta_g + \frac{3}{5}f_{NL} \zeta_g^2$, where $\zeta_g$ is a Gaussian field; the precise relation is different for different shapes of non-Gaussianity; see for instance~\cite{Barnaby-ml-2010vf} for a more precise definition.}
\begin{align}
f_{NL}^{\rm {equil}}\simeq 7.1\times 10^5\,\frac{H^6}{|\dot\phi|^3}\,\frac{e^{6\,\pi\,\xi}}{\xi^9}\,,
\label{eq:fNL}
\end{align}
so that the non-observation of scalar non-Gaussianities by the Planck experiment at cosmological scales (that correspond to frequencies of the order of $10^{-17}$~Hz) implies a strong constraint $\xi\lesssim 2.5$ at $95\%$ C.L. at those scales~\cite{Ade-ml-2015lrj, Ade-ml-2015ava}. For such small values of $\xi$, the sourced gravitational waves are very weak and unobservable, see eq.~(\ref{eq:Omega1}) and (\ref{eq:XiForLargeSR}) below.

However, the quantity $\xi$ is time dependent~\cite{Cook-ml-2011hg}, which leads to a scale dependent spectrum of photons, and thus to a scale dependent spectrum of sourced metric perturbations. Remarkably, the quantity $\xi$ will generally increase as we go to shorter scales, since $|\dot\phi|$ increases and $H$ decreases as we approach the end of inflation. It is therefore possible that $\xi$ was smaller than $2.5$ or so when CMB scales left the horizon, so that the Planck constraints on non-Gaussianity and on the growth of the power spectrum~\cite{Meerburg-ml-2012id} are satisfied~\cite{Ade-ml-2015lrj}, but then grew to larger values later, when scales probed by LISA were amplified. Note that the evolution of $\xi$ as a function of time depends on the specific form of the inflaton potential. Moreover, it will also depend on the amount of back reaction of the produced photons on the rolling inflaton.

In this section we will discuss the prospects of detectability of these gravitational waves by LISA. In subsection~\ref{sec:spectrum} we provide a quick summary of the main properties of the system. In subsection~\ref{sec:local} we will use a local approach, parametrizing the dynamics that affect \emph{only LISA scales} to determine the properties of the gravitational waves generated by this mechanism. In subsection~\ref{sec:global}, on the other hand, we will deal with the constraints that emerge by considering the dynamics of the system during the entire observable $\sim 60$ e-folds of inflation. Finally, in Subection~\ref{sec:other_constraints} we discuss some additional potential constraints on the parameter space of the model.

\subsection{The spectrum of gravitational waves\label{sec:spectrum}}

For $\xi\gtrsim {\cal O}(1)$ the spectrum of gravitational waves is well approximated by~\cite{Barnaby-ml-2010vf,Sorbo-ml-2011rz}
\begin{eqnarray}\label{eq:pGW_sourced}
P_{\rm GW} \left( k \right) &\equiv& \frac{k^3}{2 \pi^2} \sum_{i=\pm} \left \vert h_{i} \left( k \right) \right \vert^2\nonumber\\
&=& P_{\rm GW,{\rm vacuum}} \left( k \right) +  P_{\rm GW,{\rm sourced}} \left( k \right)\nonumber\phantom{\frac{1}{1}}\\
&\simeq & \frac{2\,H^2}{\pi^2\,M_{\rm Pl}^2} +  8.7 \cdot 10^{-8} \frac{H^4}{M_{\rm Pl}^4}\frac{e^{4 \pi \xi}}{\xi^6} \,.
\end{eqnarray}
In this expression, $h_{\pm}$ denote the wave functions of two helicity modes $+$ and $-$ of a gravitational wave. The quantity $k$ denotes the wave number (or, equivalently, the comoving momentum) of the mode. Although we have not indicated it explicitly, the last expressions depends on $k$ since the values of $H$ and $\xi$ need to be evaluated when a given mode left the horizon during inflation. The GWs in these models are a sum of the ``vacuum'' gravitational waves (namely, those amplified by the expansion of the Universe; this is the standard term present in any model of inflation, which has an amplitude proportional to the Hubble rate) plus the ``sourced'' gravitational waves (namely, those produced by the vector modes through a  $A+A \rightarrow \delta g$ processes). The two terms are statistically uncorrelated, so that the total GW power spectrum is the power spectra of these two terms, without interference.

Figure~\ref{fig:local-nT} shows the energy in gravitational waves as a function of frequency for $f=M_{\rm Pl}/35$ in the case of a quadratic inflaton potential. One can notice three different regimes: \emph{(i)} at large scales ($f\lesssim 10^{-5}$~Hz in figure~\ref{fig:local-nT}) the ``standard" contribution from the amplification of vacuum fluctuations of the graviton $P_{\rm GW, vacuum}$ dominates over the sourced contribution; \emph{(ii)} at intermediate scales ($10^{-5}$~Hz$ \lesssim f\lesssim 1$~Hz in figure~\ref{fig:local-nT}) the sourced gravitational waves dominate, but the back reaction of the amplified gauge modes on the inflating background is negligible, so that the time-dependence of $\dot\phi$ and $H$ is determined by the standard slow-roll equations; \emph{(iii)} at smaller scales ($f\gtrsim 1$~Hz in figure~\ref{fig:local-nT}) the back reaction of the photons on the inflaton cannot be neglected any more. Since the production of photons draws energy from the kinetic term of the inflaton, it has the effect of slowing down the increase of $|\dot\phi|$, resulting into a flattening of $\Omega_{\rm GW}\,h^2$ as a function of the frequency $f$ at smaller scales.

It is also worth noting that in the simplest scenarios ($|\dot\phi|$ monotonically increasing, $H$ monotonically decreasing) the spectrum of sourced gravitational waves is generally blue. It is however possible to consider situations where $\xi$ has a transient, resulting into a localized bump in the spectrum of tensors~\cite{Namba-ml-2015gja,Garcia-Bellido-ml-2016dkw}.

\sloppy{Besides the amplitude~(\ref{eq:pGW_sourced}) of the GW power spectrum, it is worth mentioning that this background has two very distinctive properties, namely its chirality and its \hbox{non-Gaussian~statistics:}}
\begin{itemize}

\item \emph{Parity violation.} Eq.~(\ref{eq:pGW_sourced}) gives the total power in gravitational waves. This is given by the sum of the individual powers in left- and right-handed gravitational waves, which are given by~\cite{Sorbo-ml-2011rz}
\begin{align}
P^+_{\rm GW, sourced} \simeq \;& 8.7 \cdot 10^{-8} \frac{H^4}{M_{\rm Pl}^4}\frac{e^{4 \pi \xi}}{\xi^6} \,,\nonumber\\
P^-_{\rm GW, sourced} \simeq\;&  1.8 \cdot 10^{-10} \frac{H^4}{M_{\rm Pl}^4}\frac{e^{4 \pi \xi}}{\xi^6} \,.
\end{align}
The difference in amplitude between the spectra of the left- and the right-handed gravitational waves is a sign of the parity-violating nature of the system. The resulting GW background is therefore highly chiral, which represents a very distinctive signature of this scenario. Strategies for detecting a stochastic background of chiral gravitational waves with Earth-based detectors were discussed in~\cite{Crowder-ml-2012ik}.

\item \emph{The three point function.} As discussed above, all the metric perturbations  sourced by the modes of the gauge field obey  non-Gaussian statistics. Also the spectrum of gravitational waves enjoys this property. The three point function of the gravitational wave in this scenario, in the case of constant $\xi$, was computed in~\cite{Cook-ml-2013xea}. The shape of the three point function is close to equilateral, and in the exact equilateral configuration, $|{\bf{k}}_1|=|{\bf{k}}_2|=|{\bf{k}}_3|=k$, it is given by
\begin{equation}\label{eq:threept_GW}
\langle \hat{h}_+({\bf{k}}_1)\,\hat{h}_+({\bf{k}}_2)\,\hat{h}_+({\bf{k}}_3)\rangle_{\rm {equil}}=6.1\times 10^{-10}\,\frac{H^6}{M_{\rm Pl}^6}\,\frac{e^{6\pi\xi}}{\xi^9}\,\frac{\delta^{(3)}({\bf{k}}_1+{\bf{k}}_2+{\bf{k}}_3)}{k^6}\,,
\end{equation}
which, as mentioned above, is approximately given by the two-point function, eq.~(\ref{eq:pGW_sourced}), to the power $3/2$. Constraints on the tensor bispectrum of eq.~\eqref{eq:threept_GW} have already  been obtained from the CMB bispectrum~\cite{Ade-ml-2015ava}, yielding a constraint on the parameter $\xi$ on CMB scales in agreement with the one obtained from the measurements of CMB scalar bispectrum~\eqref{eq:fNL}.
Possible detectability of non-Gaussian primordial gravitational waves at interferometers has been discussed in~\cite{Thrane-ml-2013kb}. The one-to-one correspondence of the amplitude of the signal~(\ref{eq:pGW_sourced}) and its three-point function~(\ref{eq:threept_GW}) represents a characteristic signature of this scenario.

\end{itemize}

To summarize, the mechanism of field amplification from the coupling~(\ref{eq:phiFF}) is extremely interesting since \emph{(i)} it is inherent to well motivated models of inflation, and \emph{(ii)} it naturally leads to a signal that grows with time during inflation, allowing to probe stages of inflation that occurred well after the CMB 
modes were produced, on which we currently have little or no experimental information. A stochastic GW background produced by this mechanism would have very characteristic properties that could allow to distinguish it from an astrophysical background: it is chiral, it is highly non-Gaussian, and it is characterized by a universal and scale-independent ratio between the three- and two-point function,
\begin{equation}
k^6 \, \langle h^3 \rangle_{\rm equil}' \simeq 23 P_{\rm GW}^{3/2} \;,
\end{equation}
(where prime denotes the correlator without the delta-function). Finally, it has typically a significant blue spectrum at LISA scales, as we discuss in the next section.

\subsection{Local parametrization\label{sec:local}}

As can be seen from eq.~(\ref{eq:pGW_sourced}), at large $\xi$ the sourced GWs dominate over the vacuum ones,
\begin{equation}
P_{\rm GW} \simeq P_{\rm GW,sourced,L} \simeq  8.7 \cdot 10^{-8} \frac{H^4}{M_{\rm Pl}^4}\frac{e^{4 \pi \xi}}{\xi^6} \;\;\;\;,\;\;\;\; \xi \gg 1 \,.
\end{equation}
The power spectrum is related to the fractional GW energy density by~\cite{Barnaby-ml-2011qe} $\Omega_{\rm GW} h^2 = \frac{\Omega_{\rm R,0} h^2}{24} \, P_{\rm GW}$, where
$\Omega_{R,0} h^2 \equiv \rho_{{\rm R},0} h^2 / 3 H_0^2 M_{\rm Pl}^2 \simeq 4.18 \cdot 10^{-5}$ refers to radiation nowadays (one needs to include also the neutrinos as if they were still relativistic today). Therefore
\begin{eqnarray}
\Omega_{\rm GW} h^2 & \simeq 1.5 \cdot 10^{-13} \frac{H^4}{M_{\rm Pl}^4}\frac{e^{4 \pi \xi}}{\xi^6}  \;\;\;\;,\;\;\;\; \xi \gg 1 \,.
\label{eq:Omega1}
\end{eqnarray}
In this expression, $H$ and $\xi$ need to be evaluated when a mode left the horizon, and they are therefore functions of the wavenumber $k$, or, equivalently, of the frequency $f=k/2 \pi$ of the mode. The frequency is related to the number of e-folds by\footnote{To derive this relation, we used the fact that, by definition, a mode crosses the horizon when $k = a H$, and we have taken the ratio $\frac{k_{\rm CMB}}{k_N} = \frac{a_{\rm CMB} H_{\rm CMB}}{a_N H_N}$, where $k_{N}$ and $a_{N}$ are the wavenumber and scale factor at Hubble radius crossing {\emph{N}}-folds before the end of inflation.}
\begin{eqnarray}
N &=& N_{\rm CMB} + {\rm ln} \frac{k_{\rm CMB}}{0.002 \, {\rm Mpc}^{-1}} -40.3 - {\rm ln } \left( \frac{f}{\rm Hz} \right) + {\rm ln } \left( \frac{H_N}{H_{\rm CMB}} \right)\nonumber\\
&\simeq & 19.7- {\rm ln } \left( \frac{f}{\rm Hz} \right) + {\rm ln } \left( \frac{H_N}{H_{60}} \right) \;,
\label{eq:Nf}
\end{eqnarray}
where in the second expression we have assumed that the Planck pivot scale $k_{\rm CMB} = 0.002 \, {\rm Mpc}^{-1}$ exited the horizon at $N_{\rm CMB} = 60$. At any value of the frequency we define the spectral tilt
\begin{equation}
n_T \left( f \right) \equiv \frac{d \ln \Omega_{\rm GW} h^2 }{d \ln f} \;,
\end{equation}
which gives $ \Omega_{\rm GW} h^2 \propto f^{n_T}$ in the case of constant $n_T$. Differentiating eqs.~(\ref{eq:Omega1}) and~(\ref{eq:Nf}), and using $d N = - H \, dt$, we obtain
\begin{eqnarray}
n_T &=& \left( \frac{4 \dot{H}}{H^2} + \frac{4 \pi \dot{\xi}}{H} - \frac{6 \dot{\xi}}{H \xi} \right) \left( 1 +  \frac{ \dot{H}}{H^2} \right)^{-1} \nonumber\\
&=& \frac{ - 4 \epsilon + \left(4 \pi \xi - 6 \right) \left(  \epsilon - \eta \right)}{1-\epsilon}\nonumber\\ &\simeq & - 4 \epsilon + \left(4 \pi \xi - 6 \right) \left(  \epsilon - \eta \right) \;,
\label{eq:nT}
\end{eqnarray}
where in the first expression dot denotes differentiation with respect to cosmic time, and where in the second expression we have used the slow-roll parameters
\begin{equation}\label{eq:eps_eta_def}
\epsilon \equiv - \frac{ \dot{H}}{H^2} \;\;,\;\; \eta \equiv - \frac{\ddot{\phi}}{H \dot{\phi}}   \;.
\end{equation}
In the limit of negligible negligible back reaction of the gauge field on the background dynamics,   $\epsilon$ and $\eta$ become the two combinations $\epsilon \rightarrow \epsilon_V$ and $\eta \rightarrow \eta_V - \epsilon_V$ of the slow-roll parameters $\epsilon_V \equiv \frac{M_{\rm Pl}^2}{2} \left( \frac{V'}{V} \right)^2 ,\; \eta_V \equiv M_{\rm Pl}^2 \, \frac{V''}{V}$ defined from the inflaton potential.


%
\begin{figure}
\centerline{
\includegraphics[width=12.0cm]{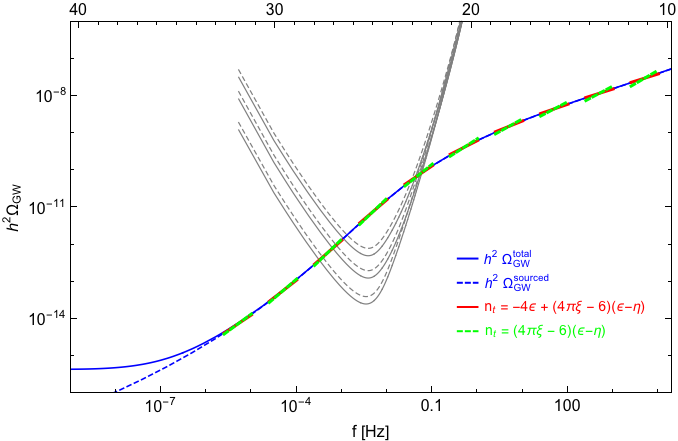}\quad
}
\caption{Spectrum of GWs today $h^2\Omega_{\rm GW}$ obtained from a numerical integration of the dynamical equations of motion (for a model of quadratic inflaton potential, with inflaton - gauge field coupling $f =  M_{\rm Pl} / 35$), versus the local parametrization $h^2\Omega_{\rm GW} \propto (f/f_*)^{n_T}$, evaluated at various pivot frequencies $f_*$ and with the spectral tilt $n_T$ obtained from successive approximations to the analytic expression~(\ref{eq:nT}).}
\label{fig:local-nT}
\end{figure}

In figure~\ref{fig:local-nT}, we compare the analytic expression~(\ref{eq:nT})  for the spectral tilt $n_T$ against the result of a numerical evolution of $\Omega_{\rm GW} h^2$. For definiteness, we choose a quadratic inflaton potential, and we fix the coupling between the gauge field and the inflaton to $f =  M_{\rm Pl} / 35$. This gives $\xi_{N=60} \simeq 2.46$ at the CMB scales. We observe from the figure that the final expression for the tilt in~(\ref{eq:nT}) provides a very good approximation (red segments in the figure) to the slope of the numerical result (blue solid line in the figure). The term $(1-\epsilon)$ in the denominator of~(\ref{eq:nT}), due to the fractional change of the Hubble rate $\dot H/H^2$, contributes to $n_T$ only to second order in slow-roll parameters, and hence we disregard it. The expression $n_T \simeq -4\epsilon + (4\pi\xi -6)(\epsilon-\eta)$ predicts correctly the slope of the numerical signal, within the LISA frequency range, to better than $\sim 4\%$. In the figure, the difference between the red segments and the true numerical signal cannot be distinguished by eye.

\begin{figure}
\centerline{
\includegraphics[width=7.0cm]{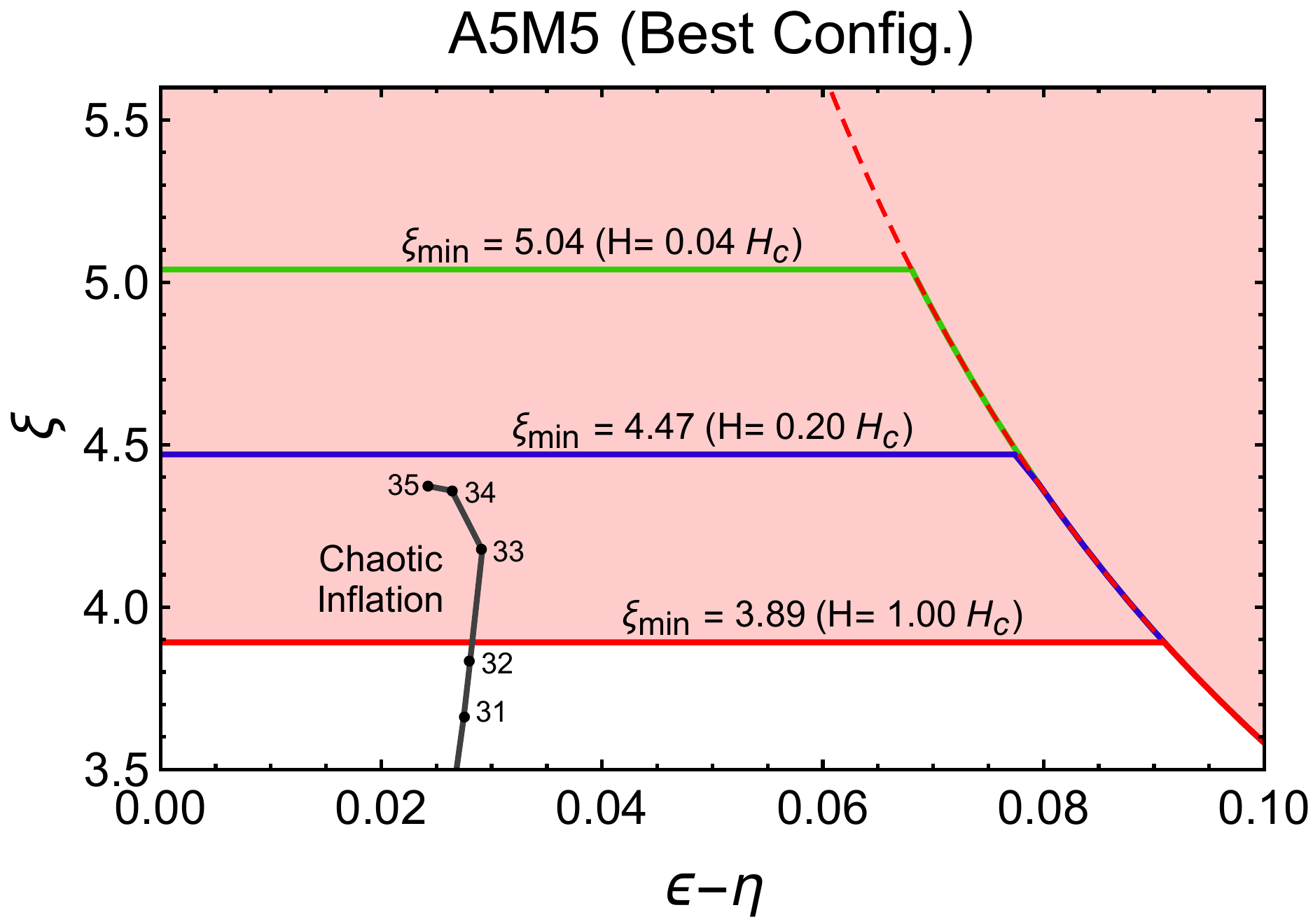}\qquad
\includegraphics[width=7.0cm]{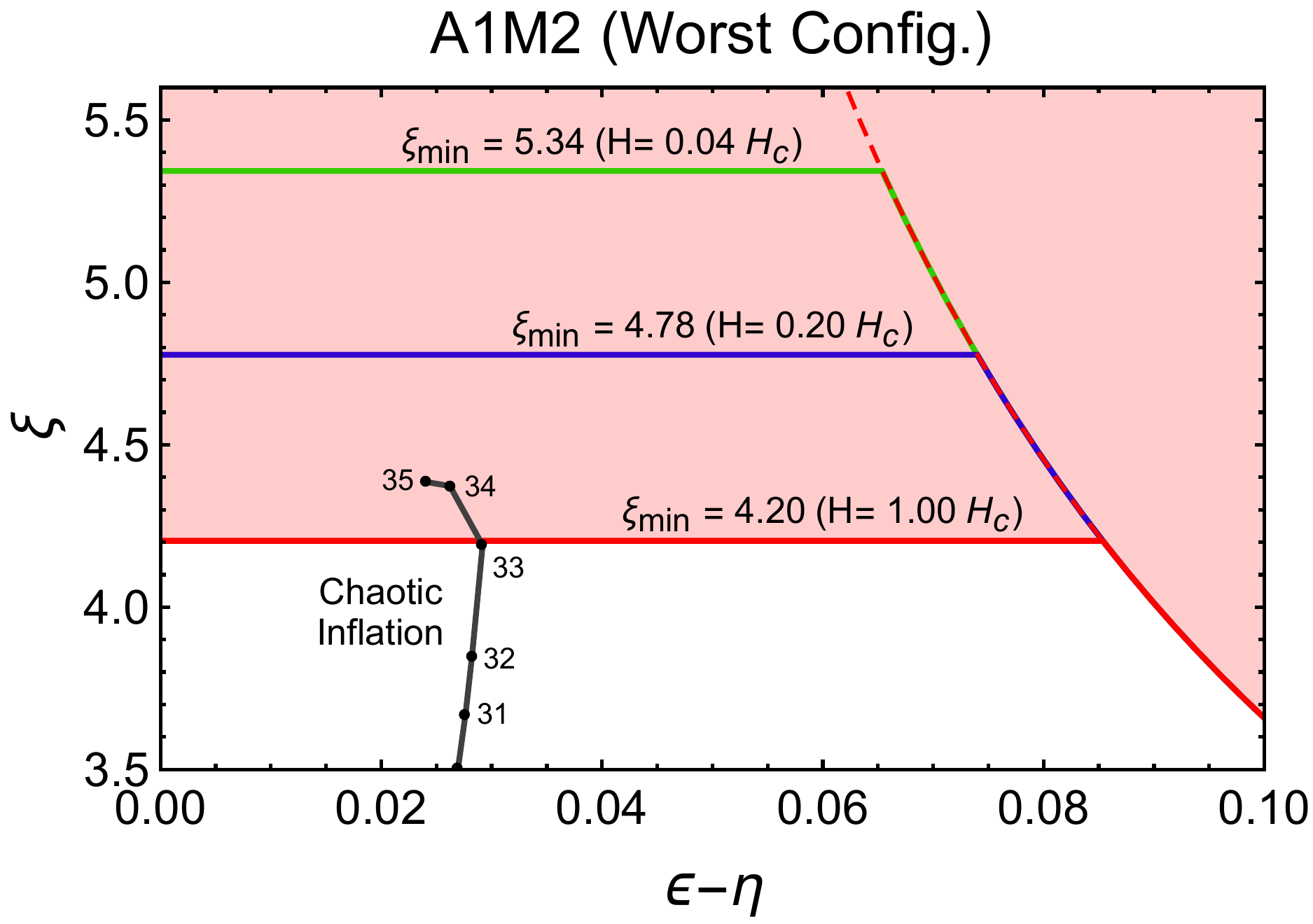}
}
\vspace{-2mm}
\caption{Region in the $(\xi,\epsilon-\eta)$ parameter space that LISA can probe, in the best configuration (left panel) and in the worst configuration (right panel). As a reference, we include the points corresponding to quadratic chaotic inflation for inflaton-gauge field coupling $M_{\rm Pl}/f = 35, 34, 33, 32$ and $31$. Note that the spectral index $n_T$, not shown in the figures to avoid to overcrowd them, is well approximated by the simple formula $n_T\simeq (4\pi\xi-6)\,(\epsilon-\eta)$.
\label{fig:Xi_vs_SR}}
\end{figure}

Let us note that for the range of $\xi$ that LISA can probe [$\xi \gtrsim 3.5$, see figure~(\ref{fig:Xi_vs_SR})], the term $-4 \epsilon$ in the final expression of~(\ref{eq:nT}) is actually negligible compared to the other terms. We can thus further approximate the expression for the tilt as $n_T \simeq  \left(4 \pi \xi - 6 \right) \left( \epsilon - \eta \right)$, which still predicts correctly the slope of the numerical signal within the LISA frequency range, for instance in the fiducial chaotic quadratic model to better than $\sim 10\%$. The advantage of using this simplified expression for the tilt is that it allows us to reduce the number of independent variables that the GW signal depends on, from $\lbrace H_N,\xi,\epsilon,\eta\rbrace$ to $\lbrace H_N,\xi,(\epsilon-\eta)\rbrace$. This simplifies our next goal, which is to obtain a model-independent parameter estimation based on the LISA sensitivity curves.

In figure~\ref{fig:Xi_vs_SR} we plot the region in the parameter space $(\xi,\epsilon-\eta)$ that LISA is capable of probing, with the left  and right panels depicting, LISA's best (A5M5) and worst (A1M2) configurations, respectively. In both panels we take as a pivot scale $f_*$ the frequency of the minimum of each LISA sensitivity curve $h^2\Omega_{\rm GW}^{(AiMj)}(f)$, with $f_{*}|_{_{A5M5}} \simeq 0.00346$ Hz and $f_{*}|_{_{A1M2}} \simeq 0.00390$ Hz. We then compute the minimum value $\xi$ required for a GW signal $h^2\Omega_{\rm GW}(\xi;f_*)$ to be above the minimum of the sensitivity curve, i.e.~$h^2\Omega_{\rm GW}(\xi \geq \xi_{\rm min};f_{*}|_{_{AiMj}}) \geq h^2\Omega_{\rm GW}^{(AiMj)}(f_{*}|_{_{AiMj}})$. For sufficiently small slow-roll parameters, $(\epsilon-\eta)\ll 0.1$, the answer is independent of the spectral tilt of the signal, and hence independent of the slow-roll parameters. This explains the horizontal lines marked as $\xi_{\rm min}$ in the plot. Of course, $\xi_{\rm min}$ depends on the inflationary Hubble scale $H_*$, evaluated at the e-fold $N_*$ corresponding to the pivot scale $f_*$, see~(\ref{eq:Nf}). In the two panels we also depict, as a reference, the $(\xi,\epsilon-\eta)$ behavior for our fiducial quadratic inflation model, evaluated numerically for $30 \lesssim M_{\rm Pl}/f \leq 35$. The Hubble rate in chaotic inflation with a quadratic potential at the e-fold $N_* \sim 25$ (corresponding to the frequencies $f_{*}|_{_{AiMj}}$) is $H_c \sim 2.6 \cdot 10^{-5} M_{\rm Pl} \simeq 6.4 \cdot 10^{13}$\,GeV\@. Taking this value as a reference, we see that LISA cannot probe any Hubble rate smaller than $\sim \mathcal{O}(10^{-2})H_c$, as a too large $\xi_{\rm min}$ is in tension with perturbativity requirements~\cite{Peloso-ml-2016gqs}. In particular, if we take $\xi_{\rm min} = 5.5$ as the maximum tolerated value at $N_* \simeq 25$, the minimum Hubble rate that can be probed by the different LISA configurations ranges from $H_{\rm min}^{(A5M5)} \simeq 6.3 \cdot 10^{11}$\,GeV to $H_{\rm min}^{(A1M2)} \simeq 1.5 \cdot 10^{12}$\,GeV.

When the slow-roll parameters are sufficiently large, it becomes possible that a GW signal with an amplitude at the pivot scale $f = f_{*}|_{_{AiMj}}$ smaller than the corresponding LISA sensitivity curve, i.e.~$h^2\Omega_{\rm GW}(f_{*}|_{_{AiMj}}) < h^2\Omega_{\rm GW}^{(AiMj)}(f_*|_{_{AiMj}})$, is yet observed thanks to its large spectral tilt $n_T$. Of course, having a scenario capable of producing such a signal becomes more and more contrived the larger the slow-roll parameters: having large slow-roll parameters at $N_* \sim 25$ requires a more complicated inflaton potential to sustain the final number e-folds of inflation, and also a GW background with such a large tilt requires a mechanism to prevent  any further growth of the GW amplitude at higher frequencies, otherwise this would violate the BBN bound [see eq.~(\ref{eq:BBNbound})]. For simplicity, we will restrict ourselves to $(\epsilon-\eta) \leq 0.1$, with $(\epsilon-\eta) = 0.1$ already quite a large value. Considering a GW signal with amplitude smaller than the LISA sensitivity curve at the corresponding frequency of the minimum $f = f_{*}|_{_{AiMj}}$, we find the minimum tilt $n_T$, and hence the minimum slow-roll parameter combination $(\epsilon-\eta)_{\rm min}$, required for the amplitude of the signal to cross the LISA curve at a higher frequency $f > f_{*}|_{_{AiMj}}$. For simplicity, we have measured the slope of the LISA sensitivity curves at a frequency $f = 10\cdot f_{*}|_{_{AiMj}}$, which we refer to as $n_{t,AiMj}$. For GW signals with amplitude $h^2\Omega_{\rm GW}(f_{*}|_{_{AiMj}}) < h^2\Omega_{\rm GW}^{(AiMj)}(f_{*}|_{_{AiMj}})$ at $f = f_{*}|_{_{AiMj}}$, we impose $n_T \geq n_{t,AiMj}$, and from the equality we obtain the dashed curve shown in both panels of figure~\ref{fig:Xi_vs_SR},
\begin{eqnarray}\label{eq:XiForLargeSR}
\xi \geq {1\over 4\pi}\left({n_{t,AiMj}\over(\epsilon-\eta)}+ 6\right)\,.
\end{eqnarray}
For $(\epsilon-\eta) = 0.1$ and for a Hubble rate as large as the one in chaotic inflation fiducial model, $H = H_c$, we deduce that $\xi \geq \xi_{\rm min,0.1}$, with $\xi_{\rm min,0.1} \simeq 3.58$ for A5M5 and $\xi_{\rm min,0.1} \simeq 3.66$ for A1M2. Note that in deriving~(\ref{eq:XiForLargeSR}) we have used the simplified expression $n_T \simeq (4\pi\xi-6)(\epsilon-\eta)$, neglecting the $-4\epsilon$ contribution. As~(\ref{eq:XiForLargeSR}) is only valid for a large slow roll parameter combination $(\epsilon-\eta) \geq 0.06$ [see curves at large $(\epsilon-\eta)$ in figure~(\ref{fig:Xi_vs_SR})], one should take the shape given in that equation only as representative indication of the effect of having a GW signal at $f_{*}|_{_{AiMj}}$ below the LISA sensitivity threshold. The exact form will depend on the exact scenario, and it is possible that the $-4\epsilon$ corrections may change to some extent the form of~(\ref{eq:XiForLargeSR}). However, at the qualitative level,~(\ref{eq:XiForLargeSR}) shows precisely what is expected, that for GW signals below the LISA sensitivity at $f = f_{*}|_{_{AiMj}}$, the minimum $\xi$ required to probe the signal with LISA is slightly smaller than the asymptotic constant $\xi_{\rm min}$ values obtained for small $(\epsilon-\eta)$ values.

\begin{figure}
\centerline{
\includegraphics[width=7.0cm]{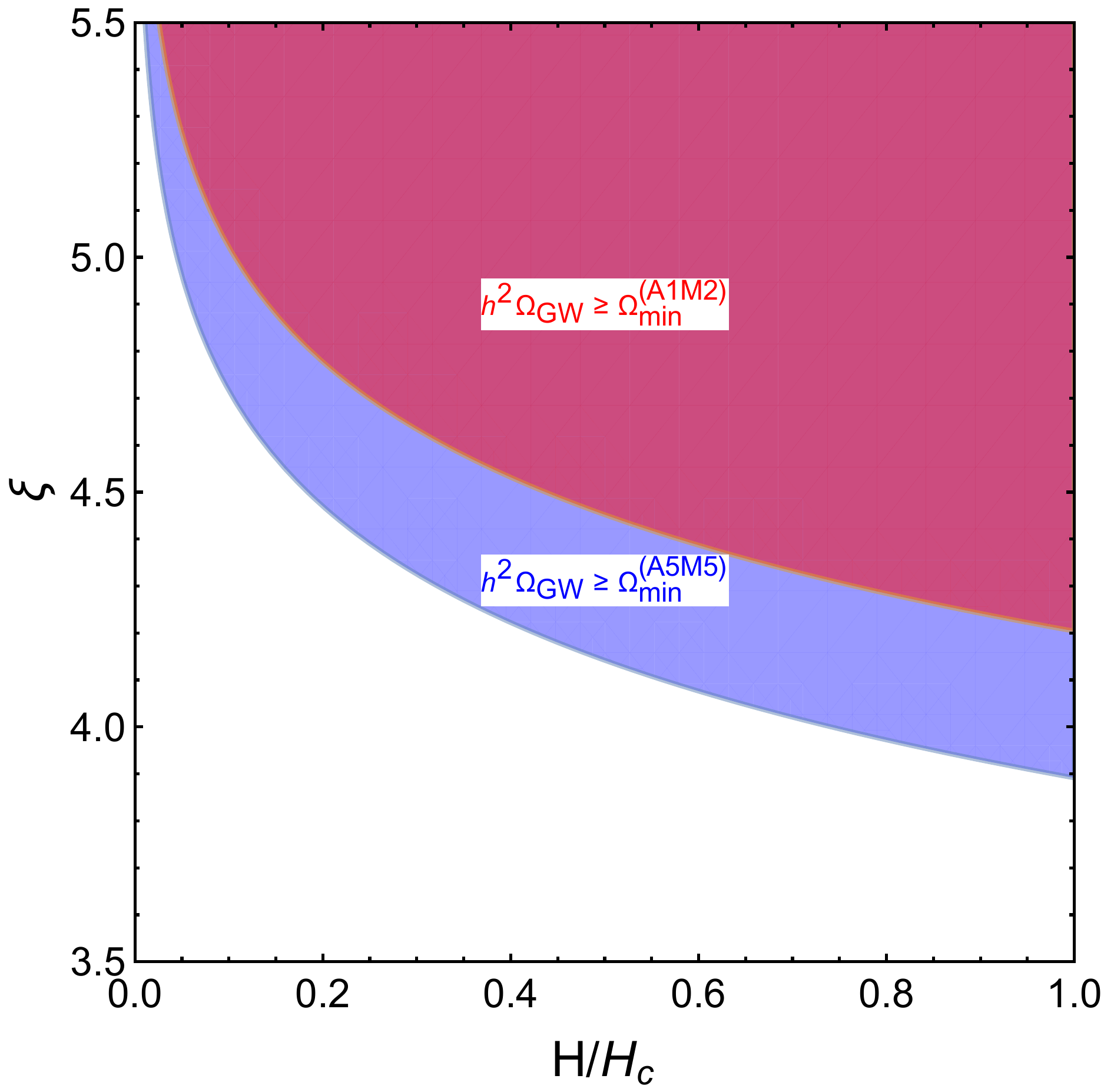}\qquad
\includegraphics[width=7.0cm]{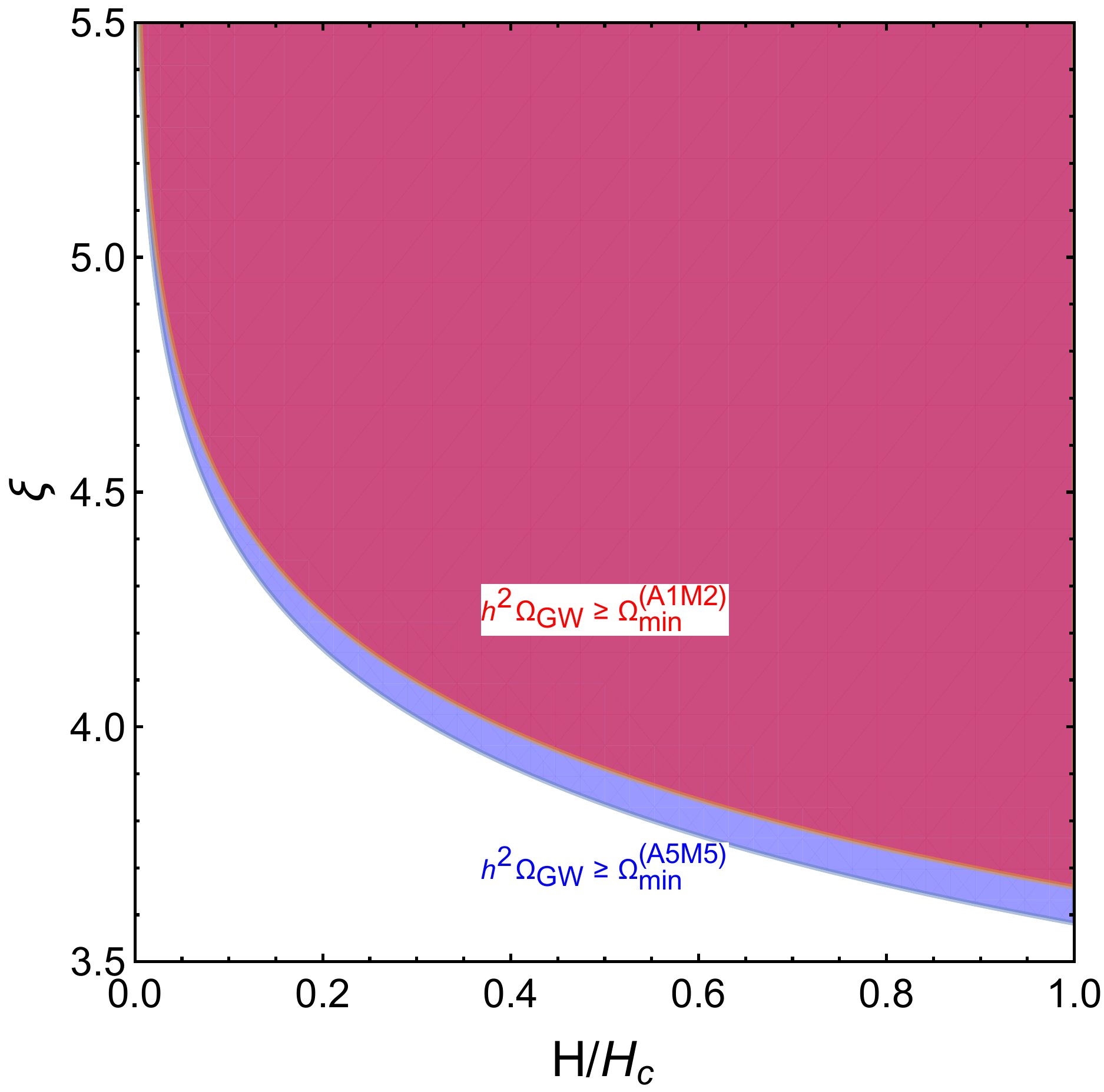}
}
\vspace{-2mm}
\caption{Regions in the $\xi-H$ parameter space that LISA can probe for the best (blue region) and worst (pink region) configurations. In the left panel we show the parameter space for $(\epsilon-\eta) = 0.02$, whereas in the right panel we show it for $(\epsilon-\eta) = 0.1$.\label{fig:Xi_vs_H}}
\end{figure}

Finally, we can also use the local parametrization to obtain contour regions in the $(\xi,H)$ plane of parameters, for fixed values of $(\epsilon-\eta)$. This does not contain more information than figure~(\ref{fig:Xi_vs_SR}), but it allows for an easy visualization of the ability of LISA to measure $\xi$ as a function of the Hubble rate $H$. We take again, as a reference, the Hubble rate $H_c \simeq 6.4\cdot 10^{13}$\,GeV at the e-fold $N_* \sim 25$ corresponding to LISA detection threshold (minimum of the sensitivity curves). We measure the Hubble rate in units of $H/H_c$. For a sufficiently small slow-roll parameter combination, say $(\epsilon-\eta) < 0.05 $ we are safely in the asymptotic regime where we just need $\xi > \xi_{\rm min}$ to guarantee a detection by LISA, independently of the actual value of $(\epsilon-\eta)$. In figure~\ref{fig:Xi_vs_H} we show the region of the $(\xi,H/H_c)$ parameter space compatible with a detection by LISA, for $(\epsilon-\eta) = 0.02$ in the left panel, and for $(\epsilon-\eta) = 0.1$ in the right panel. In each panel we show the region probed by the best and the worst configurations, A5M5 and A1M2. The two panels show clearly the degradation of LISA's ability to measure a signal for small inflationary Hubble rates, as the minimum $\xi$ required for a detection grows exponentially fast as the Hubble rate decreases.

To conclude our discussion on the local parametrization, let us note the following. When looking at figures~\ref{fig:Xi_vs_SR},~\ref{fig:Xi_vs_H}, one might be led to (misleadingly) conclude that LISA's best configuration (A5M5) does not represent much of an improvement compared to the worst configurations (A1M2), as the difference in $\xi$ probed by the two configuration is rather small, of the order $\Delta \xi \sim 0.31$. The difference in the amplitude of the GW probed by both configurations is however much larger, since the GW signal has an exponential sensitivity to $\xi$ as $\Omega_{\rm GW} \propto e^{4\pi\xi}$. A small difference $\Delta \xi \sim 0.31$ translates therefore into a GW amplitude boost factor of $\sim e^{3.9} \sim 10^2$. In other words, being capable of distinguishing small differences in $\xi$ is in fact quite relevant, as it may represent the difference between detecting and not detecting a given GW signal.

%
%



\subsection{Global parametrization \label{sec:global}}

The local parametrization discussed in the previous subsection can be used to study the phenomenology of an inflation model containing the coupling~\eqref{eq:phiFF} within a given observational window (in our case, the band of frequencies to which LISA is sensitive), while being agnostic about the inflaton potential at field values that do not impact these scales. This method focuses on what one can observe in a given experiment, making as few theoretical assumptions as possible about scales which cannot be probed in that experiment. This is for example similar to the reconstruction of the inflaton potential for a limited range of field values that can be done with CMB observations.

On the other hand, one may also choose to specify an inflaton potential, and combine the phenomenology associated to the interaction~(\ref{eq:phiFF}) at many different scales during the full observable $\sim 60$ e-folds of inflation. To analyze these effects (and to ensure that for simple inflation models one can indeed have an observable signal in the LISA band without violating any other constraints) we follow the approach of ref.~\cite{Domcke-ml-2016bkh}. Among the countless inflation models on the market, a vast amount of the single-field inflation models can be (approximately) described by the following ansatz for the first slow-roll parameter~\cite{Mukhanov-ml-2013tua},
\begin{equation}
\epsilon_V = \frac{1}{2} \left(\frac{V_{,\phi}}{V}\right)^2 = \frac{\beta}{N^p} \,,
\label{eq:eps_ansatz}
\end{equation}
thereby classifying inflation models into universality classes according to their value of $p$.
Here $\beta$ is typically an ${\cal O}(1)$ parameter, in the case of chaotic inflation it is e.g.~related to the power of the inflaton field.
This ansatz can be seen as the leading term in an expansion in the number of e-folds from the end of inflation $N$, following the observation that at the CMB-scales ($N \sim 60$) the slow-roll parameters are measured to be very small, whereas they need to become large at the end of inflation ($N = 0$) to guarantee a graceful exit. The ansatz~\eqref{eq:eps_ansatz} covers a wide range of well-motivated inflation models such as chaotic inflation ($p = 1$), supersymmetric hybrid inflation ($p = 1$), Starobinsky inflation ($p = 2$) and hilltop inflation ($p > 2$). Starting from eq.~\eqref{eq:eps_ansatz}, the corresponding scalar potential is uniquely determined~ up to a constant of integration $V_0$ which determines the overall normalizaton of the potential~\cite{Mukhanov-ml-2013tua}. As a result, starting from eq.~\eqref{eq:eps_ansatz}, the entire inflationary dynamics including the full scalar and tensor perturbation spectra are determined by four parameters only: $f$, $p$, $\beta$ and $V_0$.

In ref.~\cite{Domcke-ml-2016bkh}, it was found that among the discrete values of $p$ studied, $p = \{1,2,3,4\}$, the most promising candidates for future observations are the Starobinsky class ($p = 2$) followed by the chaotic class ($p = 1$). This is due to two competing effects which sensitively depend on $p$. Firstly, as long as the back reaction of the gauge fields is negligible,
%
%
%
\begin{equation}\label{eq:f_epsilon}
\xi \simeq \frac{M_{\rm Pl}}{\sqrt{2}\,f}\sqrt{\epsilon_V} \propto N^{-p/2} \,.
\end{equation}
Since the GW spectrum depends exponentially on the parameter $\xi$, this implies that the growth of the spectrum towards higher frequencies is much faster for larger values of $p$ - and hence these large-$p$ models are more likely to yield an observable GW signal.
However secondly, for $p > 1$, eq.~\eqref{eq:eps_ansatz} implies~\cite{Mukhanov-ml-2013tua}
\begin{equation}
n_s \simeq 1 - \frac{p}{N + 1} \,.
\end{equation}
Hence models with a large value of $p$ tend to have a too low spectral index, a situation which is aggravated in the presence of a sizable pseudoscalar coupling to gauge fields: the effective friction term in the equation of motion for the inflaton implies that the CMB observables are evaluated further down the scalar potential (compared to the situation in absence of~\eqref{eq:phiFF}), thus further decreasing the spectral index. When combining these two effects, the $p=2$ case was found to yield the largest GW signal while respecting the constraints on the CMB observables.

\begin{figure}
\centerline{\renewcommand\subfigcapskip{2mm}
\subfigure[CMB constraints and LISA sensitivity]{\includegraphics[width=.45\columnwidth]{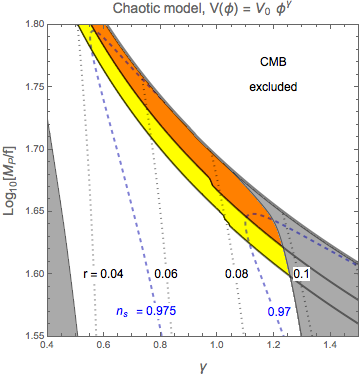}} \quad
\subfigure[Spectral tilt]{\includegraphics[width=.45\columnwidth]{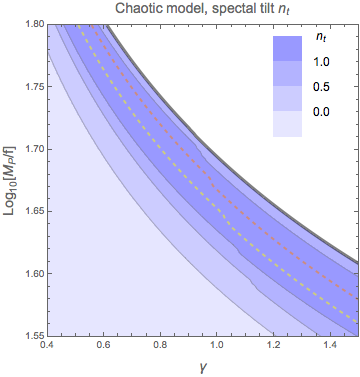}}
}
\caption{Left panel: reach of the A5M5 (yellow) and A1M2 (orange) LISA configurations, compared to region of parameter space probed by the CMB. The non-observation of non-Gaussianities  and $\mu$-distortions in the CMB excludes the region on the top right ($\xi_\text{\rm CMB} > 2.5$), whereas the gray region is is excluded at $95\%$~CL by the CMB constraints on $n_s$ and $r$. Right panel: spectral tilt of the GW spectrum in the same parameter space. For reference, the dashed yellow (orange) line show the reach of the A5M5 (A1M2) LISA configuration.}
\label{fig:chaotic}
\end{figure}

\begin{figure}
\centering\renewcommand\subfigcapskip{2mm}
\subfigure[CMB constraints and LISA sensitivity]{\includegraphics[width=.45\columnwidth]{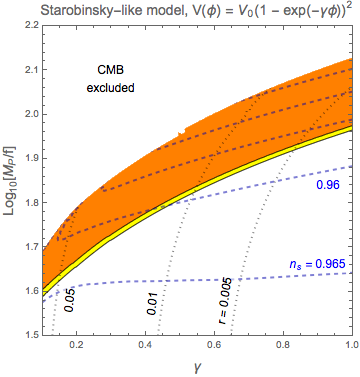}} \quad
\subfigure[Spectral tilt]{\includegraphics[width=.45\columnwidth]{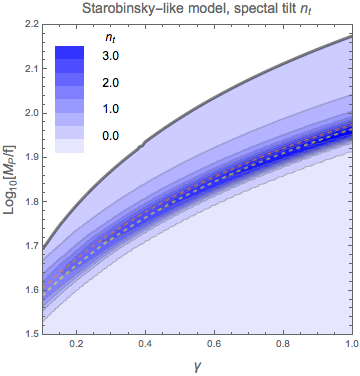}}
\caption{Left panel: reach of the best (yellow) and worst (orange) LISA configuration, compared to region of parameter space probed by the CMB. Adapted from ref.~\cite{Domcke-ml-2016bkh}. Right panel: spectral tilt of the GW spectrum in the same parameter space. Color coding as in figure~\ref{fig:chaotic}.}
\label{fig:starobinsky}
\end{figure}

In light of these results, we focus here on two representative examples: the chaotic class ($p = 1$) and the Starobinsky class ($p = 2$):
\begin{align}
\text{chaotic }(p=1):  \quad V(\phi) &= V_0 \phi^\gamma   && \rightarrow \quad \beta =  \gamma/4 \,,\\
\text{Starobinsky }(p=2): \quad V(\phi) &= V_0 (1 - e^{- \gamma \phi})^2  && \rightarrow \quad \beta =  1/(2 \gamma^2) \,.
\end{align}
Of the remaining three parameters ($f$, $\gamma$ and $V_0$) one parameter can be eliminated by imposing the COBE normalization on the amplitude of the scalar perturbation spectrum at the CMB scale. In figures~\ref{fig:chaotic} and~\ref{fig:starobinsky} we depict a selection of CMB observables as well as the amplitude and tilt of the GW spectrum in the LISA band in terms of the two remaining parameters $\gamma$ and $f$.

The relevent CMB observables are the scalar amplitude $A_s$, its spectral tilt $n_s$, the tensor to scalar ratio $r$, the equilateral non-Gaussianity parameter $f_\text{NL}^\text{equil}$ (see eq.~\eqref{eq:fNL}) and the level of $\mu$-distortion in the CMB black body spectrum.
These $\mu$-distortions are sensitive to the integrated scalar power spectrum in the range $50 \text{ Mpc}^{-1} \lesssim k \lesssim 10^4 \text{ Mpc}^{-1}$, corresponding to a frequency range of $10^{-15}~\text{ Hz} \lesssim f \lesssim 10^{-9}~\text{ Hz}$~\cite{Meerburg-ml-2012id},
\begin{equation}
\mu \simeq  \int_{k_{D}(z_i)}^{k_{D}(z_f)} d\ln k  \; A_s(k) \left[ e^{ - k/k_{D}(z)}\right]^{z_f}_{z_i} \,,
\end{equation}
with $k_D = 4 \times 10^{-6} z^{3/2} \text{ Mpc}^{-1}$ and $z_i = 2 \times 10^6$ ($z_f = 5 \times 10^4$)  denoting the redshift when the dominant inelastic (elastic) scattering processes for CMB photons freeze out.  The current bound from COBE / FIRAS constrains $\mu < 6 \times 10^{-8}$~\cite{Fixsen-ml-1996nj} at 95$\%$ CL.
The CMB observables $A_s$,  $n_s$ and $r$ are evaluated in the usual way, taking into account the reduced field excursion of the inflaton due to the gauge friction term. The GW amplitude and tilt are evaluated at the LISA peak sensitivity, $f \sim  4 \times 10^{-3}$~Hz. We have furthermore evaluated the GW amplitude in the LIGO band, however the current bound~\cite{TheLIGOScientific-ml-2016wyq} does not constrain the parameter space any further. See ref.~\cite{Domcke-ml-2016bkh} for details.

Figure~\ref{fig:chaotic} shows the results for the chaotic inflation models. The yellow/orange shaded regions in the left panel show the reach of LISA (best and worst configuration). The dotted and dashed lines show contours of the tensor-to-scalar ratio $r$ and spectral index $n_s$, respectively, with the gray shaded region disfavoured at 95$\%$~CL by the Planck data~\cite{Ade-ml-2015lrj, Ade-ml-2015ava}. In the gray region on the left of the plot this is mainly due to too large values of the spectral index, whereas the grey region on the right is mainly driven by the large values of the tensor-to-scalar ratio. The white region on the top right is excluded as it produces too large non-Gaussianities (the bound arising from $\mu$-distortions is slightly weaker). Together, this emphasizes the powerful complementarity between CMB experiments and direct gravitational wave detectors.

The right panel of figure~\ref{fig:chaotic} is dedicated to the tilt of the tensor power spectrum. For reference, we show again the region excluded by the CMB constraint on non-Gaussianity and $\mu$-distortions (white region on the top right) and the reach of the best and worst LISA configuration (dashed yellow and dashed orange). There is a clear correlation between the amplitude and the tilt of the GW spectrum in the LISA band. For example (at $\gamma = 1.0$), for a GW amplitude which is marginally detectable by the best (worst) LISA configuration, we find a prediction of $n_T \simeq 1.21$ ($1.20$), while for a GW amplitude just below the current non-Gaussianity bound, we find $n_T \simeq 0.8$. The strong correlation between the GW amplitude, the tilt and size of the non-Gaussianities can be traced back to the parameter $\xi$, which is the main parameter dependence of all these quantities.

Figure~\ref{fig:starobinsky} shows the analogous analysis for the Starobinksy class of models. As in figure~\ref{fig:chaotic}, the left panel is dedicated to contrasting the CMB constraints with the reach of LISA, whereas the right panel shows the spectral tilt of the GW spectrum. Contrary to the case discussed above, the entire parameter region shown is now within the $95\%$~CL contour of the Planck $n_s - r$ - data~\cite{Ade-ml-2015lrj}. The white region on the top left is excluded by bounds on non-Gaussianities and $\mu$-distortions, where in this case the latter yield the slighter stronger bound. Again we observe that the spectral tilt and the amplitude are highly correlated. However, due to the larger value of $p$ in this case, the growth of the GW spectrum is steeper and hence the maximal spectral tilt is larger. For example (at $\gamma = 0.5$), a spectrum marginally detectable with the best (worst) LISA configuration corresponds to a spectral tilt of $n_T \simeq 3.2$ (2.8), whereas just below the non-Gaussianity bound we find $n_T \simeq 0.2$.

%

In summary, we stress the remarkable complementarity between the CMB observables $n_s$, $r$, $f^\text{equil}_{NL}$, $\mu$ and the parameter range probed by direct GW searches. In both the chaotic and the Starobinsky class of models, this allows to constrain the parameter space from different sides. Moreover, the spectral tilt $n_T$ is found to be an approximately universal function of the amplitude $\Omega_{\rm GW}(f)$ within a given universality class, while functioning as a discriminator between different universality classes. Finally, in both cases studied, the absolute value of the spectral tilt immediately indicates if the LISA band lies in the saturation regime of strong back reaction ($n_T = {\cal O}(0.1)$), in the regime of dominated by the vacuum fluctuations ($n_T \lesssim 0$) or in the intermediate regime featuring a highly blue \hbox{spectrum~($n_T = {\cal O}(1)$).}


\subsection{Other constraints}\label{sec:other_constraints}

Beyond the GW signal and the constraints on the CMB observables $f^\text{equil}$, $n_s$, $r$ and $\mu$ discussed above, there are a number of further potentially observable features of this class of models. In this subsection, we compare the potential of these channels to probe the parameter space to the reach of LISA.
\begin{itemize}

\item Primordial black holes (PBHs). The gauge fields amplified by the interaction~(\ref{eq:phiFF}) source both gravitational waves and scalar density perturbations. Due to their non-gaussian nature, sourced scalar perturbations with amplitude
$A_s = {\rm O } \left( 10^{-4} - 10^{-3} \right)$ (the precise value depending on the scale) can lead to PBHs~\cite{Linde-ml-2012bt} in excess to the current bounds~\cite{Carr-ml-2009jm,Carr-ml-2016drx}. Ref.~\cite{Linde-ml-2012bt} showed that, for a chaotic inflationary potential, these limits force the gauge field amplification to be too small for the sourced gravitational waves to be observable at interferometers. This conclusion requires several qualifications, as we now discuss.

The study of scalar perturbations of~\cite{Linde-ml-2012bt} is based on the approximate equation~\cite{Anber-ml-2009ua}
\begin{equation}
\delta \ddot{\phi} + 3\, \beta\, H \,\delta \dot{\phi} + k^2 \delta \phi + V'' \delta \phi \simeq \frac{\vec{E} \cdot \vec{B} - \langle \vec{E} \cdot \vec{B} \rangle}{f} \;,
\label{deltaphi-beta}
\end{equation}
with $\beta \equiv 1 - \frac{2 \pi \xi}{f} \, \frac{\langle \vec{E} \cdot \vec{B} \rangle}{3 H \dot\varphi}$. The right hand side of this relation accounts for the inverse decay process $\delta A + \delta A \rightarrow \delta \phi$. The departure of $\beta$ from $1$ is due to the fact that the amplitude of the vector field depends on $\xi$, which, being proportional to $\dot{\phi}$, receives corrections from the time derivative of the inflation perturbations,  $\delta \xi = \frac{\delta \xi}{\delta \left( \delta\dot{\phi} \right)} \delta\dot{\phi}$. This introduces a damping on the growth of the inflaton perturbations, which is analogous to the friction that the gauge fields cause to the background evolution of the inflaton. This effect is significant for the values of $\xi$ necessary to produce PBHs. As remarked in~\cite{Linde-ml-2012bt}, it is possible that, when this happens, additional interactions between the inflaton and gauge field perturbations, which are  not included in~(\ref{deltaphi-beta}), become important. If it is the case,  all conclusions on the scalar perturbations in this regime are affected by a ${\cal O } \left( 1 \right)$ uncertainty, which can be enough to make the PBHs limit unimportant~\cite{Linde-ml-2012bt}.  Ref.~\cite{Ferreira-ml-2015omg} proposed some conditions for the validity of perturbative computations of the scalar perturbations in this model. The conditions were re-analyzed in~\cite{Peloso-ml-2016gqs}, which showed that these criteria are satisfied for $\xi \la 4.8$. This is parametrically close to the values necessary to generate PBHs, so that   ${\cal O } \left( 1 \right)$ corrections are certainly a possibility.

Beside the intrinsic uncertainty associated with~(\ref{deltaphi-beta}) in the $\xi \gg 1$ regime, one should also keep in mind that the PBH limit of~\cite{Linde-ml-2012bt} is enforced by modes at the smallest possible scales at which these limits exist, namely $N \simeq 10$. This is due to the fact that  $\xi \propto  \dot{\phi} / H$ continues to grow in chaotic inflation, resulting in very blue spectra for the sourced perturbations.
LISA is mostly sensitive to modes at much greater scales, with the best sensitivity for $N \sim 25$. The PBHs limit at LISA scales does not preclude the GW signal to be large enough to be visible at LISA, and it is possible that, in a different model from chaotic inflation, the inflaton potential causes the inflaton to slow down between $N = 25$ and $N=10$, so that the PBHs bounds are never evaded~\cite{Garcia-Bellido-ml-2016dkw}. Finally, the PBHs limit can be easily evaded if a number ${\cal N} > 1$ of gauge fields are amplified by this mechanism.  The different gauge fields act as an incoherent source, that amplifies the gravitational wave spectrum by a factor of ${\cal N}$. The same amplification takes place for the scalar modes. However, the parameter $\beta$ in~(\ref{deltaphi-beta}) also grows by  a factor of ${\cal N}$ in the $\xi \gg 1$ regime, giving rise to a $1/{\cal N}^2$ suppression to the scalar power spectrum. Altogether, the scalar power spectrum is therefore suppressed by a $1/{\cal N}$ factor~\cite{Anber-ml-2009ua}. As a consequence, the ratio between the tensor and scalar perturbations grows as ${\cal N}^2$, so, that, at sufficiently large ${\cal N}$, one can be sure to obtain a sufficiently large GW signal and sufficiently few PBHs. In fact, for models of chaotic (Starobinsky) inflation, already at ${\cal N} = 5$ (${\cal N} = 10$), one can obtain a visible GW at LISA without violating the PBHs limit~\cite{Domcke-ml-2016bkh,Garcia-Bellido-ml-2016dkw}.

\item The effective number of massless degrees of freedom $N_\text{eff}$ at the the time of BBN and of CMB decoupling. Since GWs contribute to the radiation energy density, the high-frequency part of the spectrum is constrained by
\begin{equation}\label{eq:BBNbound}
\int d (\ln f) \, \Omega_{\rm GW} \leq \Omega_{R,0} \frac{7}{8} \left( \frac{4}{11}\right)^{4/3} (N_\text{eff} - 3.046) \,,
\end{equation}
where the integral is performed over all frequencies $ f \gtrsim 10^{-15}$~Hz ($f \gtrsim 10^{-10}$~Hz) for the CMB (BBN)\@. The current bound from CMB reads $N_\text{eff} = 3.04 \pm 0.17$~\cite{Ade-ml-2015xua}, BBN constrains $N_\text{eff} = 3.28 \pm 0.28$~\cite{Cyburt-ml-2015mya}. For all inflation models with monotonously growing $\epsilon(\phi)$, the high-frequency end of the spectrum yields the largest contribution. The resulting bounds on the $f - \gamma$ parameter space of the Starobinsky-like models are discussed in~\cite{Domcke-ml-2016bkh}. Taken at face value, they exclude all the parameter space accessible to LISA. In the case of chaotic inflation $N_\text{eff}$ is typically smaller, but also in this case some parts of the parameter space are disfavoured by existing bounds.
However, the calculation of $N_\text{eff}$ relies sensitively on the high-frequency tail of the spectrum, far outside the LISA band. Here the strong back reaction of the gauge fields induces sizeable theoretial uncertainties, and we note that reducing the GW amplitude by an overall ${\cal O}(1)$ factor in this regime would avoid all current bounds while only marginally affecting the reach of LISA.

\end{itemize}

In summary, we stress that the observational channels listed above are powerful and highly complementary to direct GW searches with LISA. When comparing with the reach of LISA, there are however two main caveats: firstly, the constraints listed in this subsection rely on an extrapolation of the GW or scalar spectrum over many orders of magnitude of scales, which may be performed e.g.\ by assuming the validity of eq.~\eqref{eq:eps_ansatz} over these scales. However, the microphysics of inflation may be more complicated than this simple parametrization suggests, in which case this extrapolation may be misleading. Secondly, for both the PBH constraint and the $N_\text{eff}$ constraint, the main contribution arises from the high frequency tail of the spectrum, in which perturbative control of the theoretical calculation is poor at best~\cite{Peloso-ml-2016gqs}. This underlines the power of the local parametrization discussed in section~\ref{sec:local}, which focuses on the analysis purely within the LISA frequency band. On the other hand, these caveats may also be seen as features, since within this class of models, the above different observations provide access to different parts of the scalar potential of inflation, thus potentially providing a very powerful probe to learn about the microphysics of inflation.

\section{Gravitational waves from inflationary spectator fields}
\label{sec:Spectators}

Several works~\cite{Biagetti-ml-2013kwa,Biagetti-ml-2014asa,Fujita-ml-2014oba} have studied inflationary scenarios where other scalar fields, besides the inflaton, are present, even though they do not influence the inflationary background dynamics. These fields are, correspondingly, referred to as \emph{spectator fields}.\footnote{In literature, the name \emph{spectator field} is used to identify slightly different roles played by an extra scalar field. Here we refer to the one considered by~\cite{Biagetti-ml-2013kwa,Biagetti-ml-2014asa,Fujita-ml-2014oba}.} The presence of a spectator field, in particular its scalar perturbations, gives rise to a second-order source term in the equation of motion of GW, so that a classical production of tensor modes takes place\footnote{An analogous mechanism of GWs production takes place also in the curvaton scenario~\cite{Bartolo-ml-2007vp}.}~\cite{tomita,Matarrese-ml-1997ay}. Therefore, the power spectrum of the GWs created during inflation in these scenarios is given by two contributions: the irreducible contribution generated by quantum fluctuations of the gravitational field [see eqs.~(\ref{eq:omegainfl}),~(\ref{eq:GWen})], and a second contribution due to the classical production of tensor modes by the spectator field(s). From now on, we will refer to the scalar and tensor power spectra generated by a spectator field, as the \emph{sourced} power~spectra.

It has been noticed~\cite{Biagetti-ml-2013kwa} that the amplitude of the sourced GWs is strictly linked to the speed of sound of a given spectator field. More precisely, the lower the speed of sound the more efficient the production of GWs is. A number of works~\cite{Biagetti-ml-2013kwa,Biagetti-ml-2014asa,Fujita-ml-2014oba} have studied the GW production by spectator fields~\cite{Biagetti-ml-2014asa,Fujita-ml-2014oba} for specific inflationary models. They found out that the amplitude of the GWs on CMB scales, induced by the presence of a spectator field, cannot be responsible for a large value of the tensor-to-scalar ratio $r$ on such scales. The reason is that, besides the classical GW production, an extra scalar perturbation production takes place too, determined by the same parameters of the tensor counterpart. Scalar perturbations are well constrained by current CMB measurements~\cite{Ade-ml-2015lrj} and, consequently, the related GW production is bounded too.

The previous restriction attains only the amplitude of scalar and tensor power spectra on CMB scales. However, theoretical predictions allow the sourced contribution of GWs to have a blue-tilded spectrum, making this signal possibly accessible to LISA, while keeping an acceptable amplitude at the CMB scales. In this section, after reviewing the predictions about scalar and tensor power spectra in the presence of a spectator field, we will discuss its specific parameter space. Taking into account the bounds coming from current observations, we will investigate how LISA may add new information on the parameter space.

\subsection{Prediction of the gravitational wave signal}\label{subsec:GWsignalSpectFlds}

Among the models included in this framework, we consider the specific scenario investigated by~\cite{Fujita-ml-2014oba}. Compared to others~\cite{Biagetti-ml-2013kwa,Biagetti-ml-2014asa}, this model opens up the possibility for a larger production of sourced GWs, thus representing the most interesting case for our purposes. In this respect, it should be noted that, as we are investigating a specific model among those with spectator fields present, the results we are going to present are model-dependent.

Let us recapitulate first the results of~\cite{Fujita-ml-2014oba} about scalar and tensor power spectra. Let us consider the following Lagrangian,
\begin{equation}
\mathcal{L}= \frac{1}{2}M_{\rm Pl}^{2}R+\frac{1}{2}\partial_{\mu}\phi\partial^{\mu}\phi-V\left(\phi\right)+P\left(X,\sigma\right)\,,
\end{equation}
where $\phi$ is the inflaton, $\sigma$ is the spectator field, $X=\frac{1}{2}\partial_{\mu}\sigma\partial^{\mu}\sigma$ and $P$ is a generic function of $X$ and $\sigma$. We consider the inflaton to be responsible for the inflationary expansion and for the primordial scalar perturbations. On the other hand, while the spectator field does not influence the inflationary background dynamics, it creates nonetheless scalar and tensor perturbations. The spectator field $\sigma$ is characterized by a non-standard Lagrangian, with a propagation speed of its perturbations $c_{s}\equiv P_{X}/\left(P_{X}+P_{XX}\dot{\sigma}_{0}^{2}\right)$ (with $\sigma_{0}$ the background value), is different from the speed of light. In particular, we are interested in models with $c_{s}\ll 1$, as this makes the GWs production more efficient with respect to the case $c_{s} = 1$.

We also take into account the possibility that $c_{s}$ varies during inflation. This variation can be characterized by the dimensionless parameter
\begin{equation}\label{esse}
	s \equiv {\dot{c}_{s}\over H c_{s}}\neq 0\,,
\end{equation}
where the dot represents a derivative with respect to time. In the following analysis we will consider $s$ to be a small quantity, that is $\left|s\right| < 1$. The total power spectrum of the GWs generated during inflation is composed by two contributions: the irreducible part due to vacuum fluctuations given by eq.~(\ref{eq:GWen}), and the contribution generated \emph{classically} by the source term in the GW equation of motion, provided by a spectator field. Each of these contributions, turns out to be well described by a power law, i.e.~by an amplitude, referred to a pivot scale, and a spectral index (here considered as scale independent). An analogous situation takes place for the scalar perturbations, which are also sourced by vacuum and spectator field induced contributions, following as well a power law behavior. We expect therefore the total tensor and scalar power spectra to be described, respectively, by
\begin{equation}\label{powert}
	P_{\rm GW}\left(k\right)=A_{T}^{\left(v\right)}\left(k_{\ast}\right)\left(\frac{k}{k_{\ast}}\right)^{n_{T}^{\left(v\right)}}+A_{T}^{\left(\sigma\right)}\left(k_{\ast}\right)\left(\frac{k}{k_{\ast}}\right)^{n_{T}^{\left(\sigma\right)}}\,,
\end{equation}
and
\begin{equation}\label{powers}
	P_{S}\left(k\right)=A_{S}^{\left(v\right)}\left(k_{\ast}\right)\left(\frac{k}{k_{\ast}}\right)^{n_{S}^{\left(v\right)}-1}+A_{S}^{\left(\sigma\right)}\left(k_{\ast}\right)\left(\frac{k}{k_{\ast}}\right)^{n_{S}^{\left(\sigma\right)}-1}\,,
\end{equation}
where $k_{\ast}$ is a pivot scale. In both expressions the first term refers to the contribution provided by vacuum fluctuations, while the second terms are those induced by the presence of the spectator field.

The expressions for the sourced scalar and tensor power spectra are obtained from the perturbed action at third-order~\cite{Fujita-ml-2014oba}.
In the latter, a term of the form $\sim h_{ij}\delta\sigma\delta\sigma$ appears. Such a term is responsible for the generation of the sourced contribution in the GWs power spectrum. The equation of motion of tensor modes~\eqref{eq:GWeq} turns out to be~\cite{Fujita-ml-2014oba}:
\begin{equation}
	h_{ij}''+2\mathcal{H}h_{ij}'-\partial_{k}^{2}h_{ij} = \frac{2P_{\rm X}}{M_{\rm Pl}^{2}}\left\lbrace\partial_{i}\delta\sigma\partial_{j}\delta\sigma\right\rbrace^{\rm TT}\,,
\end{equation}
where $\mathcal{H}$ is the Hubble parameter in conformal time, $P_{\rm X}$ is the derivative of $P$ with respect to $X$, and $\lbrace \ldots  \rbrace^{\rm TT}$ selects the transverse and traceless part of the tensor inside the brackets. Following the calculation developed by~\cite{Fujita-ml-2014oba}, the amplitude at a given pivot scale results well approximated by
\begin{equation}\label{gwsourced}
	A_{T}^{\left(\sigma\right)}\simeq \frac{8}{15\pi c_{s}^{3}}\frac{H^{4}}{M_{\rm Pl}^{4}}\,,
\end{equation}
where $H$ and $c_{s}$ are evaluated at the pivot scale. In light of this expression it becomes clear that if $c_{s} < 1$, this enhances the sourced GW contribution with respect to the case $c_{s} = 1$.

We report also the expression for the sourced scalar perturbations, as these play a significant role in constraining the GWs production. In fact, both scalar and tensor perturbations are determined by the same parameters, and hence significant constraints on the tensor power spectrum are obtained thanks to current bounds on primordial scalar perturbations. The computation of the sourced scalar perturbations is however complicated. Several terms of the form\footnote{Where time and spatial derivative of each factor can be present.} $\sim \delta\phi\delta\sigma\delta\sigma$ appear in the action at the third order~\cite{Fujita-ml-2014oba}. The authors of~\cite{Fujita-ml-2014oba} claim that it is not clear, a priori, which term plays the main role in sourcing scalar perturbations. Based on theoretical considerations, they select one of these terms, $\delta N\left(\partial_{i}\delta\sigma\right)^{2}$ where $\delta N\sim \delta\phi$, and develop the calculations considering only such a contribution. In general, however, it is not to be excluded that other terms in the source could partially cancel each other. In the absence of a more elaborated analysis, we have decided to consider the same term taken into account in~\cite{Fujita-ml-2014oba}, developing our analysis by considering only that contribution. Later on, we will make further considerations about this point.

The amplitude of the sourced scalar contribution at the pivot scale, related to the term just mentioned, is found to be~\cite{Fujita-ml-2014oba}
\begin{equation}\label{scalar}
	A_{S}^{\left(\sigma\right)}\simeq \frac{1}{32\pi c_{s}^{7}}\frac{H^{4}}{M_{\rm Pl}^{4}}\,,
\end{equation}
where $H$ and $c_{s}$ are evaluated at the pivot scale. The amplitude of the sourced scalar power spectrum is therefore also enhanced by $c_{s} < 1$ values.

From eqs.~\eqref{gwsourced}--\eqref{scalar}, the respective spectral indexes can be obtained considering the scale dependence of $H$ and $c_{s}$. At lowest-order in the parameters $\epsilon$ and $s$, they turn out~to~be:
\begin{eqnarray}\label{indexgw}
	n_{T}^{\left(\sigma\right)}=\;&-4\epsilon-3s\,,
\\\label{indexscal}
	n_{S}^{\left(\sigma\right)}-1=\;&-4\epsilon-7s\,,
\end{eqnarray}
where $\epsilon\equiv-\dot{H}/H^{2}$ and $s$ is defined in eq.~\eqref{esse}.
For our purposes, we are interested in the scenarios in which $s\neq 0$, and in particular in the cases in which $s<0$, so that an enhancement of the GW amplitude on small scales can be obtained. Eqs.~\eqref{gwsourced}--\eqref{scalar} are the contributions to scalar and tensor perturbations induced by the presence of the spectator field. The whole scalar and tensor power spectra are obtained adding the contributions generated by vacuum oscillations of the inflaton and the gravitational field respectively.
At the end, the total power spectra of eqs.~\eqref{powert}--\eqref{powers} read
\begin{align}\label{spettrogw}
	P_{\rm GW}\left(k\right)\simeq \;&\frac{2H^{2}}{M_{\rm Pl}^{2}}\left(\frac{k}{k_{\ast}}\right)^{n_{T}^{\left(v\right)}}+\frac{8}{15\pi c_{s}^{3}}\frac{H^{4}}{M_{\rm Pl}^{4}}\left(\frac{k}{k_{\ast}}\right)^{n_{T}^{\left(\sigma\right)}}\,,
\\\label{spettri}
	P_{S}\left(k\right)\simeq \;&\frac{H^{2}}{4\epsilon M_{\rm Pl}^{2}}\left(\frac{k}{k_{\ast}}\right)^{n_{S}^{\left(v\right)}-1}+\frac{1}{32\pi c_{s}^{7}}\frac{H^{4}}{M_{\rm Pl}^{4}}\left(\frac{k}{k_{\ast}}\right)^{n_{S}^{\left(\sigma\right)}-1}\,,
\end{align}
where $H$ and $c_{s}$ are evaluated at the pivot scale $k=k_{\ast}$. The total GWs power spectrum is then given by the sum of two contributions described by two different power-laws.
The interesting fact for our purposes is that, if $c_{s}$ is sufficiently small and $s$ is negative with a sufficiently large absolute value, i.e.~$s \ll -1$, the sourced GWs could reach a sufficiently large amplitude, in principle detectable by LISA, while at the same time a small amplitude is kept at the CMB scales. Moreover, notice that, similarly to what happens for GWs sourced by gauge fields, like in section~\ref{sec:ParticleProduction}, tensor perturbations are expected to be non-Gaussian in this case too.

From eqs.~\eqref{indexgw}--\eqref{spettri}, the scalar and tensor power spectra turn out to be described by the energy scale of inflation, via the Hubble parameter $H$, by the slow-roll parameter $\epsilon$ and by the more specific quantities $c_{s}$ and $s$. We consider the parameter space $c_{s}$-$s$ evaluated at the pivot scale $k_{\ast}=0.05 {\rm Mpc}^{-1}$, for \emph{fixed values of the Hubble parameter $H$}. We find the bounds provided by CMB measurements and other observations, and we investigate which information LISA can add in such a space. Notice that the results we are going to show, lie on the assumption the expressions~\eqref{spettrogw} and~\eqref{spettri} to be valid on a wide range if frequencies, from CMB scales up to the frequencies to which laser interferometer detectors are sensitive.

\subsection{Constraints from CMB observations}

Measurements of the CMB provide several estimations and bounds on scalar and tensor perturbations at $f\sim 10^{-17}$ Hz:
\begin{itemize}

\item The scalar perturbation amplitude at the pivot scale $k_{\ast}=0.05\,{\rm Mpc}^{-1}$, $A_{0.05}=2.21\cdot 10^{-9}$ at $68\%$ C.L.~\cite{Ade-ml-2013zuv}. Notice that this bound constraints scalar perturbations without distinguishing possible different contributions.

\item An upper bound on the slow-roll parameter, $\epsilon<0.0068$ at $95\%$ C.L.~\cite{Ade-ml-2015lrj} considering \emph{\rm Planck TT+lowP}, that is temperature and low $\ell$ polarization data.

\item An upper bound on $r$ at the pivot scale $k_{\ast} = 0.05~{\rm Mpc}^{-1}$, $r_{0.05}<0.09$ at $95\%$ C.L.~\cite{Array-ml-2015xqh}, which corresponds to an upper bound of $H\simeq 8.5 \cdot 10^{13}$\,GeV at the same scale.
\end{itemize}

\begin{figure}
    \centerline{
\includegraphics[width=0.408\textwidth]{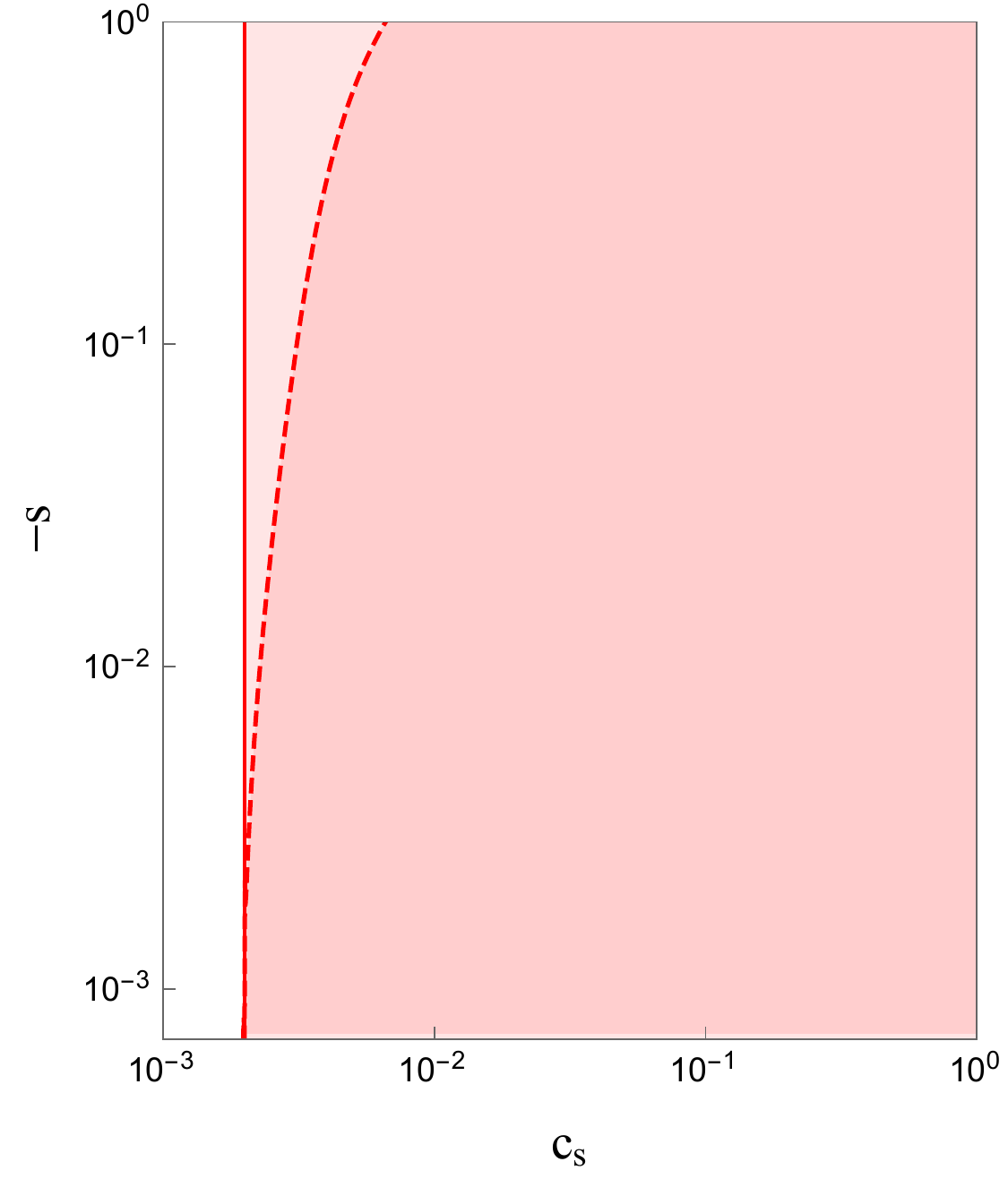}\qquad
\includegraphics[width=0.4\textwidth]{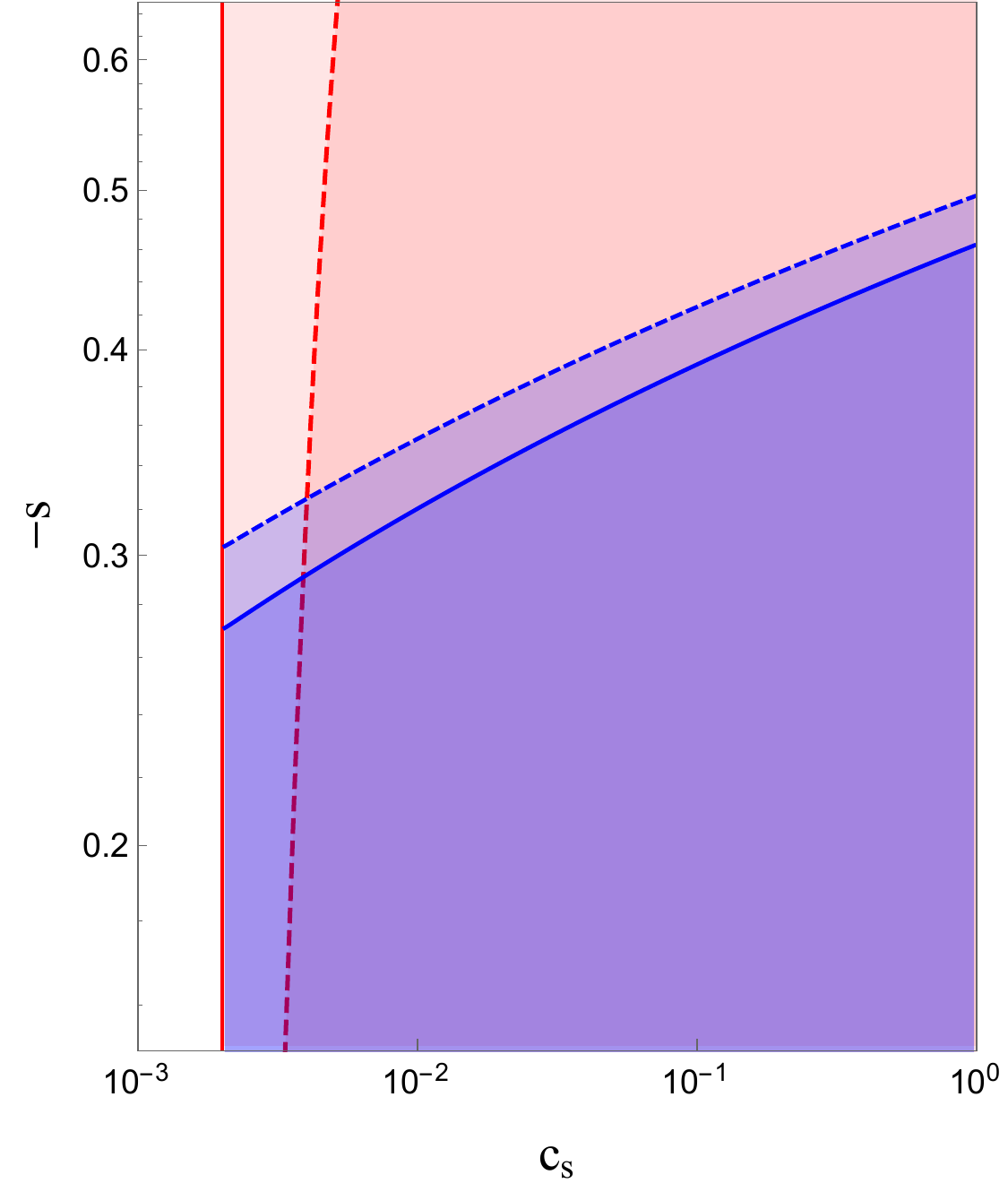}
}
\caption{Parameter space of $c_{s}$-$s$ at $k_{\ast}=0.05 {\rm Mpc}^{-1}$. In both plots, the lower bound on $c_{s}$ obtained from the upper bound on $\epsilon$ and from the estimation of the scalar amplitude on CMB scales is reported. The red region is that admitted by such constraints. The red-dashed curve is the bound obtained from CMB data at small scales, as explained in the text. Plot on the right: blue curves represent the discriminant power of LISA, the dashed curve corresponds to A$1$M$2$, the solid one to A$5$M$5$. The blue region is that left allowed by an eventual non-detection of the GW signal associated to this scenario by the respective LISA configuration. All the curves are obtained for $H=10^{12}$\,GeV.}
    \label{h}
\end{figure}

The limit on the slow-roll parameter $\epsilon$ sets a lower bound, for a fixed value of $H$, on the amplitude of the scalar perturbations due to vacuum fluctuations of the inflaton field. From the measurement of the scalar amplitude $A_{0.05}$, an upper bound on the contribution to scalar perturbations due to the spectator field is found. From the expression of the sourced scalar perturbations eq.~\eqref{scalar}, a lower limit on $c_{s}$ is obtained, indicated by the vertical line in figure~\ref{h}.

For $c_{s}$ values smaller than this limit, $\epsilon$ grows beyond the upper bound $\epsilon=0.0068$, up to a $c_{s}$ value below which the total scalar amplitude required by CMB observations cannot be obtained for a positive value of $\epsilon$. The requirement of $\epsilon$ to be positive, sets a lower bound of $c_{s}$, actually so tiny smaller with respect to the one discussed just above that in the next plots would not be distinguished with respect to the first (due to the power of $c_{s}$ in eq.~\eqref{scalar}. Since in the next analysis the value of $A_{0.05}$ given by CMB observation will be assumed in order to obtain limits of the parameter space from other experiments, we will assume directly the limit on $c_{s}$ derived from eq.~\eqref{scalar} and maintaining $\epsilon < 0.0068$. The white region on the left of each plot of figure~\ref{h} corresponds to the values of $c_s$ obtained for $\epsilon > 0.0068$. The choice $\epsilon < 0.0068$ ensures automatically that we are not considering regions of parameter space where the slow-roll condition on $\epsilon$ are significantly violated.

A large and negative value of $s$ means a positive spectral index for the sourced GWs. However, at the same time it corresponds to a positive spectral index for sourced \emph{scalar} perturbations too, see eq.~\eqref{indexgw}. Therefore, one should check the sourced scalar perturbations to be compatible with CMB data for \emph{all the range of scales} on which CMB experiments are sensitive. In particular one should keep under control the amplitude of scalar perturbations on the smaller scales to which CMB measurements are sensitive, that is $k \simeq 0.1\,{\rm Mpc}^{-1}$~\cite{Aghanim-ml-2015xee}.

We have made an estimation of this requirement considering the parametrization of the scalar power spectrum made by~\cite{Ade-ml-2015lrj}, where a spectral index, a running of the spectral index and a running of the running are admitted. We calculate the scalar amplitude at $k=0.1 {\rm Mpc}^{-1}$ with the parameter estimations provided by such analysis and we required the total amplitude of scalar power spectrum to not exceed it at the same scale.
In correspondence of positive spectral index of the sourced scalar contribution, these considerations turned out to add a more stringent lower bound on $c_{s}$, that is on the contribution of the sourced scalar perturbation to the total scalar power spectrum. Besides, an upper limit on the spectral index of the sourced scalar power spectrum (and hence on $\left|s\right|$), can also be found for a given value of the sourced scalar amplitude (for a fixed value of $H$). See figure~\ref{h}. The obtained constraints are in agreement with the general results provided by~\cite{Kinney-ml-2012ik}, which points out that CMB measurements admit a secondary contribution to the scalar power spectrum with an amplitude smaller than $10\%$ with respect to the main contribution (which corresponds to the region allowed by our estimations) without strict constraints on the related spectral index.

We find the obtained bound to be sensitive to the values of the amplitude imposed at $k=0.1 {\rm Mpc}^{-1}$ (and to the choice of this scale itself). 
We will see soon that physical observables which provide constraints on GWs at small scales split the parameter space in a different direction with respect to bounds obtained from CMB. Moreover, this bound ensures the sourced contribution to scalar perturbations to be suppressed with respect to the other one on CMB scales, and then that our choice of the estimation of the parameter $\epsilon$ referred to a single-field model, is reasonable.

Current CMB data provide significant constraints on non-Gaussianities of scalar perturbations on large scales. In particular, Planck data analysis gives an upper bound on the non-linearity parameter $f_{\rm NL}$~\cite{Ade-ml-2015ava}, which quantifies the non-Gaussianities level.
The presence of a spectator field induces the generation of extra contributions in the bispectrum of scalar perturbations, with respect to the scenario of single-field inflation (see the overview of models in~\cite{Ade-ml-2015ava}). Here $f_{\rm NL}$ is expected to be proportional to a negative power of $c_{s}$, $f_{\rm NL} \propto c_s^{-p},~p > 0$. Therefore, current constraints on non-Gaussianities are expected to set a lower bound on $c_{s}$, which eventually could shift the previous lower limit on $c_{s}$ to larger values. The only estimation~\cite{Biagetti-ml-2013kwa} available about non-Gaussianities in this type of scenarios indicate that scalar non-Gaussianities induced by the spectator field should be suppressed. In light of this consideration, LISA might turn out to be more powerful in constraining the parameter space than bounds on scalar non-Gaussianities. On the other hand, in light of the dependence of our estimations on the specific model considered, it should not be excluded that a complete computation of scalar non-Gaussianities might reduce the allowed parameter space. In this respect a complete calculation of the non-Gaussianities of scalar perturbations related to this scenario, remains as an interesting improvement of the present analysis, to be done in the~future.

A further feature that could put some constraints is a scale dependence of $f_{\rm NL}$. In fact, in the region of the parameter space interesting for LISA, the sourced scalar power spectrum can be enhanced, see eq.~\eqref{indexscal}. Therefore the contribution to the non-Gaussianity due to the presence of the spectator field could turn out to be significantly scale dependent. Current analysis of CMB data does not provide constraints on the scale dependence of $f_{\rm NL}$ suitable for the scenario considered here. If significant constraints on the non-Gaussianity scale dependence become available in the future, these could set a new bound in the parameter~space.


\subsection{Constraints from LISA}\label{subsec:ELISAconstraintsSpectFlds}

From the primordial GW power spectrum, the present-time GW spectral energy density is calculated as follows (see for example~\cite{Kuroyanagi-ml-2014nba}):
\begin{equation}
	h^{2}\Omega_{\rm GW}\left(k,\tau_{0}\right)=\frac{k^{2}}{12a_{0}^{2}H_{0}^{2}}P_{\rm GW}\left(k\right)T^{2}\left(k,\tau_{0}\right)\,,
\end{equation}
where the transfer function $T\left(k,\tau_{0}\right)$ is given by:
\begin{equation}\label{eq:transfer}
	T\left(k,\tau_{0}\right)=\frac{3\Omega_{m}j_{1}\left(k\tau_{0}\right)}{k\tau_{0}}\sqrt{1.0+1.36\left(\frac{k}{k_{eq}}\right)+2.50\left(\frac{k}{k_{eq}}\right)^{2}}\,.
\end{equation}
The  subscrit $_{0}$ indicates the present time, $k_{eq}$ is the scale of the horizon at the time of radiation-matter equality, $\Omega_{m}$ the matter density and $j_{1}$ is the Bessel function.

As the spectral index related to vacuum fluctuations is negative, in the range of LISA sensitivity the only contribution to GWs that can be relevant is the sourced one. Therefore here and in the analysis related to experiments at small scales, we will only consider such a contribution. The primordial power spectrum $P_{\rm GW}$ on LISA scales can then be characterized by the amplitude~\eqref{gwsourced} and the spectral index~\eqref{indexgw}.

At $k_{\ast}=0.05 {\rm Mpc}^{-1}$, the slow-roll parameter $\epsilon$ appearing in the spectral index can be written in terms of the parameter $c_{s}$ by requiring the whole scalar amplitude to be given by the value provided by the Planck analysis, for a fixed value of $H$, see eq.~\eqref{spettri}. So, at the end $\Omega_{\rm GW}\left(k\right)$ turns out to be parametrized by $c_{s}$ and $s$, for a fixed value of $H$. Therefore, for each value of $c_{s}$, one can identify the smaller value of $s$ for which $\Omega_{\rm GW}$ reaches a given LISA sensitivity curve. These pairs of values build up the curves represented in the right panel of figure~\ref{h}.
Notice that here we used the so-called \emph{power-law} sensitivity curves, obtained by the method provided in~\cite{Thrane-ml-2013oya}.
The region below each curve corresponds to those GW signals that, in principle, cannot be detected by the respective experiment at $95\%$ C.L. and then constitute the parameter space left open by an eventual non-detection of the GW signal associated to the current scenario.
Varying the value of the Hubble parameter moves the previous constraints as shown in the left panel of figure~\ref{bbnligo}.

As said before, we are considering the sourced scalar power spectrum generated by one of the terms that appear in the third order action. The real situation is more complicated. There could be further terms which are efficient sources for scalar perturbations. In the latter case the bounds coming from CMB would reduce the parameter space further than shown in our present plots. In any case, CMB constraints are expected to push $c_s$ towards greater values, still leaving to LISA a significant role in reducing the parameter space with respect to experiments operating at large scales.


\subsection{Other constraints}\label{subsec:OtherConsSpectFlds}

Besides CMB measurements, other bounds on primordial scalar and tensor power spectra are provided by current observations. As mentioned in section~\ref{sec:other_constraints}, Big Bang Nucleosynthesis (BBN) physics provides a stringent constraint on the integrated GW energy density, from which a significant upper limits on $\Omega_{\rm GW}$ is obtained~\cite{Pagano-ml-2015hma}, see eq.~(\ref{eq:BBNbound}). Moreover, the first aLIGO observation run set also an upper bound on $\Omega_{\rm GW}\left(f\right)$ at around $f\sim 10^{2}$ Hz~\cite{TheLIGOScientific-ml-2016wyq}.

\begin{figure}
 \centerline{
  \includegraphics[width=0.4\textwidth]{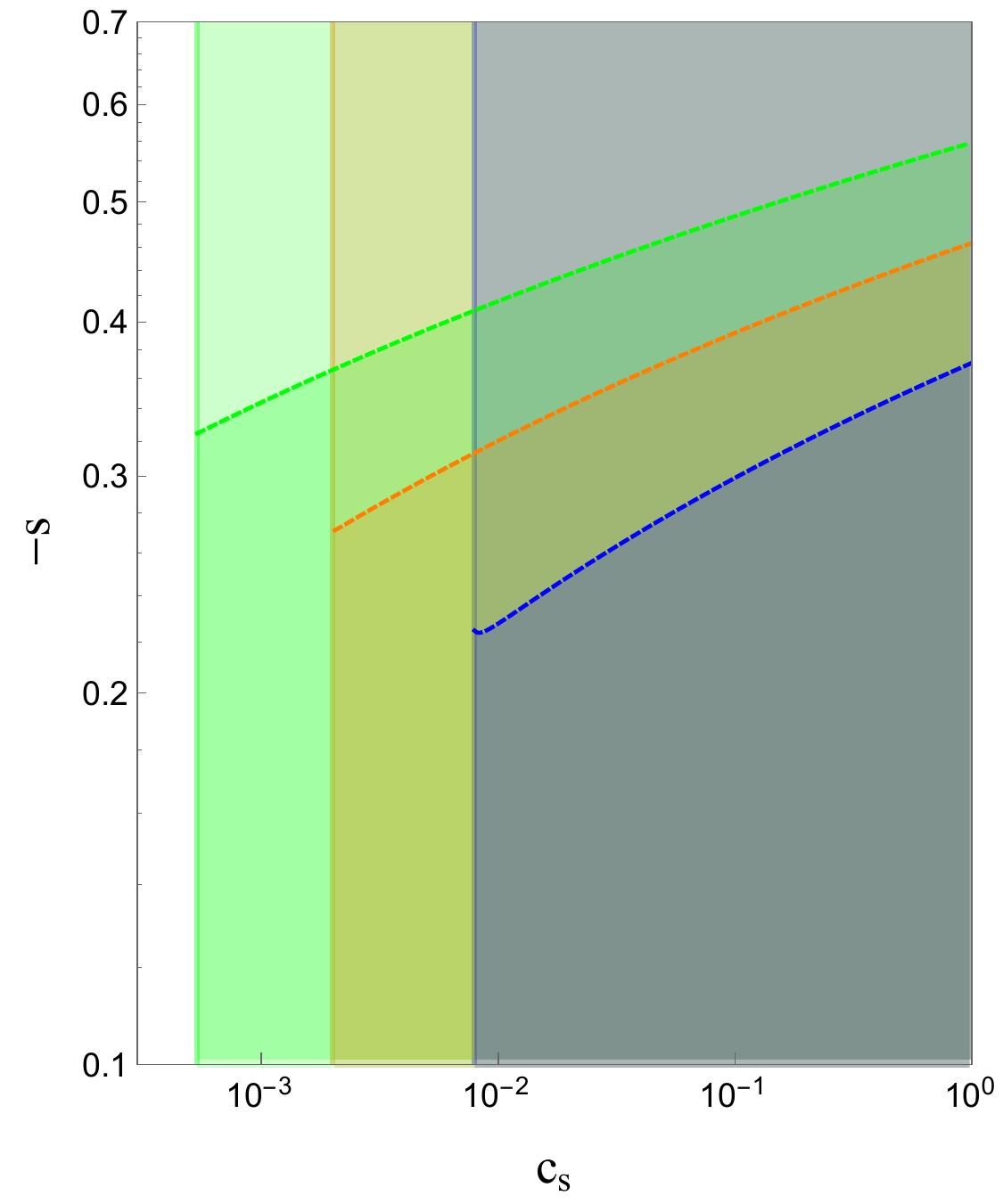}\qquad
\includegraphics[width=0.4\textwidth]{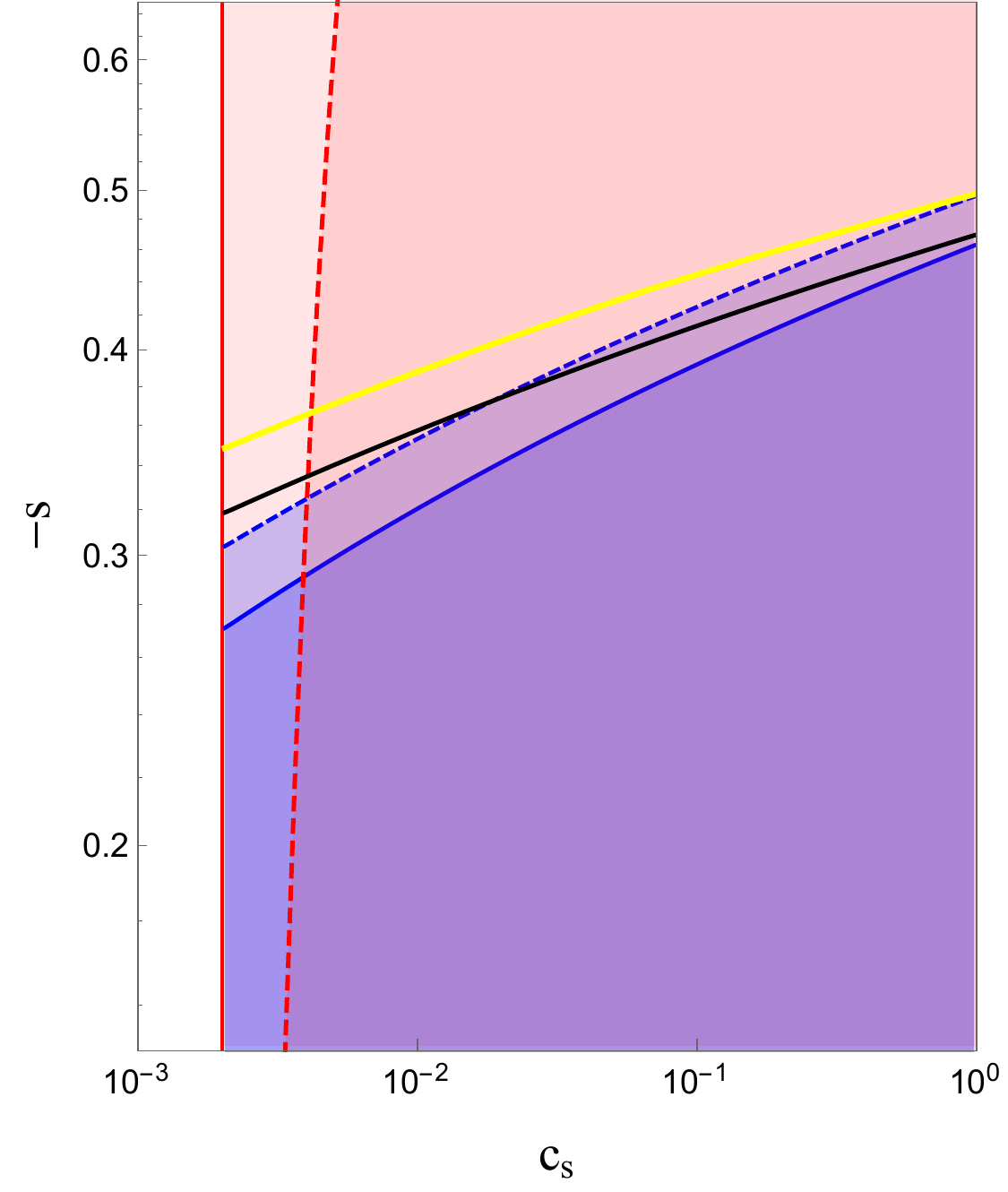}
}
\caption{Parameter space of $c_{s}$-$s$ at $k_{\ast}=0.05 {\rm Mpc}^{-1}$. On the left: bounds obtained from CMB measurements (vertical lines) and LISA  A$5$M$5$ discriminant power, for different values of the energy scale of inflation, e.g.\ of $H$. Each color corresponds to a fixed value of the Hubble parameter: $H=10^{11}$\,GeV (green), $H=10^{12}$\,GeV (orange), $H=10^{13}$\,GeV (blue). On the right:  current bounds obtained from the limit provided by~\cite{Pagano-ml-2015hma} for $T_{reh}=10^{12}$\,GeV (yellow) and aLIGO O$1\mbox{:}2015\mbox{-}16$ observation run (black).
		The energy scale of inflation is fixed at $H=10^{12}$\,GeV.    \label{bbnligo}}
\end{figure}

The Big Bang Nucleosynthesis (BBN) process provides an upper bound of the integrated GW spectral energy density for $f\gtrsim 10^{-10}$ Hz. The combination of such a limit with CMB observations, BAO and primordial Deuterium abundance measurements leads to an upper limit on the integrated GW spectral-energy density of $\Omega_{\rm GW}\lesssim 3.8 \cdot 10^{-6}$ at $95\%$ C.L.~\cite{Pagano-ml-2015hma}, see section~\ref{sec:other_constraints} for more details. Assuming a power-law shape for the GW power spectrum, the logarithmic integral of $\Omega_{\rm GW}\left(f\right)$ can be expressed in terms of the GW power spectrum amplitude at CMB scales and the spectral index. Then, employing an analogous procedure to that introduced in~\cite{Thrane-ml-2013oya}, an upper bound for $\Omega_{\rm GW}(f)$ can be found for a given range of possible values of the spectral index. Given that constraint and following the same procedure used to find the bounds related to LISA, we obtained the curves in the right panel of figure~\ref{bbnligo}.

We have checked that the obtained bound does not depend significantly on the assumed range of spectral indexes, for which we took $n_{t}\in \left(-1,3\right)$. On the other hand, it turns out that the bound on $\Omega_{\rm GW}(f)$ significantly depends on the reheating temperature of the Universe, which is an unknown quantity. In fact, the frequency at which $\Omega_{\rm GW}\left(f\right)$ decays, and hence the upper limit of the integration, strictly depends on the reheating temperature, see e.g.~\cite{Kuroyanagi-ml-2014nba}. In the right panel of figure~\ref{bbnligo} we report the curve obtained assuming instantaneous reheating, with a temperature corresponding to the inflationary energy scale. For lower reheating temperatures the curve turns out to be less restrictive.

\begin{figure}
    \centering
    \includegraphics[width=0.4\textwidth]{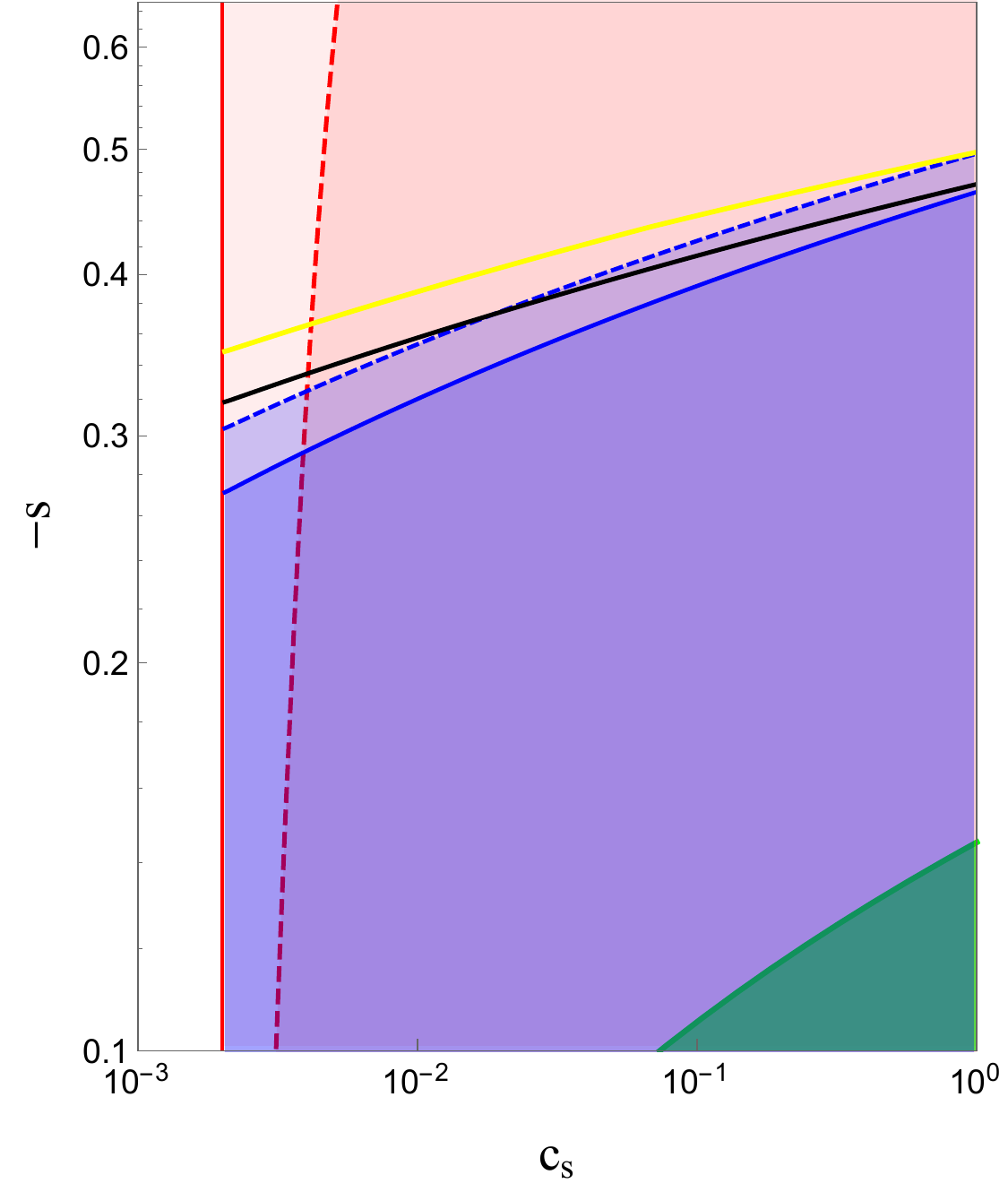}
    \caption{Parameter space of $c_{s}$-$s$ at $k_{\ast}=0.05\,{\rm Mpc}^{-1}$. Bounds obtained from considerations on PBH physics. The green region is that allowed by current considerations on such a physics. The energy scale of inflation is fixed at $H=10^{12}$\,GeV.}
		\label{d}
\end{figure}

The aLIGO O$1\mbox{:}2015\mbox{-}16$ observation run~\cite{TheLIGOScientific-ml-2016wyq} provided an upper limit on $\Omega_{\rm GW}\left(f\right)$ with respect to a power-law GW signal. Following the same procedure used for the LISA bound, the black curve in the right panel of figure~\ref{bbnligo} is obtained. For the GW signal considered here, this bound turns out to be the most stringent one we have up to now.
As introduced in~\ref{sec:other_constraints}, we know that considerations on the physics of PBH provide an upper bound on the amplitude of scalar perturbation on a broad range of scales, from very small frequencies (large scales) up to $f\sim 10^{6}$ Hz~\cite{Josan-ml-2009qn}. In most of the parameter space accessible by LISA, the spectral index of the sourced scalar contribution is expected to be positive, see eq.~\eqref{indexscal}. The upper bound on scalar perturbations provided by PBH, is expected to put an upper bound on such a quantity and hence on a combination of the parameters $s$ and $\epsilon$. As before, the latter can be written in terms of $c_{s}$ exploiting the value of $A_{0.05}$ provided by CMB measurements. Then, for a fixed value of $H$, PBH provide an upper bound on $s$ for a given $c_{s}$.
The bound plotted in figure~\ref{d} is obtained imposing that the sourced scalar perturbations get a maximal amplitude of $10^{-2}$ at $f_{\ast}=10^{6}$ Hz~\cite{Josan-ml-2009qn}. However, the physics of PBH is quite uncertain and the way in which they are modeled contains several assumptions that need to be verified. In this sense, an experiment like LISA can offer precisely the ability to test our understanding of the modeling of PBH formation. If the current modeling of PBH physics is eventually confirmed, LISA is unfortunately not expected to be able to detect GW produced during inflation by spectator fields, as we are discussing here.
The situation does not change even if we weaken the PBH constraint by one or two order of magnitude, or if we lower $f_{\ast}$ by an order of magnitude.

Ignoring PBH constraints, we find that each proposed configuration of the LISA experiment can significantly reduce the parameter space of the model with respect to CMB constraints. In fact, LISA, in particular its best configuration, is expected to sightly improve current bounds obtained from other experiments at small scales. The overall situation does not change for different values of the energy scale of inflation.

However, as we previously mentioned before a complete calculation of the sourced scalar contribution and of the related non-Gaussainities might improve the present analysis. Considering current estimations available in the references of non-Gaussianities, LISA might turn out to put stronger constraints in the parameter space than considerations on scalar non-Gaussianities from the CMB. On the other hand, it should not be excluded that a complete computation of scalar non-Gaussianities could shrink the allowed parameter space so largely, that the ability of LISA to constrain the parameter space of these scenarios could be significantly reduced with respect to the present constraints at small scales.

Anyhow, constraints obtained by data on CMB scales can be exploited for testing the origin of an eventual inflationary GW signal observed by an experiment such as LISA.
In fact, a detection by LISA of GWs identified by a point in the parameter space not allowed by CMB constraints, would indicate that the detected signal is not to be associated to the inflationary scenario considered here. In this direction, a complete calculation of the scalar power spectrum, of the related non-Gaussianities and of their dependence on the scale, could be of further improvement.

On the other hand, taking at face value the present bounds provided by PBH physics to be valid, LISA does not have the capability of detecting primordial GWs produced in this inflationary scenario. At the same time, however, LISA still has the ability to provide a validation test for the several assumptions currently made in the modeling of PBH physics. Moreover, notice that the previous constraints have been obtained assuming a constant value of $s$ on a wide range of frequencies. In particular, we are assuming the spectral index related to the sourced scalar power spectrum to be constant up to a range of frequencies higher than those related to LISA. Admitting a running of the spectral index of the sourced scalar perturbations, and then a running of the sourced GW spectral index, the scalar amplitude at small scales can be lower than in the case with $s$ constant, having at the same time a large amplitude at LISA frequencies. Therefore, admitting a running of the spectral index of the sourced scalar perturbations and a running of the sourced GW spectral index, leaves open the possibility of satisfying the PBH constraints, having at the same time a GW signal that exceeds the LISA sensitivity.

\section{GWs in the framework of EFT of broken spatial reparametrizations}
\label{sec:EFT}

 In the previous sections we examined scenarios where an enhanced primordial tensor
 spectrum is induced by fields which are usually  not responsible for  driving the inflationary expansion. Such additional fields might or might
 not interact with the inflaton field.  Their dynamics should  not
  lead to an important back reaction on the expansion of the Universe in order to avoid spoiling inflation or generating significant changes on the scalar spectrum, while their interactions should not produce large
   non-Gaussianities incompatible with current CMB data~\cite{Ade-ml-2015ava}.
   However, one may ask whether inflationary models exist, leading to  features in the GW spectrum
   similar to those described in the previous sections, while not invoking any additional fields, besides the ones driving inflation. In this section, we examine whether such scenarios,  which  have the advantage
   of avoiding back reaction issues, are plausible.

  In order to be detectable with LISA, the
   spectra must be enhanced at scales smaller than the CMB ones, thus we require a blue tensor spectrum for inflationary gravitational waves. Hence,
 such scenarios should violate
   the inflationary consistency relations  which normally lead to a red spectrum.  Scenarios
   with a blue spectrum of inflationary tensor modes can have a tensor power  spectrum whose amplitude is too small to be detected at CMB scales through the B-modes, however such amplitude increases at smaller scales, therefore maybe becoming detectable by an interferometer like LISA.

    We will use an approach based on symmetry principles, motivated by the Effective Field Theory (EFT) of inflation~\cite{Cheung-ml-2007st}. Conventional models of inflation involve scalar field(s) with  a time-dependent homogeneous profile, which
    breaks only the time-reparametrization symmetry of de Sitter space during inflation; space-reparametrizations are normally preserved. Tensor modes are transverse traceless fluctuations of  slices of constant time. Regardless of the specific model one considers,  requiring space-repara\-metrization on such slices imposes  constraints on the structure of the
    tensor action. In such a set-up,   tensors are adiabatic, massless modes during inflation,  conserved
    at superhorizon scales. The tensor power spectrum is controlled by the value of the Hubble parameter and
    the tensor sound speed; the latter can be different from the one in models
     with kinetic mixing between gravity and the inflaton, as e.g.~G-inflation~\cite{Kobayashi-ml-2010cm,Kobayashi-ml-2011nu}.
     The resulting  amplitude of tensor power spectrum decreases at smaller scales  --- corresponding to a red spectrum --- unless the tensor
     sound speed is strongly time-dependent (although this would be problematic for the scalar spectrum~\cite{Wang-ml-2014kqa}). Hence,  inflationary  scenarios breaking only time-reparametrization  do not normally lead to an enhanced tensor spectra at small
     scales, unless additional ingredients as extra fields are included, as discussed in the previous sections.

   When we break space-reparametrization during inflation, by considering inflaton fields with space-dependent
    profiles, the previously drawn conclusions may change in an interesting way.  Namely, models which break space-reparametrization can lead to  inflationary tensor modes that are massive, since there is no symmetry preventing tensor
    fluctuations  from acquiring a mass
    during inflation.
    The graviton mass, if sufficiently large, can lead to a blue spectrum for gravitational waves,
     which enhance their power  at scales much smaller than the CMB ones. At the end of inflation, the inflaton field can  arrange itself so as to recover space-reparametrization symmetry, hence making the  graviton mass equal to zero.

     There are various examples in the literature which realise such a possibility, and an effective field theory approach
     allows one to understand their features in a general  way.  Models of vector inflation, first introduced in ref.~\cite{Golovnev-ml-2008cf},
     break space-reparametrization symmetry by providing space-like {\emph{vacuum expectation values}} (\emph{vevs}) to a set of vector fields. In their original versions, they are plagued
     by a ghost mode~\cite{Himmetoglu-ml-2008zp}, but there exist scenarios, such as gauge or chromo-natural inflation~\cite{Maleknejad-ml-2011jw,Adshead-ml-2012kp}, which are free from
     such instabilities, while they lead to various distinctive signatures for the tensor spectrum (see e.g.~\cite{Dimastrogiovanni-ml-2016fuu} and references therein). Other models,
     as Solid~\cite{Endlich-ml-2013jia} and Supersolid~\cite{Koh-ml-2013msa,Cannone-ml-2014uqa,Cannone-ml-2015rra,Bartolo-ml-2015qvr} inflation, involve scalar fields only. They are described by Lagrangians which contain
     three or four scalar fields, obeying internal symmetries which ensure  the homogeneity and isotropy of the spacetime background
     configuration. The scalars have time- and space-dependent \emph{vevs} which break reparametrization symmetries of the
     background: from an EFT viewpoint,  they are the minimal scenarios with the symmetry breaking pattern one wishes
     to study. The fact that they can lead  to a blue spectrum for tensor modes, as well as to other distinctive properties
     in the tensor spectrum,  has been pointed out in ref.~\cite{Endlich-ml-2013jia}, recently studied in~\cite{Ricciardone-ml-2016aa}, and further studied in refs.~\cite{Bartolo-ml-2015qvr,Cannone-ml-2014uqa, Akhshik-ml-2014gja,Akhshik-ml-2014bla}.

     Any set-up which breaks space-reparametrization leads to several distinctive signatures also in the scalar sector, or
     in higher order interactions between scalar and tensor modes. In the next subsections~\ref{sol-pgw} and~\ref{sol-lisa} we discuss properties of the tensor sector of possible interest for LISA, and then study
     further testable predictions in subsection~\ref{sol-ft}.

\subsection{Properties of gravitational wave signals}\label{sol-pgw}

We  consider model independent features of tensor fluctuations in scenarios that break space-reparametrization during inflation. We limit our attention to  set-ups which preserve isotropy and homogeneity   of space time
at the background level, considering only tensor fluctuations $h_{ij}$ around a FLRW background, as described in the Introduction.

Under our  assumptions about the symmetry breaking pattern,
we can write the most  general form for the second order action for  tensor fluctuations~\cite{Cannone-ml-2014uqa}:
\begin{equation} \label{sol-qac}
S_{(2)}\,=\,\frac{M_{\rm Pl}^2}{8}\int dt\,d^3 x \,a^3(t)\,n(t)\,\left[\dot h_{ij}^2 -\frac{c_T^2(t)}{a^2} \,\left( \partial_l h_{ij}\right)^2
-m_h^2(t)\,h_{ij}^2\right]\,,
\end{equation}
where $n$, $c_T$ and $m_h$ are functions of $t$, determined by the model under consideration. and $a(t)$ is the scale factor of the Universe. We emphasize that in writing
the previous quadratic action, we did not  intend to commit on specific scenarios, but instead included all terms
allowed by the symmetries.  The graviton squared mass $m^2_h$ is the distinctive feature of  a set-up that breaks space-reparametrization. A tensor sound speed  $c_T\neq 1$ is normally associated with scenarios with kinetic
mixing among tensors and scalars, as in G-inflation, or more generally in inflationary models in Horndeski scalar-tensor theories~\cite{Horndeski-ml-1974wa}.
Notice that a disformal transformation can  be applied on the metric, so to  set $c_T=1$~\cite{Creminelli-ml-2014wna}.
 Such transformation
would also change the graviton mass,  modifying  the scale dependence of the tensor spectrum and affecting
also the scalar sector.
  For this reason, we prefer  not to consider it here. Also the overall factor $n(t)$, which can be thought as renormalizing the  Planck mass, has to be
expected in scenarios with kinetic mixing   among gravity and scalar fields.

Let us comment on the mass term: $m_h^2$ can be both positive and negative. The Higuchi bound suggests $m_h^2\,<\,0$, since it states that tensor modes can not have a mass in the interval $0\,<m_h^2\,\le \,2 H$ in pure
de Sitter space, when considering theories that are Lorentz invariant~\cite{Higuchi-ml-1986py,Arkani-Hamed-ml-2015bza}.  However,  inflation does not occur in a
pure  de Sitter space, so a small and positive $m_h^2$ might be allowed in theories which preserve Lorentz symmetry at the fundamental level, but spontaneously
break de Sitter symmetry. Moreover,  considering theories which break Lorentz symmetry,
e.g.\ Horava-Lifshitz scenarios~\cite{Horava-ml-2009uw}, the Higuchi bound does not necessarily apply. We refer the reader to ref.~\cite{Blas-ml-2009my} for a formulation of Lorentz violating
massive gravity in cosmological backgrounds.   In our case, we keep agnostic and allow also for a positive
$m_h^2$.

In order to find the power spectrum of gravitational waves, we  need to have information about the time-dependence
of the functions entering eq.~\eqref{sol-qac}. The computation of the tensor spectrum normally requires a careful numerical analysis. Analytical considerations can be carried out  in a quasi de Sitter space, $\epsilon\ll1$ for $c_T$ and $m_h$ only mildly depending on time.  Focusing on the approximation of pure de Sitter, $H=$ const.,  for $c_T$ and $m_h$ constants and setting $n=1$, we
can follow the standard procedure of quantizing the tensor modes around  de Sitter background.\footnote{As far as we are aware,  more general analysis of dynamics and  quantization of
tensor modes
with  arbitrary time dependence of tensor mass and sound speed during inflation have not been carried on so far in the literature.} Tensor modes
in Fourier space are indicated with $\tilde h_{ij}(t,\,\vec k)$. Starting from the  equal-time two point function for tensor modes, which we expressed as
\begin{equation}
\langle \tilde h_{ij}(\vec k)\,  \tilde h_{k l}(\vec k') \rangle\,=\,(2 \pi)^3\,\delta^{(3)}(\vec k+\vec k')
\,{\cal P}_{ij,\,kl}\,,
\end{equation}
we define the primordial tensor power spectrum evaluated at comoving scale $k$, as
\begin{equation}
{\cal P}_h\,=\,\frac{k^3}{2\,\pi^2}\,{\cal P}_{ij,\,ij}~.
\end{equation}
In our set-up, this leads to
%
\begin{equation}  \label{eq:PSHankel}
{\cal P}_h\,=\,\frac{H^2}{4 \pi\,M_{\rm Pl}^2} \,
\left(
\frac{k^3}{  \,k_{*}^3}\right)\,\Big| H^{(1)}_\nu\left( \frac{c_T\,k}{k_*}\right)\Big|^2\,,
\end{equation}
with $H_\nu^{(1)}(x)$  the Hankel function of first kind,   $k_*$ a reference scale, and
\begin{equation}
\nu\,=\,\frac32\,\sqrt{1-\frac{4 m_h^2}{9 H^2}}\,.
\end{equation}
A simplification occurs in the limit of small graviton mass, $|m_h/H|\ll1$, for which the power spectrum becomes a power law for
sufficiently large scales:
\begin{equation}\label{eq:PS}
{\cal P}_h\,=\,\frac{H^2}{2\pi^2\,M_{\rm Pl}^2\,c_T^3}\,\left( \frac{k}{k_*}\right)^{n_T}\,,
\end{equation}
with (recall we are considering an approximation of pure de Sitter space)
\begin{equation}\label{eq:tensortilt}
n_T\,=\,\frac23\,\frac{m_h^2}{H^2}\,.
\end{equation}
Notice that a blue spectrum, $n_T>0$, requires a positive $m_h^2$. This is the case we are most interested to,
since it enhances the tensor spectrum at small scales, and can lead to a signal detectable with LISA. We present our study in the
next subsection. For simplicity, we use the representative formulae of eq.~\eqref{eq:PS} for the power spectrum
and eq.~\eqref{eq:tensortilt} for the spectral~tilt.

\subsection{Parameter analysis based on the LISA sensitivity curves}\label{sol-lisa}
Our framework is  convenient  for carrying out a model-independent analysis of the consequences of breaking space-reparametrization symmetry
during inflation,  and connecting it with a possible detection of the GW signal in the range of frequencies and energy densities probed by LISA.
The power spectrum of primordial tensor modes in~\eqref{eq:PS} depends on the energy scale of inflation $H$,
on the tensor speed of sound $c_{T}$ that can varies in the interval $0\le c_{T}\le1$ (in unity of speed of light) and finally on the mass of the graviton $m_{h}$.

\begin{figure}
\centerline{\includegraphics[width=13cm]{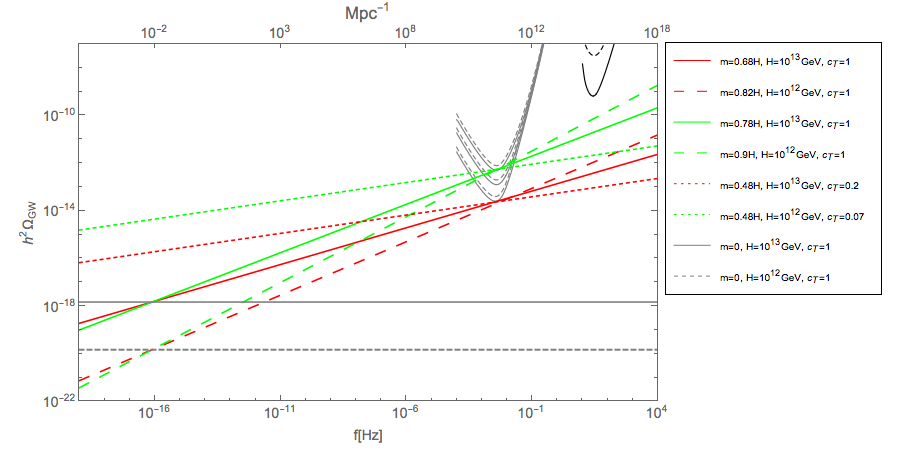}}
\vspace{-2mm}
\caption{Spectrum of GWs energy density $h^{2}\Omega_{gw}$ for different values of the effective mass of the graviton $m_{h}$, Hubble rate during inflation $H$, and tensor sound speed $c_{T}$, compared with the sensitivity of LISA (grey curves) and LIGO (black curves) detectors. We use $k_{*}=0.05~{\rm{Mpc^{-1}}}$ as a pivot scale.\label{fig:Sens_vs_signal}}
\end{figure}

 In this section, we investigate the ability of the different LISA configurations, mentioned in section~\ref{sec:LISAcurves}, in constraining the space of such parameters. In particular the Hubble parameter $H$ and the tensor sound speed $c_T$ control
 the amplitude of the power spectrum, while the dimensionless combination $m_h/H$ characterizes the tensor spectral index $n_{T}$.
We emphasize that, for simplicity, we use the representative eqs.~\eqref{eq:PS} and~\eqref{eq:tensortilt} for  investigating the combined effects of  graviton mass and tensor sound speed during inflation.
We fix the pivot scale at $k_{*}=0.05~{\rm{Mpc^{-1}}}$ and we use  $M_{\rm Pl}\,=\,2.4 \times 10^{18}$ {\rm{GeV}} for the (reduced) Planck mass. For simplicity in our analysis we assume that both the tensor sound speed and the graviton mass are time independent; their time dependence could be subject of a further analysis, for example along the lines of the recent works~\cite{Cai-ml-2015yza,Cai-ml-2016ldn}. In figure~\ref{fig:Sens_vs_signal} we plot the GWs energy density for some representative values of the mass of the graviton and speed of sound, for two different energy scales of inflation. The fractional GW energy density is related to the power spectrum~\eqref{eq:PS} by the transfer function, defined in~\eqref{eq:transfer}. One can easily notice that the effect of the mass is an enhancement of the power on small scales. Then this opens the possibility to extract limits on the (minimum) mass of the graviton during inflation in order to have a signal detectable by LISA. We show the ability of the  ``best'' (A5M5) and the ``worst'' (A1M2) LISA configurations in putting a lower bound on the mass of the graviton $m_{h}$, fixing the Hubble parameter and the tensor speed of sound.
 From figure~\ref{fig:Sens_vs_signal} we see that a 6 links, 1 million km arm-length, 2 years of observation with LISA (A1M2) will be able to probe the effects of a graviton with $m_{h} \simeq 0.78\, H$, for an energy scale of inflation of $H=10^{13}{\rm{GeV}}$ and $c_{T}=1$,              while the  ``best" LISA configuration (A5M5), still with 6 links, but with 5 million km arm-length, 5 years of observation will put a smaller lower bound on the mass, $m_{h}\simeq 0.68 \,H$ for the same inflationary energy scale and speed of sound. Lowering the energy scale of inflation to $H=10^{12}\,{\rm{GeV}}$ allows to probe higher valuer of the effective mass.
In the same figure we also show the effect of fixing the value of the graviton mass to some representative values and  plot the  ability of LISA configurations  to put bounds on the tensor speed of sound $c_{T}$. In particular we find that the best LISA configuration (A5M5) for a mass of the graviton $m_{h} \simeq 7.8 \times 10^{12}\,{\rm{GeV}}$ will be able to test speed of sound at the level of 20\% of the speed of light, while lowering the energy scale of inflation allows to reach 1\% level. For completness we also draw the predicted energy density for a complete scale-invariant case $(m_{h}=0)$.
For comparison we also plot the aLIGO sensitivities: in particular we see that neither aLIGO O1 nor the future aLIGO/adVirgo O5 will be able to probe such a signal characteristic of this kind of scenario.

\begin{figure}
\centerline{
\includegraphics[width=7.0cm]{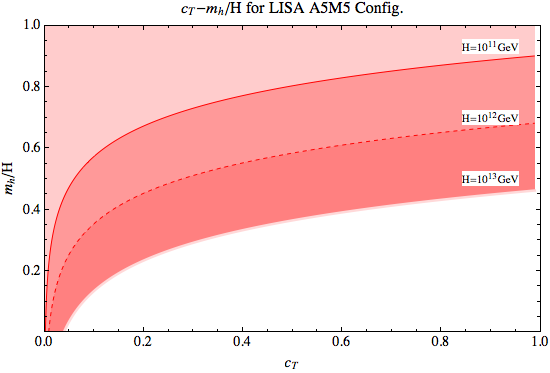}\qquad
\includegraphics[width=7.0cm]{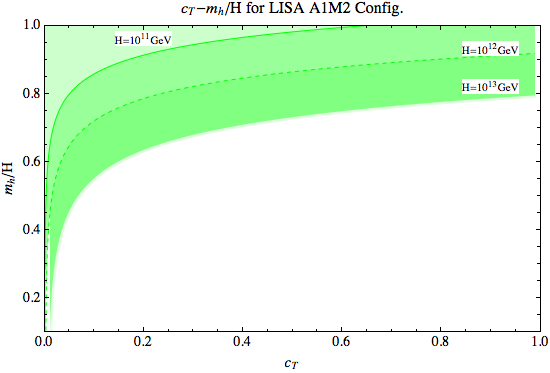}
}
\caption{Region in the  ($c_{T}-(m_{h}/H)$) parameter space of the theory that LISA can probe, for the ``best'' (A5M5) configuration (left panel)  and the ``worst'' (A1M2) configuration (right panel) at different Hubble rate during inflation.\label{fig:massspeedt}}
\end{figure}

\begin{figure}{t}
\centerline{
\includegraphics[width=7.0cm]{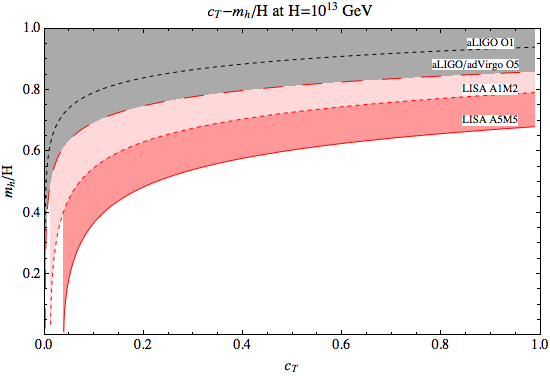}\qquad
\includegraphics[width=7.0cm]{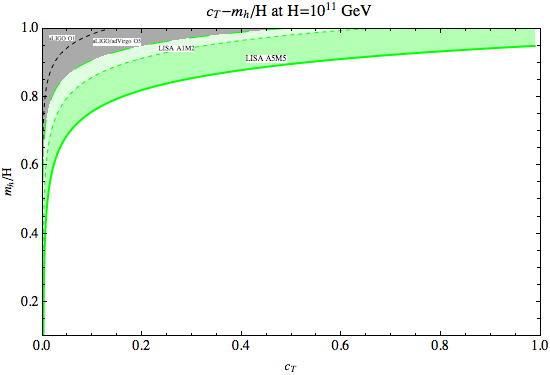}
}
\caption{Region of the  ($c_{T}-(m_{h}/H)$) parameter space constrained by the LISA ``best'' (A5M5)  and ``worst'' configuration (A1M2) and the (a)LIGO detector, for two different Hubble rate: $H=10^{13} \rm{GeV}$ on the left and $H=10^{11} \rm{GeV}$ on the right.
In the figure are shown also the upper limits from a non-detection of such GW signal by the LIGO detector and eventually by the future aLIGO/adVirgo~one.\label{fig:massspeedt2}}
\end{figure}

The interplay among the three independent parameters $\{H, c_{T}, m_{h}\}$ allows us to put constraints on the space of two of them fixing the third one. In particular in figure~\ref{fig:massspeedt} we show the perspective of LISA in constraining the parameter space $(c_{T} - \,m_{h}/H)$ for the best configuration: A5M5 (left panel) and the worst configuration: A1M2 (right panel) at different energy scales. In both panels we have fixed the pivot scale $k_{*}=0.05{\rm{Mpc}^{-1}}$ and have used the following strategy: for a fixed value of the tensor speed of sound $c_{T}$ in the interval   $0\le c_{T}\le1$ we have computed the minimum of the graviton mass $m_{h}/H$ in order to have a signal above the minimum of the two LISA sensitivity curves. We can see, on the left panel, how the LISA best configuration is able to probe a larger range of masses for the graviton, which increases for a higher value of the energy scale of inflation.
This first result highlights the importance of longer duration of the mission and longer arm-lengths.
Our  analysis shows the ability of the LISA mission in excluding  a large region of the  space  of parameters that we
are considering.
With the same aformentioned strategy, we fix the energy scale of inflation to two representative  values: $10^{13}{\rm{GeV}}$ on the left and $10^{11}{\rm{GeV}}$ on the right panels of figure~\ref{fig:massspeedt2}  and we draw the contour plot in the $(c_{T}-m_{h}/H)$ parameter space, for the best and worst LISA configurations, adding the current upper bound on the primordial GWs energy density coming from the recent LIGO detection of GWs coming from the merging of BHs~\cite{Abbott-ml-2016blz,Abbott-ml-2016nmj}. In this way we shrink the parameter space of the theory since a non-detection by LIGO of this signal puts an upper bound on the graviton mass and the sound speed of tensor perturbations during inflation.

\begin{figure}{h}
\centerline{
\includegraphics[width=7.6cm]{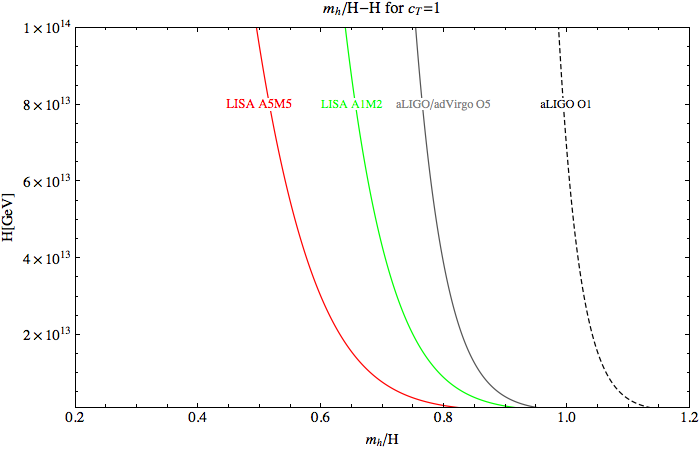}\quad
\includegraphics[width=6.8cm]{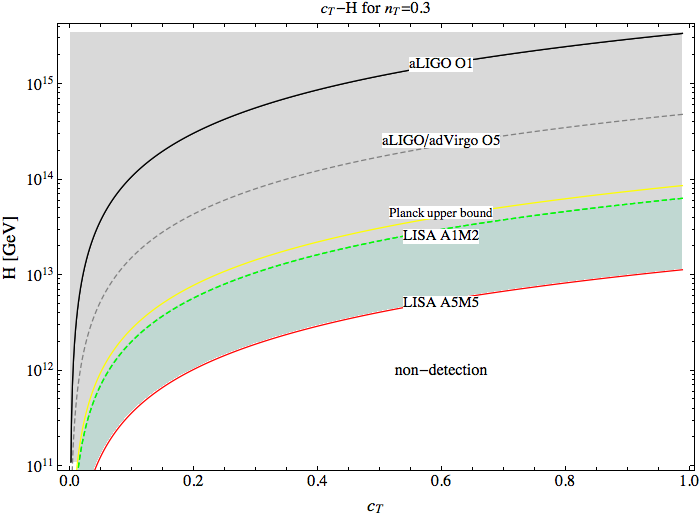}}
\caption{{\emph{Left:}} lower limits on the mass of the graviton  in the  ($\frac{m_{h}}{H}-H$) parameter space for the  ``best'' (A5M5), the ``worst'' (A1M2) LISA configurations and the (a)LIGO detectors, for a tensor speed of sound $c_{T}=1$. {\emph{Right:}} lower limit on the Hubble rate during inflation for a GW signal detectable by LISA with a tilt $n_{T}=0.3$, compatible with the maximum ability of LISA in constraining such a parameter, as explained in section 2. The yellow curve represents the current Planck upper bound on the Hubble rate during inflation, extrapolated at values $c_{T}<1$.
\label{fig:Hvsct}}
\end{figure}

The ability of LISA to put constraints on the GW spectral index $n_{T}$, explained in section 2, translates in this scenario in the possibility to scan the $(m_{h}/H-H)$ parameter space for a given tensor speed of sound. Assuming the standard case of a graviton travelling at the speed of light, from the left panel of figure~\ref{fig:Hvsct}, we can read the lower masses of the graviton to which the LISA best (red curve) and worst (green curve) and aLIGO (gray curve) and aLIGO/adVirgo (black dashed curve) are sensitive.

From a different perspective we can use the same information from the $r-n_{T}$ plot~\ref{fig:rvsnt} to show the probed region in the  $(c_{T}-H)$ parameter space for a value of the tensor spectral tilt $n_{T}$. Choosing for istance $n_{T}=0.3$, value in the range of LISA possibilities, in figure~\ref{fig:Hvsct} we see that the best LISA configuration (A5M5) can be sensitive to energy scales of inflation around $10^{12} {\rm{GeV}}$. Of course a higher value of the spectral tilt, or a tensor sound speed lower than the speed of light, would allow to reach lower inflationary energy scales. We have plotted also the current Planck upper bound on the Hubble rate during inflation, extrapolated at values $c_{T}\le1$ (yellow curve).

In future investigations, it will be interesting to extend the analysis of consequences of inflationary gravitational mass and tensor sound speeds to richer scenarios, besides the ones that can be parametrized by our equations~\eqref{eq:PS} and~\eqref{eq:tensortilt}.
%

\subsection{Further constraints on GWs from other observables}\label{sol-ft}
Scenarios which break space-reparametrization symmetry during inflation can have several other distinctive  observational consequences,
besides the small scale enhancement of tensor modes. Such features are complementary to the ones
discussed above, and
 make these scenarios
 distinguishable from  other set ups with small scale enhancements of
   tensor modes, as the ones discussed in the previous sections. The details  of these observational consequences  are model-dependent, but we can list common features  related to the fact that tensors have a mass during inflation (while become massless after inflation ends).
\begin{itemize}
\item
Inflationary  tensor modes are  not adiabatic; in general, this implies that inflation
is not an efficient isotropic attractor, and large scale anisotropies  can be produced~\cite{Bartolo-ml-2013msa,Bartolo-ml-2014xfa,Bordin-ml-2016ruc}.
 This is not necessarily a bad
feature, since it can lead  to distinctive consequences associated with  modulations of the scalar two point function~\cite{Bartolo-ml-2014xfa,Bordin-ml-2016ruc},  a property that can be (and is) constrained  by CMB experiments as Planck~\cite{Ade-ml-2013nlj,Ade-ml-2015hxq}.

\item
Primordial non-Gaussianities can be enhanced (but not at a level to be excluded by current constraints) and
have distinctive features as an angular dependent squeezed limit of bispectrum for the scalar three-point function~\cite{Endlich-ml-2012pz,Endlich-ml-2013jia,Bartolo-ml-2015qvr}.
 Such properties
can be tested through CMB~\cite{Ade-ml-2013ydc,Ade-ml-2015ava} (although no dedicated templates have been implemented so far for examining current CMB data); in the future,
galaxy surveys can also offer opportunities for testing the angular dependence of squeezed \hbox{bispectra~\cite{Schmidt-ml-2015xka,Chisari-ml-2016xki}.}

\item
Tensor-scalar-scalar three-point functions can also be enhanced, to a level which can be tested through B-modes searches of tensor
non-Gaussianity~\cite{Cannone-ml-2014uqa,Meerburg-ml-2016ecv}. Such
 effects also enhance the scalar 4pt function, leading to a peculiar angular dependence  distinguishable from a trispectrum amplitude
   $\tau_{\rm NL}$ signal, as recently pointed out in ref.~\cite{Bordin-ml-2016ruc}.
\end{itemize}
Any one of these additional effects, if detected, would provide strong hints in favour  of scenarios breaking space-reparametrization during inflation. With the same spirit of the EFT of inflation~\cite{Cheung-ml-2007st}, our constraints on the physical observables of the theory, automatically translate to operators appearing in the quadratic action for the tensor field. In this sense, LISA has the possibility to test symmetry breaking patterns and models of the early Universe at smaller scales than the ones probed by CMB.
%
\section{Gravitational wave background from merging PBHs}
\label{sec:PBH}


It has been known for some time~\cite{GarciaBellido-ml-1996qt} that large peaks in the matter power spectrum can give rise to primordial black holes (PBH) with masses of order a few solar masses, when those scales reenter the horizon during the radiation era. They typically give rise to a population of isolated PBH that could act as cold dark matter, see also~\cite{Nakamura-ml-1997sm}. Recently, however, it has been pointed out that in very generic hybrid models with long-lasting waterfall regimes, the peak in the spectrum is not only large but very broad~\cite{Clesse-ml-2015wea}. In those cases, the production of massive PBH occurs in clusters, which subsequently merge during the matter era, soon after recombination, creating a stochastic background of GWs that lie, for large PBH masses ($M_{\rm PBH} \sim 10^2 - 10^4 M_\odot$), just within the range of LISA sensitivity~\cite{Clesse-ml-2016ajp}. These PBHs could constitute today all of the DM, and act as seeds for early galaxy formation, thus making the detection of the GWs background from merging of PBHs at high redshift a tantalizing signature of structure formation~\cite{Clesse-ml-2015wea}. It has been pointed out that these PBH could have already been detected by AdvLIGO~\cite{Bird-ml-2016dcv,Clesse-ml-2016vqa,Sasaki-ml-2016jop,Cholis-ml-2016kqi}. In the near future we may be able to measure the broad PBH mass distribution with ground and space interferometers~\cite{Clesse-ml-2016vqa,Clesse-ml-2016ajp}. Moreover, the full mass range constraints on PBH as DM were recently described in ref.~\cite{Carr-ml-2016drx}.

\subsection{Waterfall hybrid inflation model}

Hybrid inflation models end inflation when a symmetry breaking field $\psi$ triggers a quantum phase transition, while the inflaton $\phi$ is still in slow-roll. The original hybrid potential, with a constant plus a quadratic term, predicts a blue-tilted scalar spectrum which is ruled out by Planck, but other closely related models, like inverted hybrid, can be constructed, see also~\cite{Renaux-Petel-ml-2015mga}. Here we have considered a more general form for the effective potential close to the critical point of instability: it exhibits a negative curvature in order to generate a red spectrum, plus a linear term in $\phi$ to control the duration of the waterfall phase.  The two-field potential~reads
\begin{equation} \label{eq:potential}
V(\phi,\psi) = \Lambda \left[ \left(1-\frac{\psi^2}{v^2}\right)^2 + \frac{(\phi-\phi_c)}{\mu_1}
-  \frac{(\phi-\phi_c)^2}{\mu_2^2}+ \frac{2 \phi^2 \psi^2}{\phi_c^2 v^2} \right] \,.
\end{equation}
Initially, inflation takes place along the valley $\psi = 0$.  Below the critical value $\phi_c$, this potential develops a tachyonic instability, forcing the field trajectories to reach one of the global minima, located at $\phi = 0$, $\psi = \pm v$.   Apart from the negative curvature and the additional linear term, the potential is identical to the one of the original hybrid model.

The slope and the curvature of the potential at the critical point are thus controlled respectively by the mass parameters $\mu_1$ and $\mu_2$.  We assume that $\mu_1$ is sufficiently large compared to $\mu_2$ for the slope along the valley to be constant over the range of scales going from scales relevant to CMB anisotropies down do scales that exit the Hubble radius at the critical instability point.
At $\phi = \phi_c$, the slow-roll approximation is valid and the first and second Hubble-flow slow-roll parameters are given by
\begin{eqnarray}
{\epsone}_{\phi_c} &=& \frac{M_{\rm Pl}^2}{2} \left( \frac{V'}{V} \right)^2 = \frac{M_{\rm Pl}^2}{2 \mu_1^2},
\\
{\epstwo}_{\phi_c} &=& 2 M_{\rm Pl}^2  \left[ \left( \frac{V'}{V}  \right)^2 - \frac{V''}{V}  \right]  = 2 M_{\rm Pl}^2 \left( \frac{1}{\mu_1^2} + \frac{2}{\mu_2^2}   \right).
\end{eqnarray}
where $\Mpl$ is the reduced Planck mass and a prime denotes the derivative with respect to the field $\phi$.   In the regime of interest, $\mu_1 \gg \mu_2$ and the scalar spectral index, given by
\be
\ns = 1- 2 {\epsone}_* - {\epstwo}_* \simeq 1 - \frac{4 M_{\rm Pl}^2}{\mu_2^2}
\ee
is dominated by the contribution of the second slow-roll parameter.   The star index denotes quantities evaluated at the time $t_* $ when $k_* = a(t_*) H(t_*)$ with $k_*= 0.05~\Mpc^{-1}$ being the pivot scale used by Planck.  If the scalar spectral index is given by the best fit value from Planck (2015), $\ns \simeq 0.967$, one obtains
\be
\mu_2 = \frac{2 M_{\rm Pl}}{\sqrt{1-\ns}} \simeq 10 M_{\rm Pl}~.
\ee
The scalar power spectrum amplitude is also measured by Planck and is given at the pivot scale by
\begin{equation}
\mathcal P_\zeta(k_*)  =  \ \frac{H_*^2}{8 \pi^2 M_{\rm Pl}^2 {\epsone}_*}
\simeq \ \frac{\Lambda \mu_1^2}{12 \pi^2 M_{\rm Pl}^6} \left( \frac{k_*}{k_{\phi_c}}  \right)^{\ns - 1}  =  3.06 \times 10^{-9}.
\end{equation}
The second equality is derived by using the Friedmann equation in slow-roll $H^2 \simeq V /3 \Mpl^2$.  This leads to a relation between the $\Lambda$ and $\mu_1$ parameters,
\be \label{eq:Lambda}
\Lambda \simeq 3.06 \times 10^{-9} \times
\frac{12 \pi^2 M_{\rm Pl}^6}{\mu_1 ^2} \,,
\ee
where we have considered for simplicity the case of instantaneous reheating~\cite{GarciaBellido-ml-1997wm}.  Thus via the CMB, $\Lambda$ and $\mu_1$ are closely related.
On the other hand, the amplitude of tensor fluctuations at CMB scales from this model is characterized by $H^2 \simeq \Lambda/3\Mpl^2$ and
\be
r = \frac{8 M_{\rm Pl}^2}{\mu_1^2} \lesssim 0.07 \hspace{5mm} \Longrightarrow
\hspace{5mm}  \mu_1 \gtrsim 10 \,M_{\rm Pl}\,.
\ee
Therefore, CMB observables fix some of the parameters, but leave others free. We will see that observations of a GW background associated with a peak in the matter power spectrum at scales well below those of CMB and LSS will be able to further constrain the model. Let us explore here the waterfall regime responsible for such a peak in the scalar fluctuation~spectrum.

\subsection{Waterfall phase}

Due to the tachyonic instability, the symmetry breaking field $\psi$ will go through a process of quantum diffusion at the critical point $\phi_c$, starting the waterfall phase. This process may take less than an e-fold, and end inflation like the usual hybrid scenario, or take a few e-folds, during which backreaction on the scalar produces a sharp peak in the scalar metric fluctuations, responsible for PBH production at reentry, as first suggested in ref.~\cite{GarciaBellido-ml-1996qt}. If the diffusion process is slow it may take several e-folds to complete and the peak will be broad, as first discussed in ref.~\cite{Clesse-ml-2015wea}. We will consider here scenarios of the last type.

During diffusion we can treat the symmetry breaking field as a Gaussian random variable whose width at the critical point of instability can be calculated by integrating the quantum stochastic dynamics of $\psi$~\cite{GarciaBellido-ml-1996qt,Clesse-ml-2010iz,Clesse-ml-2013jra},
\be \label{eq:psi0}
\psi_0 \equiv \sqrt{\langle \psi^2 \rangle} = \left( \frac{\Lambda  }{96 \pi^{3/2} v \sqrt{2 \phi_c \mu_1} }   \right)^{1/2} ,
\ee
where brackets denote averaging in real space. Note that quantum diffusion only plays a role very close to the critical instability point and that classical dynamics is quickly recovered.  Quantum effects taking place after crossing $\phi_c $ actually influence only marginally the waterfall dynamics, but we have taken into account that different classical evolutions can emerge from various initial values $\psi_i$.  For each parameter set, the classical two-field dynamics is integrated numerically over many realizations of $\psi_i$, distributed according to a Gaussian of width $\psi_0$.  Then the mean scalar power spectrum is obtained by averaging over all realizations.  Since each realization can be more or less efficient in producing PBH (which is a non-linear process), the same averaging procedure is applied for the calculation of the fraction of the Universe that collapses into PBHs.

Note also that the inflaton field $\phi$ remains classical during the waterfall phase and drives the expansion of the Universe in the regime where the stochastic dynamics of $\psi$ is important. One can solve the two-field dynamics during the waterfall regime by introducing the notation
\be
\phi \equiv \phi_c {\rm e}^\xi, \hspace{1cm} \psi \equiv \psi_0 {\rm e}^\chi\,.
\ee
During the waterfall, as long as the slow-roll approximation is valid, we have $| \xi| \ll 1$.   One therefore has $\phi \simeq \phi_c (1 + \xi)$ and the evolution equations for the scalar fields in the slow-roll approximation reduce to
\begin{eqnarray}  \label{eq:KG1}
3 H \dot \xi &\simeq& - \frac{2 \Lambda}{\mu_1^2} \left(1+\frac{2 \mu_1^2 \psi^2}{v^2 \phi_c^2}  \right),
\\ \label{eq:KG2}
3 H \dot \chi  &\simeq& - \frac{4 \Lambda}{v^2} \left( 2 \xi + \frac{\psi^2}{v^2}  \right)\,.
\end{eqnarray}
During phase-1, the second term of eqs.~(\ref{eq:KG1}) and~(\ref{eq:KG2}) are by definition negligible.  At some time, the second term in the r.h.s.\ of eq.~(\ref{eq:KG1}) becomes larger than unity and the dynamics enters into phase-2. Integrating the slow-roll equations in phase-1 gives the field trajectories $\xi^2 = \chi\,v^2/4 \mu_1\,\phi_c$. The total number of $e$-folds in phase-1 is given by
\be
N_1 = \frac{v\,\sqrt{\chi_2\,\mu_1\,\phi_c}}{2 M_{\rm Pl}^2} = \frac{\sqrt{\chi_2}}{2}\,\Pi\,,
\ee
where
\be\label{PI}
\Pi \equiv \frac{v\,\sqrt{\mu_1\,\phi_c}}{M_{\rm Pl}^2}\,,
\ee
and $$\chi_2 = \ln \left(\frac{v\,\phi_c^{1/2}}{2  \mu_1^{1/2} \psi_0}\right) \simeq 10 + \frac{3}{2}\ln \Pi$$
is the value of $\chi$ at the transition point between phase-1 and phase-2.    The duration of the second phase is well approximated by
\be
N_2 = \frac{v\, \mu_1^{1/2} \phi_c^{1/2} }{4 M_{\rm Pl}^2 \chi_2^{1/2}}
= \frac{\Pi}{4\sqrt{\chi_2}}\,,
\ee
which determines the time of the transition.
Finally, inflation ends when the slow-roll approximation breaks down for the field $\psi$, at $\xi_{\rm end}  = - v^2/8\Mpl^2$. The final number of e-folds from the waterfall to the end of inflation is
\be\label{Nc}
N_c = N_1 + N_2 = \frac{\Pi}{2}\left(\sqrt{\chi_2}+ \frac{1}{2\sqrt{\chi_2}}\right) =
\frac{\Pi}{2}\sqrt{10+3\ln\sqrt\Pi}\left(1+ \frac{1}{20+3\ln\Pi}\right) \simeq 2\Pi \,,
\ee
which is just a function of $\Pi$.

\subsection{Matter power spectrum and formation of primordial black holes}

The power spectrum of scalar perturbations can be calculated numerically by integrating both the classical exact homogeneous dynamics and the linear perturbations. We have checked that these results coincide with the analytical expressions we obtained with the $\delta N$ formalism. The power spectrum at the peak can be written as~\cite{Clesse-ml-2015wea}
\be
{\cal P}_\zeta (k)  = \frac{\Pi^3}{\sqrt{2\pi}} \times \exp \left[-\frac{{4(N_c - N_k)}^2}{\Pi^2}  \right]\,,
\label{P:exponential}
\ee
which is maximal at the critical point of instability, $N_c$ e-folds before the end of inflation. The mild waterfall therefore induces a broad peak in the scalar power spectrum for modes leaving the horizon in phase-1, just before the critical point. Depending of the model parameters, the scalar perturbations can exceed a threshold value, leading to the formation of PBH.

\begin{figure}
\centering
\includegraphics[height=50mm]{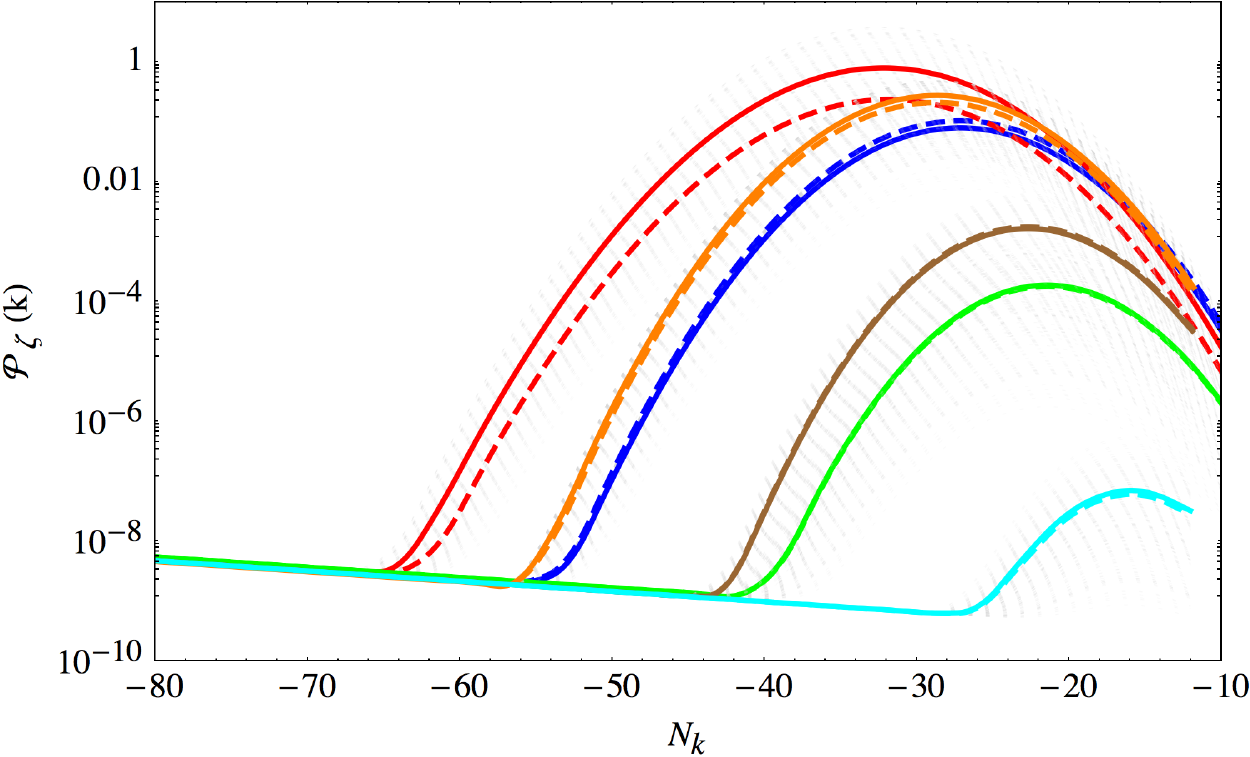}
\caption{\label{fig:Pzeta}Power spectrum of scalar perturbations for parameters values $M = 0.1 \Mpl$, $\mu_1 = 3 \times 10^5 \Mpl$ and $\phi_c = 0.125 \Mpl $ (red), $\phi_c = 0.1 \Mpl $ (blue) and $\phi_c = 0.075 \Mpl $ (green), $\phi_c = 0.1 \Mpl $ (blue) and $\phi_c = 0.05 \Mpl $ (cyan).  Those parameters correspond respectively to $\Pi^2 = 375 / 300 / 225/150$.  The power spectrum is degenerate for lower values of $M,\phi$ and larger values of $\mu_1$, keeping the combination $\Pi^2$ constant.  For larger values of $M, \phi_c$ the degeneracy is broken: power spectra in orange and brown are obtained respectively for $ M = \phi_c = \Mpl$ and $ \mu_1 = 300 \Mpl / 225 \Mpl$.  Dashed lines assume $\psi_c = \psi_0$ whereas solid lines are obtained after averaging over 200 power spectra obtained from initial conditions on $\psi_c$ distributed according to a Gaussian of width $\psi_0$.  The power spectra corresponding to these realizations are plotted in dashed light gray for illustration.  The $\Lambda$ parameter has been fixed so that the amplitude of the spectrum on CMB scales is in agreement with Planck data.  The parameter $\mu_2 = 10 \Mpl$ so that the scalar spectral index on those scales is given by $\ns = 0.96$.
}
\end{figure}

In figure~\ref{fig:Pzeta}  the power spectrum of scalar perturbations has been plotted for different values of the parameters.  This shows the strong enhancement of power not only for the modes exiting the Hubble radius in phase-1, but also for modes becoming super-horizon before field trajectories have crossed the critical point.   One can observe that if the waterfall lasts for about 35 $e$-folds then the modes corresponding to $ 35 \lesssim N_k \lesssim 50 $ are also affected.   As expected one can see also that the combination of parameters $\Pi$ drives the modifications of the power spectrum.  We find that it is hard to modify independently the width, the height and the position of the peak in the scalar power spectrum, since they are all correlated.

Peaks in the matter power spectrum collapse to form black holes when scalar fluctuations of large amplitude re-enter the horizon during the matter era. For gravitational collapse to end in the formation of a black hole one needs the amplitude of the fluctuation to be above a certain critical value $\zeta_c$ that has been evaluated both analytically and numerically. A recent analysis suggests $\zeta_c \simeq 0.03 - 0.3$. We will take, for definiteness, $\zeta_c = 0.1$.

Assuming that the probability distribution of density perturbations are Gaussian, one can evaluate the fraction $\beta$ of the Universe collapsing into primordial black holes of mass $M$ at the time of formation $t_M$ as
\be
\beta^{\rm form}(M)  \equiv   \left. \frac{\rho_{\rm PBH} (M) }{\rho_{\rm tot}} \right|_{t=t_M}
  =   \int_{\zeta_c}^{\infty} \frac{{\rm d} \zeta}{\sqrt{2 \pi} \sigma}\, e^{- \frac{\zeta^2}{2 \sigma^2}}  =  \frac{1}{2} \, {\rm erfc} \left( \frac{\zeta_c}{\sqrt 2 \sigma}  \right) \label{eq:betaform}.
\ee
In the limit where $\sigma \ll \zeta_c$, one gets
\be \label{eq:betaapprox}
\beta^{\rm form}(M) = \frac{ \sigma}{ \sqrt{2\pi}\, \zeta_c} \, e^{- \frac{\zeta_c^2}{2 \sigma^2 }}.
\ee
The variance $\sigma$ of scalar perturbations is related to the power spectrum through $\langle \zeta^2 \rangle = \sigma^2 = \mathcal P_\zeta (k_M) $, where $k_M$ is the wavelength of the mode re-entering inside the Hubble radius at time $t_M$.

In our scenario of mild waterfall, the peak in the power spectrum of scalar perturbations is broad and covers several order of magnitudes in  wavenumber.  Therefore, instead of a   distribution of black holes that would be close to monochromatic, which is easy to evolve in the radiation era, one expects that PBH have a broad mass spectrum and form at different times in the radiation era.   Since the energy density associated to PBH of mass $M$ decreases like $\sim a^{-3}$ due to expansion,  the contribution of PBH to the total energy density in the radiation era grows like $\sim a$.  As a result, at the end of the radiation era, PBH with low masses, forming earlier, contribute more importantly to the total energy density than more massive ones, forming later, given identical values of $\beta^{\rm form}$.

Taking into account those considerations, during the radiation-dominated era, the fraction of the Universe that has collapsed into primordial black-holes of mass $M_k$  evolves as
\be \label{eq:dbetadN}
\frac{{\rm d} \beta(M_{k},N(t))}{{\rm d} N} =  \beta(M_k,N(t))  \,.
\ee
Note that we have neglected evaporation through Hawking radiation since it is relevant only for PBH with very low masses that are formed immediately after inflation.   These are very subdominant in our model due to the duration of the waterfall.   In order to get $\beta^{\rm eq} \equiv \beta(M_k,N(t_{\rm eq}))$, this equation must be integrated over cosmic history, from the time of PBH formation until matter-radiation equality.   For all the considered scalar power spectra, the formation of PBH stops before $N_{\rm eq}$ (corresponding to $\ln (a_{\rm eq} / a_0) \simeq - 8 $), since the variance of scalar perturbations can be close or overpass the threshold value only in the range \hbox{$ -40 \lesssim - N_k \lesssim 10 $.}

The total density of PBHs at radiation-matter equality is obtained by integrating $\beta^{\rm eq}$ over masses:
\be \label{eq:OmPBH}
\Omega_{\rm PBH}(z_{\rm eq})  = \int_0^{M_{t_{\rm eq}}} \beta(M,N_{\rm eq} )  {\rm d} \ln M .
\ee eqs.~(\ref{eq:dbetadN}) and~(\ref{eq:OmPBH}) have been solved numerically using bins $\Delta N = 1$, corresponding to $\Delta \ln M = 2 $.  At matter-radiation equality one has
$\Omega_M(t_{\rm eq}) = 0.5 $ and PBH constitute the totality of the dark matter if $\Omega_{\rm PBH}(t_{\rm eq}) \simeq 0.42$, the rest coming from baryons.  For simplicity we have neglected the matter contribution to the Universe expansion in the radiation era.  This effect is only important close to matter-radiation equality, when all PBH are formed, and it is expected to be compensated by a small variation of $\zeta_c$.

For the parameter sets considered in figure~\ref{fig:Pzeta}, we have found the value of $\zeta_c$ that give rise to the right amount of dark matter.   They are reported in table~\ref{tab:omegaPBH}.  This must not be seen as an accurate result, because the matter contribution to the Universe's expansion is not accounted for in eq.~(\ref{eq:dbetadN}) even though it is not negligible in the last few $e$-folds before reaching matter-radiation equality.   This effect reduces the value of $\beta^{\rm eq} $, which must be compensated by a lower value of $\zeta_c $ to get the right amount of dark matter (thus values $\zeta_c /\zeta_{c,\,{\rm fid} } $ of a few tens can still be seen as realistic).

\begin{table}
\centering\renewcommand\arraystretch{1.4}
\begin{tabular}{|c|c|c|c|c|c|}
\hline
 $\Pi^2$ & $(\mu_1, v, \phi_c)$, in $\Mpl$  & $\zeta_c / \zeta_{c, {\rm fid}} $   \\
\hline
 375 & $(3 \times 10^5, 0.1, 0.125) $ & $88.06 $  \\
 300 & $(3 \times 10^5, 0.1, 0.1) $ & $18.96 $ \\
 300 & $(3 \times 10^8, 0.01, 0.01) $ & $17.37 $ \\
 300 & $(3 \times 10^2, 1.0, 1.0) $ & $ 49.60$  \\
 225 & $(3 \times 10^5, 0.1, 0.075)  $ & $2.009$   \\
 225 & $(2.25 \times 10^2, 1.0, 1.0) $ & $5.211$  \\
 150 & $(3 \times 10^5, 0.1, 0.05) $  & $0.0487$  \\
\hline
\end{tabular}
\caption{ Critical value $\zeta_c$ of scalar fluctuation (2nd column) leading to PBH formation with $\Omega_{\rm PBH} (z_{\rm eq}) = 0.42$ at matter radiation equality, for several sets of the model parameters (1st column).  The fiducial value is $\zeta_{c,\, {\rm fid}} = 0.1$.  }
 \label{tab:omegaPBH}
\end{table}

The masses of PBH can be computed very approximately by the mass within the horizon at the time the large fluctuation re-enters during the radiation era. This gives
\be\label{MPBH}
M_{\rm PBH}(N) = \gamma\,\frac{4\pi M_{\rm Pl}^2}{H_N}\,e^{2N} =
0.65\,g\,\frac{\gamma\,\mu_1}{M_{\rm Pl}}\,e^{2N}
\simeq 30\,\msun\,\frac{\mu_1}{10M_{\rm Pl}}\, e^{2(N-39.9)}\,,
\ee
where $\gamma\simeq0.2$ is an unknown factor describing the efficiency of collapse, and we have used eq.~(\ref{eq:Lambda}).

The mass range for PBHs is very broad, $10^{-20} \msun \lesssim M_{\rm PBH} \lesssim 10^{5} \msun$.  But given one set of parameters, the mass spectrum typically covers 3-5 orders of magnitudes at matter-radiation equality.  Given $\Pi^2$, we find that PBH can be made arbitrarily massive by increasing $\mu_1$ and reducing $v$ and $\phi_c$.  This lowers the energy scale of inflation and thus increases PBH masses, but this does not affect importantly the shape of the mass spectrum.  Therefore it is easy to find parameters for which the mass spectrum peaks in the range where there is no solid observational constraints.  It is also possible that the peak in the mass spectrum is located on planet-like masses at recombination (so that CMB distortion constraints are satisfied), but evade micro-lensing limits of PBHs abundances if merging induces their growth by more than two or three orders of magnitudes during cosmic history.  Finally, the width of the peak in  $ \beta^{\rm eq}$ is reduced for lower values of $\Pi^2$, as expected given that it is related to the broadness of the peak in the scalar power spectrum.  It is therefore possible, in principle, to control this width, but note that the range where $\Pi^2$ can vary is rather limited by the value of  $\zeta_c $, which needs to be realistic.

\subsection{Gravitational waves from inspiralling PBHs}
\label{sec:PBHs}

\begin{figure}
\centerline{
\includegraphics[width=7.0cm]{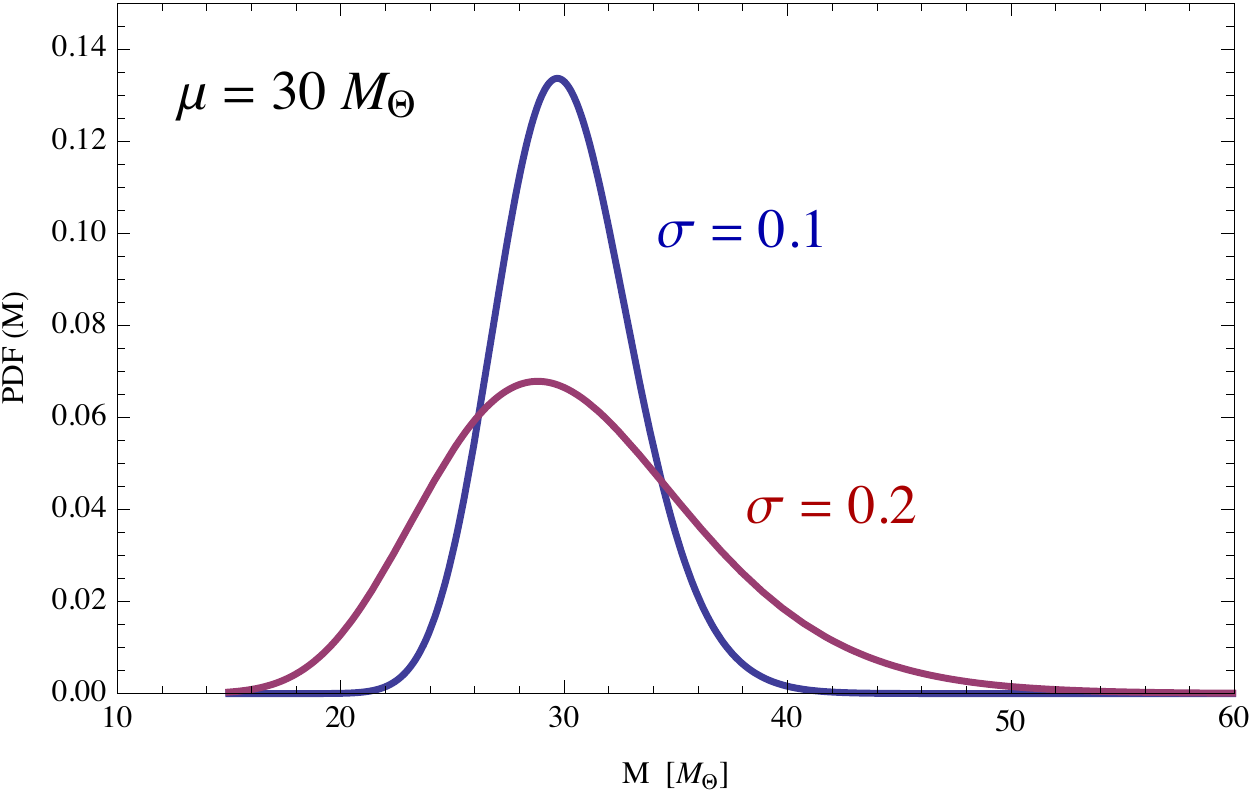}\qquad
\includegraphics[width=7.15cm]{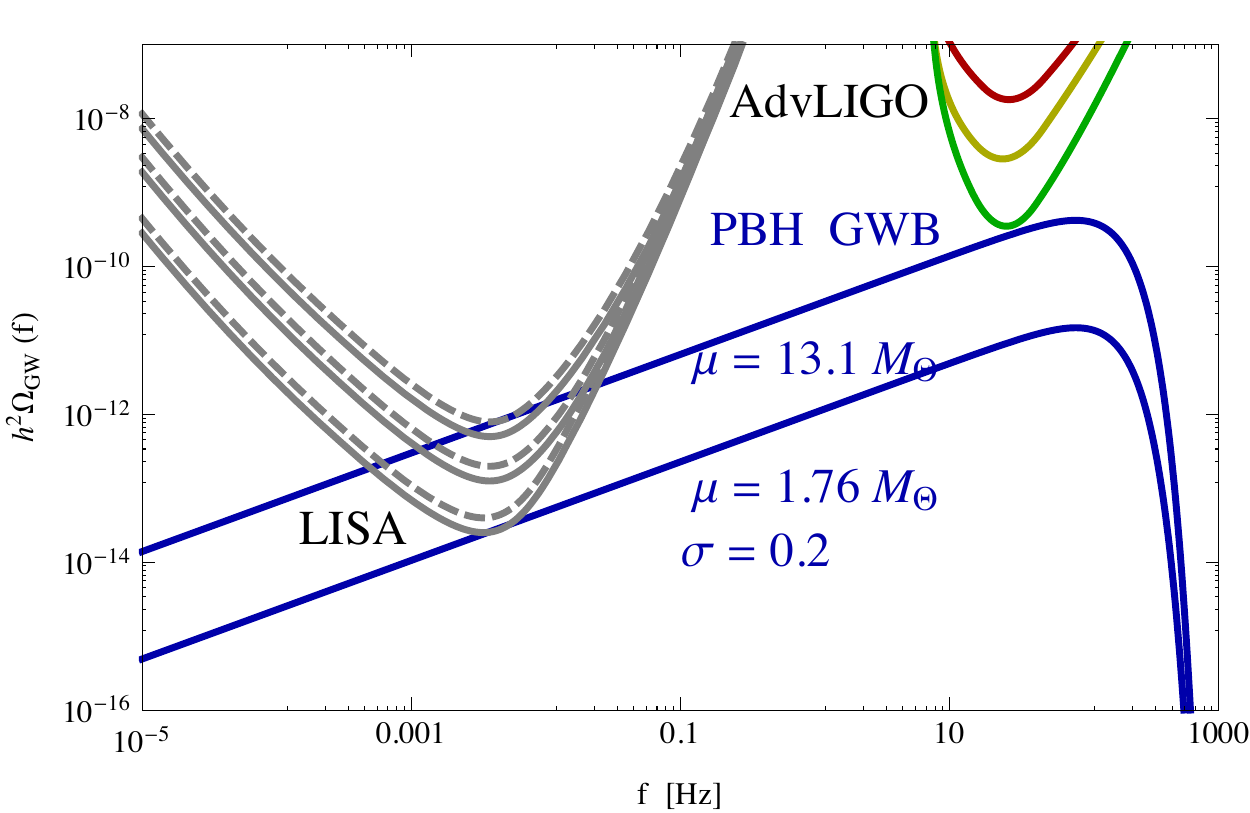}
}
\caption{Stochastic Gravitational Wave Background from inspiraling PBHs since recombination. The amplitude depends on the mass distribution of PBHs, see left panel, mainly through the mean and the width of the distribution. Here we assumed a fixed merger rate of 50 events per year and Gpc$^3$.}
\label{fig:PDF}
\end{figure}

Here we will assume that PBHs are distributed as a broad lognormal distribution
\begin{equation}
PDF(m) =  \frac{1}{m\sqrt{2\pi\sigma^2}} \exp\left(-
\frac{\log^2(m/\mu)}{2\sigma^2}\right)\,,
\label{eq:PDF}
\end{equation}
see figure~\ref{fig:PDF}a, coming from peaks in the power spectrum produced during inflation (e.g.\ during slow-waterfall hybrid inflation), which reenter inside the horizon during the radiation epoch and collapse to form black holes of different masses that are clustered and start to coalesce after recombination.

For the mild-waterfall hybrid inflation model of the previous section, the mean mass $\mu$ is given by eq.~(\ref{MPBH}) at
$N=N_c$, the number of e-folds to the end of inflation~(\ref{Nc}), which depends on the model parameter $\mu_1$, while the dispersion $\sigma$ is simply given by $\Pi$ in~(\ref{PI}).

The gravitational wave background from inspiraling black holes can be obtained straightforwardly from the GWs emission of binary systems, see ref.~\cite{Phinney-ml-2001di},
\begin{equation}
\Omega_{\rm GW}(f) =  \frac{2\pi^2}{3H_0^2}\,f^2\,h_c^2(f)
\equiv \frac{1}{\rho_c}\,\frac{d\,\rho_{\rm GW}}{d\,\ln f}\,,
\label{eq:GWB}
\end{equation}
where
\begin{equation}
\frac{d\,\rho_{\rm GW}}{d\,\ln f} = \int_0^\infty\frac{dz}{1+z}\,\frac{dn}{dz}\,
\frac{\pi^{2/3}}{3c^2}\,{\cal M}_c^{5/3}\,(G\,f_r)^{2/3}\,,
\label{eq:rhoGW}
\end{equation}
with ${\cal M}_c^{5/3}=m_1m_2(m_1+m_2)^{-1/3}$ the chirp mass and
$f_r=f(1+z)$ the restmass frequency at the source. The number density of GW events within the redshift interval $[z,z+dz]$ is given in terms of the merger rate in a comoving volume $\tau_{\rm merger}$ as
\begin{equation}
\frac{dn}{dz} = \tau_{\rm merger}\,\frac{dt}{dz} =
\frac{\tau_{\rm merger}}{H(z)(1+z)}\,,
\label{eq:tauM}
\end{equation}
with the Hubble rate given by $\Lambda$CDM, $H^2(z) = H_0^2\Big[\Omega_{\rm M}(1+z)^3 + \Omega_\Lambda\Big]$. Assuming a constant merger rate as a function of redshift, and doing the integral over redshift one finds an amplitude of GWs from inspiraling PBHs
\begin{equation}
h_c(f) = 1.14\times10^{-25} \ \tau_{\rm merger}^{1/2}
\left(\frac{f}{{\rm Hz}}\right)^{-2/3}
\left(\frac{{\cal M}_c}{M_\odot}\right)^{5/6}\,,
\label{eq:hcf}
\end{equation}
with typical values are $\tau_{\rm merger}\simeq 50$ yr$^{-1}$Gpc$^{-3}$ in the AdvLIGO detectors.

\begin{figure}
\centering
\includegraphics[width=7.10cm]{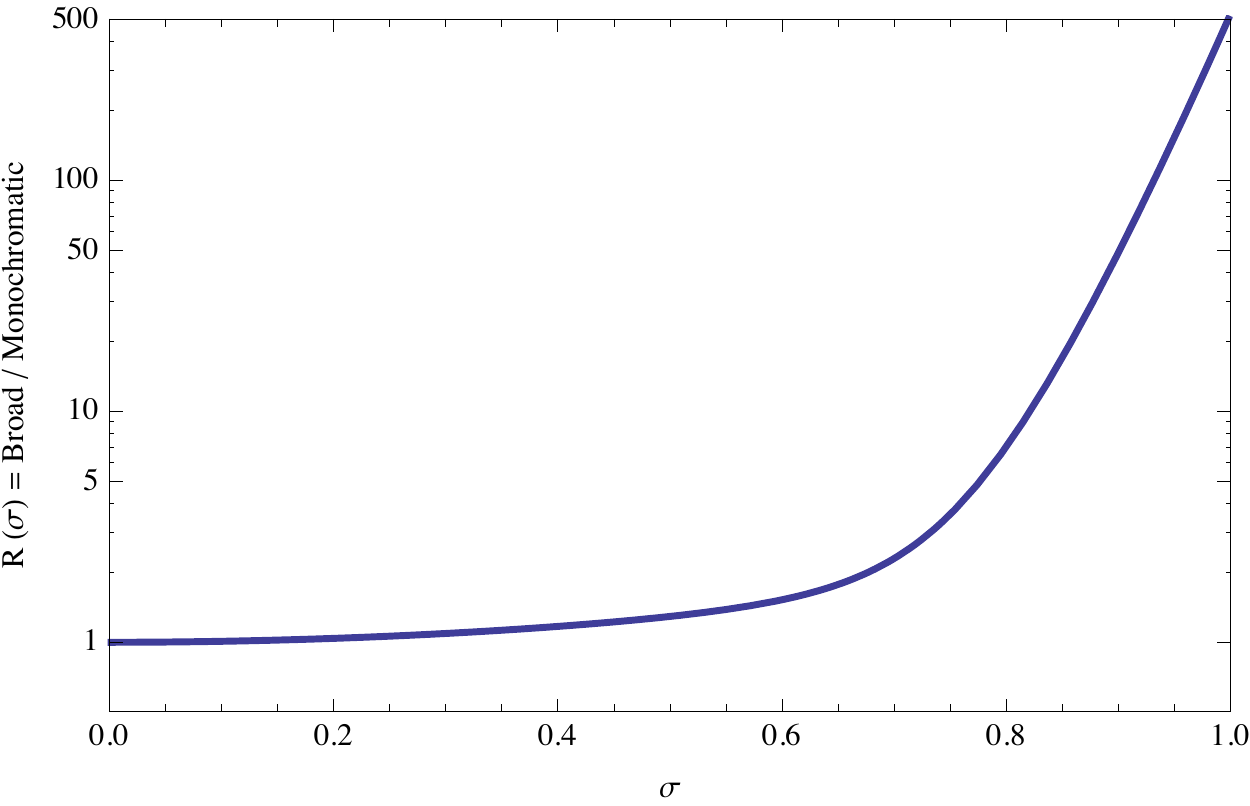}\quad
\includegraphics[width=7.10cm]{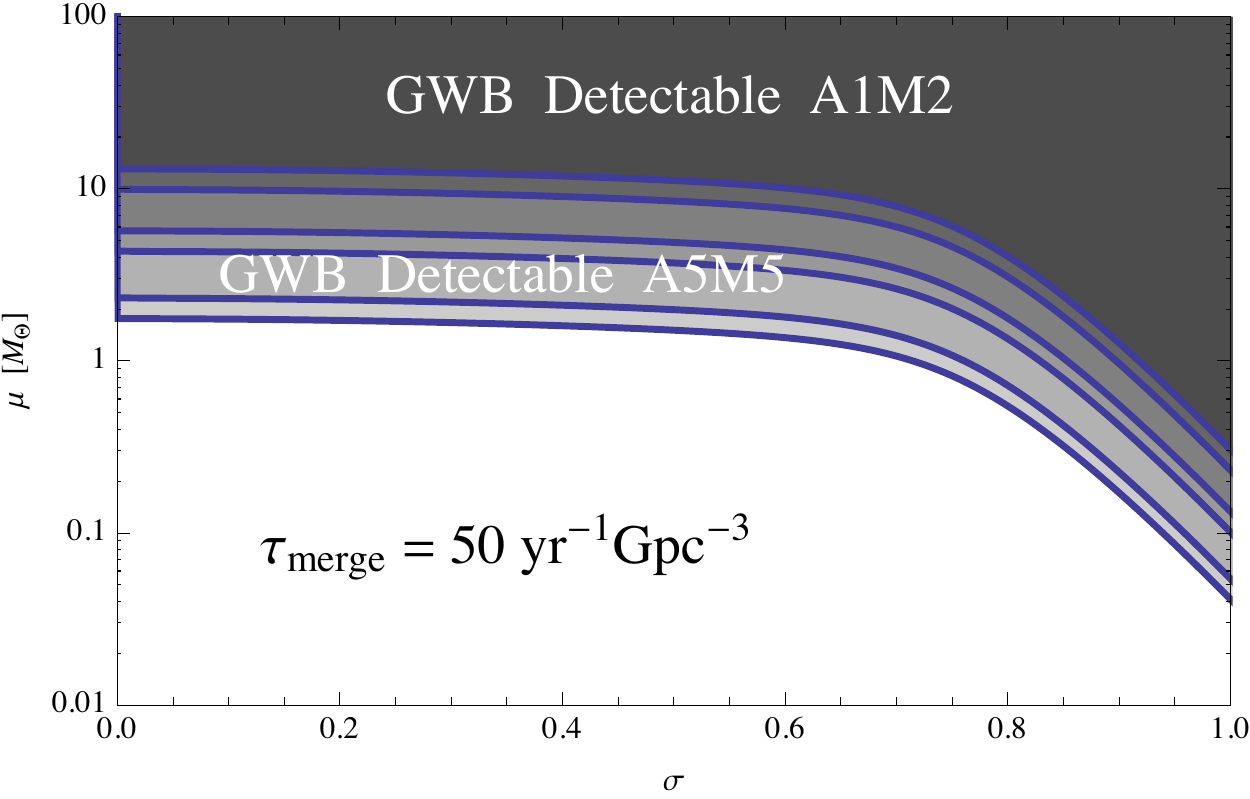}
\caption{The ratio of the stochastic GW background due to the existence of a broad mass spectrum of PBH versus a monochromatic spectrum. Right panel: regions in the $\sigma-\mu$ parameter space where LISA could detect the GW background from PBHs, for the six different sensitivity curves of LISA. We have assumed here a merger rate of 50 events per year and Gpc$^3$.\label{fig:ratio}}
\end{figure}

We can now integrate over masses with a broad mass distribution like~(\ref{eq:PDF}) for both $m_1$ and $m_2$, with the same parameters $(\mu,\,\sigma)$. This gives the final expression~\cite{Clesse-ml-2016vqa}
\begin{align}\label{ratio}
h^2\,\Omega_{\rm GW}(f) &= \ 8.15\times10^{-15} \ \tau_{\rm merger} \,
\left(\frac{f}{{\rm Hz}}\right)^{2/3}
\left(\frac{\mu}{M_\odot}\right)^{5/3} R(\sigma) \,, \\
R(\sigma) &= \frac{e^{\frac{793}{882}\,\sigma^2}}{1245889}
\left(639009 + 583443\,e^{\frac{2}{21}\,\sigma^2} + 30429\,
e^{\frac{40}{21}\,\sigma^2} - 9177\,e^{\frac{122}{21}\,\sigma^2}
+ 2185\,e^{\frac{82}{7}\,\sigma^2}\right)\,, \nonumber
\end{align}
which becomes $R(\sigma=0) = 1$ for a monochromatic spectrum with mass $M=\mu$, see figure~\ref{fig:ratio}a. The width of the mass spectrum is extremely important and can give a tremendous boost to the GWs background amplitude.

One can see in figure~\ref{fig:PDF}b a concrete case of the stochastic GW background from PBHs with $\mu=13.1\,M_\odot$ and $\sigma=0.2$ which could easily be detected by LISA.

It is clear that, from the point of view of parameter space, that there is a degeneracy between the merger rate and the mean mass $\mu$, which satisfies $\tau_{\rm merger}\times\mu^{5/3}(M_\odot)=$ const. Therefore, we will choose here to fix the merger rate to the middle of the range found by AdvLIGO, $\tau_{\rm merger} = 50$ events/yr/Gpc$^3$, and leave the mass $\mu$ free, together with the width $\sigma$ of the PDF~(\ref{eq:PDF}).

Now, taking into account the LISA sensitivity from the various configurations (A5M5 to A1M2), we notice that these particular GWs from inspiraling PBHs, with tilt $n_T=2/3$, can be detectable by LISA in a very wide range of parameters of the model. We plot the possible parameter range in figure~\ref{fig:ratio}b.

\section{Discussion and conclusions}
\label{sec:conclusions}

We have investigated the potential of the LISA space-based interferometer to detect the stochastic gravitational wave background produced from different mechanisms during inflation. We have focused on well-motivated scenarios which produce GW backgrounds with a large amplitude and tilt, very differently from the almost scale-invariant irreducible background due to vacuum tensor modes. We have studied the resulting GW signal from particle production during inflation (section~\ref{sec:ParticleProduction}), inflationary spectator fields with varying speed of sound (section~\ref{sec:Spectators}), effective field theories of inflation with new symmetry patterns (section~\ref{sec:EFT}) and inflationary models leading to the formation of primordial black holes (section~\ref{sec:PBH}). We have used the projected sensitivities of LISA in a model-independent way for various detector designs and configurations, demonstrating that LISA is capable of probing these well-motivated inflationary scenarios.


In the case of particle production during inflation, we have considered a broad class of well-motivated inflation models, where the inflaton $\phi$ is coupled to gauge fields via $\frac{\phi}{f}\,F^{\mu\nu}\,\tilde{F}_{\mu\nu}$. This operator, generally expected to be present in shift-symmetric models of inflation, leads to the amplification of the vacuum fluctuations of the gauge field, which in their turn are a source of GWs. The parity-violating and highly non-Gaussian nature of these gravitational waves is the smoking gun of this mechanism. We have presented two different ways of characterizing the detectability by LISA of the GWs generated this way. First, we have focused only on the dynamics of the system at LISA scales, connecting the amplitude and tilt of the signal to the parameter $\xi=\frac{\dot\phi}{2\,f\,H}$, that characterizes the strength of the inflaton-gauge coupling, and to the parameters that describe the inflationary dynamics at those scales. Our findings, summarized in figures~\ref{fig:Xi_vs_SR} and~\ref{fig:Xi_vs_H}, show that models with a Hubble rate $H\gtrsim 5 \cdot 10^{11}$ {\rm{GeV}} can produce a GW signal within LISA's reach if the parameter $\xi$ takes a value in the range $4\lesssim\xi\lesssim 5.5$. Then we have considered, in a less model independent but more powerful way, the global dynamics of the system, accounting for the constraints from observations at scales that are much larger (CMB) and  much smaller (PBHs, effective number of neutrinos) than the LISA ones. This mechanism can thus provide a powerful probe of the dynamics of inflation during the $\sim 30$ e-folds that separate CMB scales to LISA ones.

In the case of having inflationary spectator fields present during inflation,  classical production of GWs also takes place. The amplitude and spectral index of such a GW background turn out to be determined by the sound speed of the spectator field(s), as well as by the time variation of the latter. Interestingly, this GW background can be expected to be blue tilted and to exceed the sensitivity of LISA. Considering the parameter space which describes the spectator sector, we found that LISA is expected to add information which are complementary to the constraints provided by current CMB measurements. At the same time, the best configuration of LISA is expected to slightly improve current bounds obtained from others GW experiments at small scales. We notice that comparing contraints at CMB scales with bounds at smaller scales, provides the possibility of discriminating between different inflationary GW signals. Let us recall that a complete computation of the scalar power spectrum and its related non-Gaussianity is still missing and this may impact on the results of our analysis, in particular changing 
the conclusions about PBH bounds.

\enlargethispage{\baselineskip}

We have considered scenarios where space-reparametrization can be  spontaneously broken during inflation. Then, there is no symmetry preventing the graviton from having a mass during inflation. We examined this possibility using an approach based on EFT of inflation, including   scenarios where the tensors can have generic sound speed.  These   properties  influence the amplitude and scale dependence of primordial tensor spectrum, allowing for  a blue  tensor tilt, and a  spectrum enhanced and detectable at LISA frequency  scales.    After discussing explicit examples of  models with   these   properties, we focussed our analysis   on  a simple, representative case. We  showed that, in order for being detectable with LISA, the graviton mass during inflation should lie within certain ranges,  depending on the tensor sound speed, and the value of the inflationary Hubble parameter. We then  presented plots  with the allowed  regions in the space of available parameters  for ensuring a detection with LISA. We compared with LIGO detectors,  showing that  LISA can probe regions of  larger size in   parameter space. Finally, we discussed specific  predictions of these scenarios in the scalar inflationary sector, which make  models with broken space-reparametrization distinguishable from other inflationary scenarios with small scale enhancements of the tensor spectrum.

Finally, in the case of certain models of inflation, like the mild-waterfall hybrid model, or due to particle production well before the end of inflation, large peaks appear in the matter power spectrum, that later collapse to form primordial black holes, at horizon reentry during the radiation era. These PBHs are strongly clustered, and merge within the age of the Universe, generating a stochastic background of GWs that could be detected by LISA. Some of these late mergings may have already been observed by AdvLIGO. Furthermore, for certain parameters of the models, these PBHs could constitute all of the dark matter in the~Universe.

In summary, in this paper we have addressed the capability of the LISA mission for extracting information from the inflationary era, studying the parameter space compatible with a detection/non-detection of a GW signal with LISA. We have quantified the ability of LISA to probe inflation, focussing in the above four well motivated family of inflationary scenarios. Our study clearly assesses that  LISA will be able to test the latest stages of the inflationary period, to probe the couplings of the inflaton to other degrees of freedom, or simply the presence of extra fields besides the inflaton, and to probe the degree of violation of the inflationary consistency relation. We have combined our results for LISA with independent constraints coming from other probes at different scales. From our analysis we argue that measurements of a GW signal on the small scales accessible to LISA, will become of fundamental importance in order to provide constraints on tensor perturbations complementary to the CMB.

\acknowledgments

We thank Germano Nardini for helping in this project. We thank the University of Stavanger for hosting the second eLISA workshop, where this work has started. M.C.G. and N.B. thank Matteo Fasiello for useful correspondence. V.D. acknowledges the financial support of the UnivEarthS Labex program at Sorbonne Paris Cit\'e (ANR-10-LABX-0023 and ANR-11-IDEX-0005-02) and the Paris Centre for Cosmological Physics. The work of J.G-B. is partially supported by the Research Project of the Spanish MINECO, FPA2013-47986-03-3P, and the Centro de Excelencia Severo Ochoa Program SEV-2012-0249. M.C.G. thanks financial support from a Cariparo foundation grant. N.B., M.L. and S.M. acknowledge partial financial support by the ASI/INAF Agreement I/072/09/0 for the Planck LFI Activity of Phase E2. The work of M.P. is partially supported from the DOE grant DE-SC0011842 at the University of Minnesota. The work of L.S. is partially supported by the US NSF grant PHY-1520292. The work of G.T. is partially supported by STFC grant ST/N001435/1.


\begin{thebibliography}{100}

\bibitem{Abbott-ml-2016blz}
%
{\scshape Virgo, LIGO Scientific} collaboration, B.P. Abbott et~al.,
   \emph{Observation of gravitational waves from a binary black hole merger},
\href{http://dx.doi.org/10.1103/PhysRevLett.116.061102}{\emph{Phys.\ Rev.\
  Lett.} {\bf 116} (2016) 061102} [\arXivid{1602.03837}] [\inspire{EPRINT+arXiv:1602.03837}].

\bibitem{Abbott-ml-2016nmj}
%
{\scshape Virgo, LIGO Scientific} collaboration, B.P. Abbott et~al.,
   \emph{GW151226: observation of gravitational waves from a $22$-solar-mass
  binary black hole coalescence},
\href{http://dx.doi.org/10.1103/PhysRevLett.116.241103}{\emph{Phys.\ Rev.\
  Lett.} {\bf 116} (2016) 241103} [\arXivid{1606.04855}] [\inspire{EPRINT+arXiv:1606.04855}].

\bibitem{Harry-ml-2010zz}
%
{\scshape LIGO Scientific} collaboration, G.M. Harry,  \emph{Advanced LIGO:
  the next generation of gravitational wave detectors},
\href{http://dx.doi.org/10.1088/0264-9381/27/8/084006}{\emph{Class.\ Quant.\
  Grav.} {\bf 27} (2010) 084006}
[\inspire{J+\%22Class.Quant.Grav.,27,084006\%22}].

\bibitem{Acernese-ml-2015gua}
%
{\scshape Virgo} collaboration, F.~Acernese,  \emph{The advanced Virgo
  detector},
\href{http://dx.doi.org/10.1088/1742-6596/610/1/012014}{\emph{J.\
  Phys.\ Conf.\ Ser.} {\bf 610} (2015) 012014}
[\inspire{J+\%22J.Phys.Conf.Ser.,610,012014\%22}].

\bibitem{Somiya-ml-2011np}
%
{\scshape KAGRA} collaboration, K.~Somiya,  \emph{Detector configuration of
  KAGRA: The Japanese cryogenic gravitational-wave detector},
\href{http://dx.doi.org/10.1088/0264-9381/29/12/124007}{\emph{Class.\ Quant.\
  Grav.} {\bf 29} (2012) 124007} [\arXivid{1111.7185}] [\inspire{EPRINT+arXiv:1111.7185}].

\bibitem{LIGO-India}
\href{http://gw-indigo.org/tiki-index.php?page=LIGO-India/}{http://gw-indigo.org/tiki-index.php?page=LIGO-India/}.

\bibitem{Sathyaprakash-ml-2012jk}
%
B.~Sathyaprakash et~al.,  \emph{Scientific objectives of einstein telescope},
\href{http://dx.doi.org/10.1088/0264-9381/29/12/124013}{\emph{Class.\ Quant.\
  Grav.} {\bf 29} (2012) 124013} [\erratum{30}{2013}{079501}]  [\arXivid{1206.0331}] [\inspire{J+\%22Class.Quant.Grav.,29,124013\%22}].

\bibitem{AmaroSeoane-ml-2012km}
%
P.~Amaro-Seoane et~al.,  \emph{eLISA/NGO: astrophysics and cosmology in the
  gravitational-wave millihertz regime},
{\emph{GW Notes} {\bf 6} (2013) 4} [\arXivid{1201.3621}] [\inspire{EPRINT+arXiv:1201.3621}].

\bibitem{Khlebnikov-ml-1997di}
%
S.Y. Khlebnikov and I.I. Tkachev,  \emph{Relic gravitational waves produced
  after preheating},
\href{http://dx.doi.org/10.1103/PhysRevD.56.653}{\emph{Phys.\ Rev.} {\bf D 56}
  (1997) 653} [\hepph{9701423}] [\inspire{EPRINT+hep-ph/9701423}].

\bibitem{Easther-ml-2006gt}
%
R.~Easther and E.A. Lim,  \emph{Stochastic gravitational wave production after
  inflation},
\jcap{04}{2006}{010} [\astroph{0601617}] [\inspire{EPRINT+astro-ph/0601617}].

\bibitem{Easther-ml-2006vd}
%
R.~Easther, J.T. Giblin, Jr. and E.A. Lim,  \emph{Gravitational wave
  production at the end of inflation},
\href{http://dx.doi.org/10.1103/PhysRevLett.99.221301}{\emph{Phys.\ Rev.\
  Lett.} {\bf 99} (2007) 221301} [\astroph{0612294}] [\inspire{EPRINT+astro-ph/0612294}].

\bibitem{GarciaBellido-ml-2007dg}
%
J.~Garc\'ia-Bellido and D.G. Figueroa,  \emph{A stochastic background of
  gravitational waves from hybrid preheating},
\href{http://dx.doi.org/10.1103/PhysRevLett.98.061302}{\emph{Phys.\ Rev.\
  Lett.} {\bf 98} (2007) 061302} [\astroph{0701014}] [\inspire{EPRINT+astro-ph/0701014}].

\bibitem{GarciaBellido-ml-2007af}
%
J.~Garc\'ia-Bellido, D.G. Figueroa and A.~Sastre,  \emph{A gravitational wave
  background from reheating after hybrid inflation},
\href{http://dx.doi.org/10.1103/PhysRevD.77.043517}{\emph{Phys.\ Rev.} {\bf D 77} (2008) 043517} [\arXivid{0707.0839}] [\inspire{EPRINT+arXiv:0707.0839}].

\bibitem{Dufaux-ml-2007pt}
%
J.F. Dufaux, A.~Bergman, G.N. Felder, L.~Kofman and J.-P. Uzan,  \emph{Theory
  and numerics of gravitational waves from preheating after inflation},
\href{http://dx.doi.org/10.1103/PhysRevD.76.123517}{\emph{Phys.\ Rev.} {\bf D 76} (2007) 123517} [\arXivid{0707.0875}] [\inspire{EPRINT+arXiv:0707.0875}].

\bibitem{Dufaux-ml-2008dn}
%
J.-F. Dufaux, G.~Felder, L.~Kofman and O.~Navros,  \emph{Gravity waves from
  tachyonic preheating after hybrid inflation},
\jcap{03}{2009}{001} [\arXivid{0812.2917}] [\inspire{EPRINT+arXiv:0812.2917}].

\bibitem{Figueroa-ml-2011ye}
%
D.G. Figueroa, J.~Garc\'ia-Bellido and A.~Rajantie,  \emph{On the
  transverse-traceless projection in lattice simulations of gravitational wave
  production},
\jcap{11}{2011}{015} [\arXivid{1110.0337}] [\inspire{EPRINT+arXiv:1110.0337}].

\bibitem{Bethke-ml-2013aba}
%
L.~Bethke, D.G. Figueroa and A.~Rajantie,  \emph{Anisotropies in the
  gravitational wave background from preheating},
\href{http://dx.doi.org/10.1103/PhysRevLett.111.011301}{\emph{Phys.\ Rev.\
  Lett.} {\bf 111} (2013) 011301} [\arXivid{1304.2657}] [\inspire{EPRINT+arXiv:1304.2657}].

\bibitem{Bethke-ml-2013vca}
%
L.~Bethke, D.G. Figueroa and A.~Rajantie,  \emph{On the anisotropy of the
  gravitational wave background from massless preheating},
\jcap{06}{2014}{047} [\arXivid{1309.1148}] [\inspire{EPRINT+arXiv:1309.1148}].

\bibitem{Enqvist-ml-2012im}
%
K.~Enqvist, D.G. Figueroa and T.~Meriniemi,  \emph{Stochastic background of
  gravitational waves from fermions},
\href{http://dx.doi.org/10.1103/PhysRevD.86.061301}{\emph{Phys.\ Rev.} {\bf D 86} (2012) 061301} [\arXivid{1203.4943}] [\inspire{EPRINT+arXiv:1203.4943}].

\bibitem{Figueroa-ml-2013vif}
%
D.G. Figueroa and T.~Meriniemi,  \emph{Stochastic background of gravitational
  waves from fermions --- Theory and applications},
\jhep{10}{2013}{101} [\arXivid{1306.6911}] [\inspire{EPRINT+arXiv:1306.6911}].

\bibitem{Figueroa-ml-2014aya}
%
D.G. Figueroa,  \emph{A gravitational wave background from the decay of the
  standard model Higgs after inflation},
\jhep{11}{2014}{145} [\arXivid{1402.1345}] [\inspire{EPRINT+arXiv:1402.1345}].

\bibitem{Figueroa-ml-2016ojl}
%
D.G. Figueroa, J.~Garc\'{\i}a-Bellido and F.~Torrent\'{\i},  \emph{Gravitational wave
  production from the decay of the standard model Higgs field after
  inflation},
\href{http://dx.doi.org/10.1103/PhysRevD.93.103521}{\emph{Phys.\
  Rev.} {\bf D 93} (2016) 103521} [\arXivid{1602.03085}] [\inspire{EPRINT+arXiv:1602.03085}].

\bibitem{Antusch-ml-2016con}
%
S.~Antusch, F.~Cefala and S.~Orani,  \emph{Gravitational waves from oscillons
  after inflation},
\arXivid{1607.01314}
[\inspire{EPRINT+arXiv:1607.01314}].

\bibitem{Kosowsky-ml-1991ua}
%
A.~Kosowsky, M.S. Turner and R.~Watkins,  \emph{Gravitational radiation from
  colliding vacuum bubbles},
\href{http://dx.doi.org/10.1103/PhysRevD.45.4514}{\emph{Phys.\ Rev.} {\bf D 45} (1992) 4514}
[\inspire{J+\%22Phys.Rev.,D45,4514\%22}].

\bibitem{Kosowsky-ml-1992rz}
%
A.~Kosowsky, M.S. Turner and R.~Watkins,  \emph{Gravitational waves from first
  order cosmological phase transitions},
\href{http://dx.doi.org/10.1103/PhysRevLett.69.2026}{\emph{Phys.\ Rev.\
  Lett.} {\bf 69} (1992) 2026}
[\inspire{J+\%22Phys.Rev.Lett.,69,2026\%22}].

\bibitem{Kosowsky-ml-1992vn}
%
A.~Kosowsky and M.S. Turner,  \emph{Gravitational radiation from colliding
  vacuum bubbles: envelope approximation to many bubble collisions},
\href{http://dx.doi.org/10.1103/PhysRevD.47.4372}{\emph{Phys.\ Rev.} {\bf D 47} (1993) 4372} [\astroph{9211004}] [\inspire{EPRINT+astro-ph/9211004}].

\bibitem{Kamionkowski-ml-1993fg}
%
M.~Kamionkowski, A.~Kosowsky and M.S. Turner,  \emph{Gravitational radiation
  from first order phase transitions},
\href{http://dx.doi.org/10.1103/PhysRevD.49.2837}{\emph{Phys.\ Rev.} {\bf D 49} (1994) 2837} [\astroph{9310044}] [\inspire{EPRINT+astro-ph/9310044}].

\bibitem{Grojean-ml-2006bp}
%
C.~Grojean and G.~Servant,  \emph{Gravitational waves from phase transitions at
  the electroweak scale and beyond},
\href{http://dx.doi.org/10.1103/PhysRevD.75.043507}{\emph{Phys.\ Rev.} {\bf D 75} (2007) 043507} [\hepph{0607107}] [\inspire{EPRINT+hep-ph/0607107}].

\bibitem{Caprini-ml-2007xq}
%
C.~Caprini, R.~Durrer and G.~Servant,  \emph{Gravitational wave generation from
  bubble collisions in first-order phase transitions: An analytic approach},
\href{http://dx.doi.org/10.1103/PhysRevD.77.124015}{\emph{Phys.\ Rev.} {\bf D 77} (2008) 124015} [\arXivid{0711.2593}] [\inspire{EPRINT+arXiv:0711.2593}].

\bibitem{Huber-ml-2008hg}
%
S.J. Huber and T.~Konstandin,  \emph{Gravitational wave production by
  collisions: more bubbles},
\jcap{09}{2008}{022} [\arXivid{0806.1828}] [\inspire{EPRINT+arXiv:0806.1828}].

\bibitem{Kahniashvili-ml-2008pf}
%
T.~Kahniashvili, A.~Kosowsky, G.~Gogoberidze and Y.~Maravin,
   \emph{Detectability of gravitational waves from phase transitions},
\href{http://dx.doi.org/10.1103/PhysRevD.78.043003}{\emph{Phys.\ Rev.} {\bf D 78} (2008) 043003} [\arXivid{0806.0293}] [\inspire{EPRINT+arXiv:0806.0293}].

\bibitem{Kahniashvili-ml-2009mf}
%
T.~Kahniashvili, L.~Kisslinger and T.~Stevens,  \emph{Gravitational radiation
  generated by magnetic fields in cosmological phase transitions},
\href{http://dx.doi.org/10.1103/PhysRevD.81.023004}{\emph{Phys.\ Rev.} {\bf D 81} (2010) 023004} [\arXivid{0905.0643}] [\inspire{EPRINT+arXiv:0905.0643}].

\bibitem{Caprini-ml-2009fx}
%
C.~Caprini, R.~Durrer, T.~Konstandin and G.~Servant,  \emph{General properties
  of the gravitational wave spectrum from phase transitions},
\href{http://dx.doi.org/10.1103/PhysRevD.79.083519}{\emph{Phys.\ Rev.} {\bf D 79} (2009) 083519} [\arXivid{0901.1661}] [\inspire{EPRINT+arXiv:0901.1661}].

\bibitem{Child-ml-2012qg}
%
H.L. Child and J.T. Giblin Jr.,  \emph{Gravitational radiation from
  first-order phase transitions},
\jcap{10}{2012}{001} [\arXivid{1207.6408}] [\inspire{EPRINT+arXiv:1207.6408}].

\bibitem{Hindmarsh-ml-2013xza}
%
M.~Hindmarsh, S.J. Huber, K.~Rummukainen and D.J. Weir,  \emph{Gravitational
  waves from the sound of a first order phase transition},
\href{http://dx.doi.org/10.1103/PhysRevLett.112.041301}{\emph{Phys.\ Rev.\
  Lett.} {\bf 112} (2014) 041301} [\arXivid{1304.2433}] [\inspire{EPRINT+arXiv:1304.2433}].

\bibitem{Giblin-ml-2014qia}
%
J.T. Giblin and J.B. Mertens,  \emph{Gravitional radiation from first-order
  phase transitions in the presence of a fluid},
\href{http://dx.doi.org/10.1103/PhysRevD.90.023532}{\emph{Phys.\ Rev.} {\bf D 90} (2014) 023532} [\arXivid{1405.4005}] [\inspire{EPRINT+arXiv:1405.4005}].

\bibitem{Hindmarsh-ml-2015qta}
%
M.~Hindmarsh, S.J. Huber, K.~Rummukainen and D.J. Weir,  \emph{Numerical
  simulations of acoustically generated gravitational waves at a first order
  phase transition},
\href{http://dx.doi.org/10.1103/PhysRevD.92.123009}{\emph{Phys.\ Rev.} {\bf D 92} (2015) 123009} [\arXivid{1504.03291}] [\inspire{EPRINT+arXiv:1504.03291}].

\bibitem{Kisslinger-ml-2015hua}
%
L.~Kisslinger and T.~Kahniashvili,  \emph{Polarized gravitational waves from
  cosmological phase transitions},
\href{http://dx.doi.org/10.1103/PhysRevD.92.043006}{\emph{Phys.\ Rev.} {\bf D 92} (2015) 043006} [\arXivid{1505.03680}] [\inspire{EPRINT+arXiv:1505.03680}].

\bibitem{Caprini-ml-2015zlo}
%
C.~Caprini et~al.,  \emph{Science with the space-based interferometer eLISA.
  II: gravitational waves from cosmological phase transitions},
\jcap{04}{2016}{001} [\arXivid{1512.06239}] [\inspire{EPRINT+arXiv:1512.06239}].

\bibitem{Weir-ml-2016tov}
%
D.J. Weir,  \emph{Revisiting the envelope approximation: gravitational waves
  from bubble collisions},
\href{http://dx.doi.org/10.1103/PhysRevD.93.124037}{\emph{Phys.\ Rev.} {\bf D 93} (2016) 124037} [\arXivid{1604.08429}] [\inspire{EPRINT+arXiv:1604.08429}].

\bibitem{Vachaspati-ml-1984gt}
%
T.~Vachaspati and A.~Vilenkin,  \emph{Gravitational radiation from cosmic
  strings},
\href{http://dx.doi.org/10.1103/PhysRevD.31.3052}{\emph{Phys.\
  Rev.} {\bf D 31} (1985) 3052}
[\inspire{J+\%22Phys.Rev.,D31,3052\%22}].

\bibitem{Caldwell-ml-1996en}
%
R.R. Caldwell, R.A. Battye and E.P.S. Shellard,  \emph{Relic gravitational
  waves from cosmic strings: updated constraints and opportunities for
  detection},
\href{http://dx.doi.org/10.1103/PhysRevD.54.7146}{\emph{Phys.\
  Rev.} {\bf D 54} (1996) 7146} [\astroph{9607130}] [\inspire{EPRINT+astro-ph/9607130}].

\bibitem{Damour-ml-2000wa}
%
T.~Damour and A.~Vilenkin,  \emph{Gravitational wave bursts from cosmic
  strings},
\href{http://dx.doi.org/10.1103/PhysRevLett.85.3761}{\emph{Phys.\
  Rev.\ Lett.} {\bf 85} (2000) 3761} [\grqc{0004075}] [\inspire{EPRINT+gr-qc/0004075}].

\bibitem{Damour-ml-2001bk}
%
T.~Damour and A.~Vilenkin,  \emph{Gravitational wave bursts from cusps and
  kinks on cosmic strings},
\href{http://dx.doi.org/10.1103/PhysRevD.64.064008}{\emph{Phys.\ Rev.} {\bf D 64} (2001) 064008} [\grqc{0104026}] [\inspire{EPRINT+gr-qc/0104026}].

\bibitem{Damour-ml-2004kw}
%
T.~Damour and A.~Vilenkin,  \emph{Gravitational radiation from cosmic
  (super)strings: bursts, stochastic background and observational windows},
\href{http://dx.doi.org/10.1103/PhysRevD.71.063510}{\emph{Phys.\ Rev.} {\bf D 71} (2005) 063510} [\hepth{0410222}] [\inspire{EPRINT+hep-th/0410222}].

\bibitem{Siemens-ml-2006yp}
%
X.~Siemens, V.~Mandic and J.~Creighton,  \emph{Gravitational wave stochastic
  background from cosmic (super)strings},
\href{http://dx.doi.org/10.1103/PhysRevLett.98.111101}{\emph{Phys.\ Rev.\
  Lett.} {\bf 98} (2007) 111101} [\astroph{0610920}] [\inspire{EPRINT+astro-ph/0610920}].

\bibitem{Siemens-ml-2006vk}
%
X.~Siemens, J.~Creighton, I.~Maor, S.~Ray~Majumder, K.~Cannon and J.~Read,
   \emph{Gravitational wave bursts from cosmic (super)strings: quantitative
  analysis and constraints},
\href{http://dx.doi.org/10.1103/PhysRevD.73.105001}{\emph{Phys.\ Rev.} {\bf D 73} (2006) 105001} [\grqc{0603115}] [\inspire{EPRINT+gr-qc/0603115}].

\bibitem{JonesSmith-ml-2007ne}
%
K.~Jones-Smith, L.M. Krauss and H.~Mathur,  \emph{A nearly scale invariant
  spectrum of gravitational radiation from global phase transitions},
\href{http://dx.doi.org/10.1103/PhysRevLett.100.131302}{\emph{Phys.\ Rev.\
  Lett.} {\bf 100} (2008) 131302} [\arXivid{0712.0778}] [\inspire{EPRINT+arXiv:0712.0778}].

\bibitem{Fenu-ml-2009qf}
%
E.~Fenu, D.G. Figueroa, R.~Durrer and J.~Garc\'ia-Bellido,  \emph{Gravitational
  waves from self-ordering scalar fields},
\jcap{10}{2009}{005} [\arXivid{0908.0425}] [\inspire{EPRINT+arXiv:0908.0425}].

\bibitem{Figueroa-ml-2012kw}
%
D.G. Figueroa, M.~Hindmarsh and J.~Urrestilla,  \emph{Exact scale-invariant
  background of gravitational waves from cosmic defects},
\href{http://dx.doi.org/10.1103/PhysRevLett.110.101302}{\emph{Phys.\ Rev.\
  Lett.} {\bf 110} (2013) 101302} [\arXivid{1212.5458}] [\inspire{EPRINT+arXiv:1212.5458}].

\bibitem{Olmez-ml-2010bi}
%
S.~Olmez, V.~Mandic and X.~Siemens,  \emph{Gravitational-wave stochastic
  background from kinks and cusps on cosmic strings},
\href{http://dx.doi.org/10.1103/PhysRevD.81.104028}{\emph{Phys.\ Rev.} {\bf D 81} (2010) 104028} [\arXivid{1004.0890}] [\inspire{EPRINT+arXiv:1004.0890}].

\bibitem{Dufaux-ml-2010cf}
%
J.-F. Dufaux, D.G. Figueroa and J.~Garc\'ia-Bellido,  \emph{Gravitational waves
  from abelian gauge fields and cosmic strings at preheating},
\href{http://dx.doi.org/10.1103/PhysRevD.82.083518}{\emph{Phys.\ Rev.} {\bf D 82} (2010) 083518} [\arXivid{1006.0217}] [\inspire{EPRINT+arXiv:1006.0217}].

\bibitem{Binetruy-ml-2012ze}
%
P.~Binetruy, A.~Bohe, C.~Caprini and J.-F. Dufaux,  \emph{Cosmological
  backgrounds of gravitational waves and eLISA/NGO: phase transitions, cosmic
  strings and other sources},
\jcap{06}{2012}{027} [\arXivid{1201.0983}] [\inspire{EPRINT+arXiv:1201.0983}].

\bibitem{Sanidas-ml-2012ee}
%
S.A. Sanidas, R.A. Battye and B.W. Stappers,  \emph{Constraints on cosmic
  string tension imposed by the limit on the stochastic gravitational wave
  background from the European Pulsar Timing Array},
\href{http://dx.doi.org/10.1103/PhysRevD.85.122003}{\emph{Phys.\ Rev.} {\bf D 85} (2012) 122003} [\arXivid{1201.2419}] [\inspire{EPRINT+arXiv:1201.2419}].

\bibitem{Starobinsky-ml-1979ty}
%
A.A. Starobinsky,  \emph{Spectrum of relict gravitational radiation and the
  early state of the universe},
{\emph{JETP Lett.} {\bf 30} (1979) 682}
[\inspire{J+\%22JETPLett.,30,682\%22}].

\bibitem{Guzzetti-ml-2016mkm}
%
C.~Guzzetti, M., N.~Bartolo, M.~Liguori and S.~Matarrese,  \emph{Gravitational
  waves from inflation},
\href{http://dx.doi.org/10.1393/ncr/i2016-10127-1}{\emph{Riv.\ Nuovo Cim.}
  {\bf 39} (2016) 399} [\arXivid{1605.01615}] [\inspire{EPRINT+arXiv:1605.01615}].

\bibitem{Ade-ml-2015lrj}
%
{\scshape Planck} collaboration, P.A.R. Ade et~al.,  \emph{Planck 2015
  results. XX. Constraints on inflation},
\href{http://dx.doi.org/10.1051/0004-6361/201525898}{\emph{Astron.\
  Astrophys.} {\bf 594} (2016) A20} [\arXivid{1502.02114}] [\inspire{EPRINT+arXiv:1502.02114}].

\bibitem{Liddle-ml-1992wi}
%
A.R. Liddle and D.H. Lyth,  \emph{COBE, gravitational waves, inflation and
  extended inflation},
\href{http://dx.doi.org/10.1016/0370-2693(92)91393-N}{\emph{Phys.\ Lett.}
  {\bf B 291} (1992) 391} [\astroph{9208007}] [\inspire{EPRINT+astro-ph/9208007}].

\bibitem{Kosowsky-ml-1994cy}
%
A.~Kosowsky, \emph{Cosmic microwave background polarization}, \href{http://dx.doi.org/10.1006/aphy.1996.0020}{\emph{Annals Phys.} {\bf 246} (1996) 49} [\astroph{9501045}].


\bibitem{Hu-ml-1997hv}
%
W.~Hu and M.J. White,  \emph{A CMB polarization primer},
\href{http://dx.doi.org/10.1016/S1384-1076(97)00022-5}{\emph{New Astron.}
  {\bf 2} (1997) 323} [\astroph{9706147}] [\inspire{EPRINT+astro-ph/9706147}].

\bibitem{Corbin-ml-2005ny}
%
V.~Corbin and N.J. Cornish,  \emph{Detecting the cosmic gravitational wave
  background with the big bang observer},
\href{http://dx.doi.org/10.1088/0264-9381/23/7/014}{\emph{Class.\ Quant.\
  Grav.} {\bf 23} (2006) 2435} [\grqc{0512039}] [\inspire{EPRINT+gr-qc/0512039}].

\bibitem{Kawamura-ml-2011zz}
%
S.~Kawamura et~al.,  \emph{The Japanese space gravitational wave antenna:
  DECIGO},
\href{http://dx.doi.org/10.1088/0264-9381/28/9/094011}{\emph{Class.\ Quant.\
  Grav.} {\bf 28} (2011) 094011}
[\inspire{J+\%22Class.Quant.Grav.,28,094011\%22}].

\bibitem{Gasperini-ml-2016gre}
%
M.~Gasperini,  \emph{Observable gravitational waves in pre-big bang cosmology:
  an update},
\arXivid{1606.07889}
[\inspire{EPRINT+arXiv:1606.07889}].

\bibitem{Seoane-ml-2013qna}
%
{\scshape eLISA} collaboration, P.A. Seoane et~al.,  \emph{The gravitational
  universe},
\arXivid{1305.5720}
[\inspire{EPRINT+arXiv:1305.5720}].

\bibitem{Klein-ml-2015hvg}
%
A.~Klein et~al.,  \emph{Science with the space-based interferometer eLISA:
  supermassive black hole binaries},
\href{http://dx.doi.org/10.1103/PhysRevD.93.024003}{\emph{Phys.\ Rev.} {\bf D 93} (2016) 024003} [\arXivid{1511.05581}] [\inspire{EPRINT+arXiv:1511.05581}].

\bibitem{Sesana-ml-2016ljz}
%
A.~Sesana,  \emph{Prospects for multiband gravitational-wave astronomy after
  GW150914},
\href{http://dx.doi.org/10.1103/PhysRevLett.116.231102}{\emph{Phys.\ Rev.\
  Lett.} {\bf 116} (2016) 231102} [\arXivid{1602.06951}] [\inspire{EPRINT+arXiv:1602.06951}].

\bibitem{Barausse-ml-2016eii}
%
E.~Barausse, N.~Yunes and K.~Chamberlain,  \emph{Theory-agnostic constraints on
  black-hole dipole radiation with multiband gravitational-wave astrophysics},
\href{http://dx.doi.org/10.1103/PhysRevLett.116.241104}{\emph{Phys.\ Rev.\
  Lett.} {\bf 116} (2016) 241104} [\arXivid{1603.04075}] [\inspire{EPRINT+arXiv:1603.04075}].

\bibitem{Nelemans-ml-2013yg}
%
G.~Nelemans,  \emph{Galactic binaries with eLISA},
{\emph{ASP Conf.\ Ser.}
  {\bf 467} (2013) 27} [\arXivid{1302.0138}] [\inspire{EPRINT+arXiv:1302.0138}].

\bibitem{Gair-ml-2012vi}
%
J.R. Gair and E.K. Porter,  \emph{Observing EMRIs with eLISA/NGO},
{\emph{ASP Conf.\ Ser.} {\bf 467} (2013) 173} [\arXivid{1210.8066}] [\inspire{EPRINT+arXiv:1210.8066}].

\bibitem{Tamanini-ml-2016zlh}
%
N.~Tamanini, C.~Caprini, E.~Barausse, A.~Sesana, A.~Klein and A.~Petiteau,
   \emph{Science with the space-based interferometer eLISA. III: probing the
  expansion of the Universe using gravitational wave standard sirens},
\jcap{04}{2016}{002} [\arXivid{1601.07112}] [\inspire{EPRINT+arXiv:1601.07112}].

\bibitem{Bonvin-ml-2016qxr}
%
C.~Bonvin, C.~Caprini, R.~Sturani and N.~Tamanini,  \emph{The effect of matter
  structure on the gravitational waveform},
\arXivid{1609.08093}
[\inspire{EPRINT+arXiv:1609.08093}].

\bibitem{Armano-ml-2016bkm}
%
M.~Armano et~al.,  \emph{Sub-Femto- g free fall for space-based gravitational
  wave observatories: LISA pathfinder results},
\href{http://dx.doi.org/10.1103/PhysRevLett.116.231101}{\emph{Phys.\ Rev.\
  Lett.} {\bf 116} (2016) 231101}
[\inspire{J+\%22Phys.Rev.Lett.,116,231101\%22}].

\bibitem{antoine}
%
A.~Petiteau, in preparation.

\bibitem{Adams-ml-2010vc}
%
M.R. Adams and N.J. Cornish,  \emph{Discriminating between a stochastic
  gravitational wave background and instrument noise},
\href{http://dx.doi.org/10.1103/PhysRevD.82.022002}{\emph{Phys.\ Rev.} {\bf D 82} (2010) 022002} [\arXivid{1002.1291}] [\inspire{EPRINT+arXiv:1002.1291}].

\bibitem{Adams-ml-2013qma}
%
M.R. Adams and N.J. Cornish,  \emph{Detecting a stochastic gravitational wave
  background in the presence of a galactic foreground and instrument noise},
\href{http://dx.doi.org/10.1103/PhysRevD.89.022001}{\emph{Phys.\ Rev.} {\bf D 89} (2014) 022001} [\arXivid{1307.4116}] [\inspire{EPRINT+arXiv:1307.4116}].

\bibitem{Thrane-ml-2013oya}
%
E.~Thrane and J.D. Romano,  \emph{Sensitivity curves for searches for
  gravitational-wave backgrounds},
\href{http://dx.doi.org/10.1103/PhysRevD.88.124032}{\emph{Phys.\ Rev.} {\bf D 88} (2013) 124032} [\arXivid{1310.5300}] [\inspire{EPRINT+arXiv:1310.5300}].

\bibitem{Meerburg-ml-2015zua}
%
P.D. Meerburg, R.~Hlo\v{z}ek, B.~Hadzhiyska and J.~Meyers,  \emph{Multiwavelength
  constraints on the inflationary consistency relation},
\href{http://dx.doi.org/10.1103/PhysRevD.91.103505}{\emph{Phys.\ Rev.} {\bf D 91} (2015) 103505} [\arXivid{1502.00302}] [\inspire{EPRINT+arXiv:1502.00302}].

\bibitem{Array-ml-2015xqh}
%
{\scshape BICEP2, Keck Array} collaboration, P.A.R. Ade et~al.,
   \emph{Improved constraints on cosmology and foregrounds from BICEP2 and Keck
  Array Cosmic Microwave Background data with inclusion of $95$\,GHz band},
\href{http://dx.doi.org/10.1103/PhysRevLett.116.031302}{\emph{Phys.\ Rev.\
  Lett.} {\bf 116} (2016) 031302} [\arXivid{1510.09217}] [\inspire{EPRINT+arXiv:1510.09217}].

\bibitem{Lasky-ml-2015lej}
%
P.D. Lasky et~al.,  \emph{Gravitational-wave cosmology across $29$ decades in
  frequency},
\href{http://dx.doi.org/10.1103/PhysRevX.6.011035}{\emph{Phys.\
  Rev.} {\bf X 6} (2016) 011035} [\arXivid{1511.05994}] [\inspire{EPRINT+arXiv:1511.05994}].

\bibitem{Cabass-ml-2015jwe}
%
G.~Cabass, L.~Pagano, L.~Salvati, M.~Gerbino, E.~Giusarma and A.~Melchiorri,
   \emph{Updated constraints and forecasts on primordial tensor modes},
\href{http://dx.doi.org/10.1103/PhysRevD.93.063508}{\emph{Phys.\ Rev.} {\bf D 93} (2016) 063508} [\arXivid{1511.05146}] [\inspire{EPRINT+arXiv:1511.05146}].

\bibitem{Anber-ml-2006xt}
%
M.M. Anber and L.~Sorbo,  \emph{$N$-flationary magnetic fields},
\jcap{10}{2006}{018} [\astroph{0606534}] [\inspire{EPRINT+astro-ph/0606534}].

\bibitem{Barnaby-ml-2010vf}
%
N.~Barnaby and M.~Peloso,  \emph{Large non-gaussianity in axion inflation},
\href{http://dx.doi.org/10.1103/PhysRevLett.106.181301}{\emph{Phys.\ Rev.\
  Lett.} {\bf 106} (2011) 181301} [\arXivid{1011.1500}] [\inspire{EPRINT+arXiv:1011.1500}].

\bibitem{Ade-ml-2015ava}
%
{\scshape Planck} collaboration, P.A.R. Ade et~al.,  \emph{Planck 2015
  results. XVII. Constraints on primordial non-Gaussianity},
\href{http://dx.doi.org/10.1051/0004-6361/201525836}{\emph{Astron.\
  Astrophys.} {\bf 594} (2016) A17} [\arXivid{1502.01592}] [\inspire{EPRINT+arXiv:1502.01592}].

\bibitem{Cook-ml-2011hg}
%
J.L. Cook and L.~Sorbo,  \emph{Particle production during inflation and
  gravitational waves detectable by ground-based interferometers},
\href{http://dx.doi.org/10.1103/PhysRevD.86.069901}{\emph{Phys.\ Rev.} {\bf D 85} (2012) 023534} [\erratum{D 86}{2012}{069901}] [\arXivid{1109.0022}] [\inspire{EPRINT+arXiv:1109.0022}].

\bibitem{Meerburg-ml-2012id}
%
P.D. Meerburg and E.~Pajer,  \emph{Observational constraints on gauge field
  production in axion inflation},
\jcap{02}{2013}{017} [\arXivid{1203.6076}] [\inspire{EPRINT+arXiv:1203.6076}].

\bibitem{Sorbo-ml-2011rz}
%
L.~Sorbo,  \emph{Parity violation in the Cosmic Microwave Background from a
  pseudoscalar inflaton},
\jcap{06}{2011}{003} [\arXivid{1101.1525}] [\inspire{EPRINT+arXiv:1101.1525}].

\bibitem{Namba-ml-2015gja}
%
R.~Namba, M.~Peloso, M.~Shiraishi, L.~Sorbo and C.~Unal,  \emph{Scale-dependent
  gravitational waves from a rolling axion},
\jcap{01}{2016}{041} [\arXivid{1509.07521}] [\inspire{EPRINT+arXiv:1509.07521}].

\bibitem{Garcia-Bellido-ml-2016dkw}
%
J.~Garc\'ia-Bellido, M.~Peloso and C.~Unal,  \emph{Gravitational waves at
  interferometer scales and primordial black holes in axion inflation},
\arXivid{1610.03763}
[\inspire{EPRINT+arXiv:1610.03763}].

\bibitem{Crowder-ml-2012ik}
%
S.G. Crowder, R.~Namba, V.~Mandic, S.~Mukohyama and M.~Peloso,
   \emph{Measurement of parity violation in the early universe using
  gravitational-wave detectors},
\href{http://dx.doi.org/10.1016/j.physletb.2013.08.077}{\emph{Phys.\ Lett.}
  {\bf B 726} (2013) 66} [\arXivid{1212.4165}] [\inspire{EPRINT+arXiv:1212.4165}].

\bibitem{Cook-ml-2013xea}
%
J.L. Cook and L.~Sorbo,  \emph{An inflationary model with small scalar and
  large tensor non-Gaussianities},
\jcap{11}{2013}{047} [\arXivid{1307.7077}] [\inspire{EPRINT+arXiv:1307.7077}].

\bibitem{Thrane-ml-2013kb}
%
E.~Thrane,  \emph{Measuring the non-Gaussian stochastic gravitational-wave
  background: a method for realistic interferometer data},
\href{http://dx.doi.org/10.1103/PhysRevD.87.043009}{\emph{Phys.\ Rev.} {\bf D 87} (2013) 043009} [\arXivid{1301.0263}] [\inspire{EPRINT+arXiv:1301.0263}].

\bibitem{Barnaby-ml-2011qe}
%
N.~Barnaby, E.~Pajer and M.~Peloso,  \emph{Gauge field production in axion
  inflation: consequences for monodromy, non-gaussianity in the CMB and
  gravitational waves at interferometers},
\href{http://dx.doi.org/10.1103/PhysRevD.85.023525}{\emph{Phys.\ Rev.} {\bf D 85} (2012) 023525} [\arXivid{1110.3327}] [\inspire{EPRINT+arXiv:1110.3327}].

\bibitem{Peloso-ml-2016gqs}
%
M.~Peloso, L.~Sorbo and C.~Unal,  \emph{Rolling axions during inflation:
  perturbativity and signatures},
\jcap{09}{2016}{001} [\arXivid{1606.00459}] [\inspire{EPRINT+arXiv:1606.00459}].

\bibitem{Domcke-ml-2016bkh}
%
V.~Domcke, M.~Pieroni and P.~Bin\'etruy,  \emph{Primordial gravitational waves
  for universality classes of pseudoscalar inflation},
\jcap{06}{2016}{031} [\arXivid{1603.01287}] [\inspire{EPRINT+arXiv:1603.01287}].

\bibitem{Mukhanov-ml-2013tua}
%
V.~Mukhanov,  \emph{Quantum cosmological perturbations: predictions and
  observations},
\href{http://dx.doi.org/10.1140/epjc/s10052-013-2486-7}{\emph{Eur.\ Phys.\
  J.} {\bf C 73} (2013) 2486} [\arXivid{1303.3925}] [\inspire{EPRINT+arXiv:1303.3925}].

\bibitem{Fixsen-ml-1996nj}
%
D.J. Fixsen, E.S. Cheng, J.M. Gales, J.C. Mather, R.A. Shafer and E.L.
  Wright,  \emph{The Cosmic Microwave Background spectrum from the full COBE
  FIRAS data set},
\href{http://dx.doi.org/10.1086/178173}{\emph{Astrophys.\
  J.} {\bf 473} (1996) 576} [\astroph{9605054}] [\inspire{EPRINT+astro-ph/9605054}].

\bibitem{TheLIGOScientific-ml-2016wyq}
%
{\scshape Virgo, LIGO Scientific} collaboration, B.P. Abbott et~al.,
   \emph{GW150914: implications for the stochastic gravitational wave
  background from binary black holes},
\href{http://dx.doi.org/10.1103/PhysRevLett.116.131102}{\emph{Phys.\ Rev.\
  Lett.} {\bf 116} (2016) 131102} [\arXivid{1602.03847}] [\inspire{EPRINT+arXiv:1602.03847}].

\bibitem{Linde-ml-2012bt}
%
A.~Linde, S.~Mooij and E.~Pajer,  \emph{Gauge field production in supergravity
  inflation: Local non-Gaussianity and primordial black holes},
\href{http://dx.doi.org/10.1103/PhysRevD.87.103506}{\emph{Phys.\ Rev.} {\bf D 87} (2013) 103506} [\arXivid{1212.1693}] [\inspire{EPRINT+arXiv:1212.1693}].

\bibitem{Carr-ml-2009jm}
%
B.J. Carr, K.~Kohri, Y.~Sendouda and J.~Yokoyama,  \emph{New cosmological
  constraints on primordial black holes},
\href{http://dx.doi.org/10.1103/PhysRevD.81.104019}{\emph{Phys.\ Rev.} {\bf D 81} (2010) 104019} [\arXivid{0912.5297}] [\inspire{EPRINT+arXiv:0912.5297}].

\bibitem{Carr-ml-2016drx}
%
B.~Carr, F.~Kuhnel and M.~Sandstad,  \emph{Primordial black holes as dark
  matter},
\href{http://dx.doi.org/10.1103/PhysRevD.94.083504}{\emph{Phys.\
  Rev.} {\bf D 94} (2016) 083504} [\arXivid{1607.06077}] [\inspire{EPRINT+arXiv:1607.06077}].

\bibitem{Anber-ml-2009ua}
%
M.M. Anber and L.~Sorbo,  \emph{Naturally inflating on steep potentials
  through electromagnetic dissipation},
\href{http://dx.doi.org/10.1103/PhysRevD.81.043534}{\emph{Phys.\ Rev.} {\bf D 81} (2010) 043534} [\arXivid{0908.4089}] [\inspire{EPRINT+arXiv:0908.4089}].

\bibitem{Ferreira-ml-2015omg}
%
R.Z. Ferreira, J.~Ganc, J.~Nore\~na and M.S. Sloth,  \emph{On the validity of
  the perturbative description of axions during inflation},

\bibitem{Ade-ml-2015xua}
%
{\scshape Planck} collaboration, P.A.R. Ade et~al.,  \emph{Planck 2015
  results. XIII. Cosmological parameters},
\href{http://dx.doi.org/10.1051/0004-6361/201525830}{\emph{Astron.\
  Astrophys.} {\bf 594} (2016) A13} [\arXivid{1502.01589}] [\inspire{EPRINT+arXiv:1502.01589}].

\bibitem{Cyburt-ml-2015mya}
%
R.H. Cyburt, B.D. Fields, K.A. Olive and T.-H. Yeh,  \emph{Big Bang
  nucleosynthesis: 2015},
\href{http://dx.doi.org/10.1103/RevModPhys.88.015004}{\emph{Rev.\ Mod.\
  Phys.} {\bf 88} (2016) 015004} [\arXivid{1505.01076}] [\inspire{EPRINT+arXiv:1505.01076}].

\bibitem{Biagetti-ml-2013kwa}
%
M.~Biagetti, M.~Fasiello and A.~Riotto,  \emph{Enhancing inflationary tensor
  modes through spectator fields},
\href{http://dx.doi.org/10.1103/PhysRevD.88.103518}{\emph{Phys.\ Rev.} {\bf D 88} (2013) 103518} [\arXivid{1305.7241}] [\inspire{EPRINT+arXiv:1305.7241}].

\bibitem{Biagetti-ml-2014asa}
%
M.~Biagetti, E.~Dimastrogiovanni, M.~Fasiello and M.~Peloso,
   \emph{Gravitational waves and scalar perturbations from spectator fields},
\jcap{04}{2015}{011} [\arXivid{1411.3029}] [\inspire{EPRINT+arXiv:1411.3029}].

\bibitem{Fujita-ml-2014oba}
%
T.~Fujita, J.~Yokoyama and S.~Yokoyama,  \emph{Can a spectator scalar field
  enhance inflationary tensor mode?},
\href{http://dx.doi.org/10.1093/ptep/ptv037}{\emph{PTEP} {\bf 2015} (2015)
  043E01} [\arXivid{1411.3658}] [\inspire{EPRINT+arXiv:1411.3658}].

\bibitem{Bartolo-ml-2007vp}
%
N.~Bartolo, S.~Matarrese, A.~Riotto and A.~Vaihkonen,  \emph{The maximal amount
  of gravitational waves in the curvaton scenario},
\href{http://dx.doi.org/10.1103/PhysRevD.76.061302}{\emph{Phys.\ Rev.} {\bf D 76} (2007) 061302} [\arXivid{0705.4240}] [\inspire{EPRINT+arXiv:0705.4240}].

\bibitem{tomita}
%
K.~Tomita, \emph{Non-linear theory of gravitational instability in the expanding universe}, \href{http://dx.doi.org/10.1143/PTP.37.831}{\emph{Prog. Theor. Phys.} {\bf 37} (1967) 831}.


\bibitem{Matarrese-ml-1997ay}
%
S.~Matarrese, S.~Mollerach and M.~Bruni,  \emph{Second order perturbations of
  the Einstein-de~Sitter universe},
\href{http://dx.doi.org/10.1103/PhysRevD.58.043504}{\emph{Phys.\ Rev.} {\bf D 58} (1998) 043504} [\astroph{9707278}] [\inspire{EPRINT+astro-ph/9707278}].

\bibitem{Ade-ml-2013zuv}
%
{\scshape Planck} collaboration, P.A.R. Ade et~al.,  \emph{Planck 2013
  results. XVI. Cosmological parameters},
\href{http://dx.doi.org/10.1051/0004-6361/201321591}{\emph{Astron.\
  Astrophys.} {\bf 571} (2014) A16} [\arXivid{1303.5076}] [\inspire{EPRINT+arXiv:1303.5076}].

\bibitem{Aghanim-ml-2015xee}
%
{\scshape Planck} collaboration, N.~Aghanim et~al.,  \emph{Planck 2015 results.
  XI. CMB power spectra, likelihoods and robustness of parameters},
\href{http://dx.doi.org/10.1051/0004-6361/201526926}{\emph{Astron.\
  Astrophys.} {\bf 594} (2016) A11} [\arXivid{1507.02704}] [\inspire{EPRINT+arXiv:1507.02704}].

\bibitem{Kinney-ml-2012ik}
%
W.H. Kinney, A.M. Dizgah, B.A. Powell and A.~Riotto,  \emph{Inflaton or
  curvaton? Constraints on bimodal primordial spectra from mixed
  perturbations},
\href{http://dx.doi.org/10.1103/PhysRevD.86.023527}{\emph{Phys.\ Rev.} {\bf D 86} (2012) 023527} [\arXivid{1203.0693}] [\inspire{EPRINT+arXiv:1203.0693}].

\bibitem{Kuroyanagi-ml-2014nba}
%
S.~Kuroyanagi, T.~Takahashi and S.~Yokoyama,  \emph{Blue-tilted tensor spectrum
  and thermal history of the universe},
\jcap{02}{2015}{003} [\arXivid{1407.4785}] [\inspire{EPRINT+arXiv:1407.4785}].

\bibitem{Pagano-ml-2015hma}
%
L.~Pagano, L.~Salvati and A.~Melchiorri,  \emph{New constraints on primordial
  gravitational waves from Planck 2015},
\href{http://dx.doi.org/10.1016/j.physletb.2016.07.078}{\emph{Phys.\ Lett.}
  {\bf B 760} (2016) 823} [\arXivid{1508.02393}] [\inspire{EPRINT+arXiv:1508.02393}].

\bibitem{Josan-ml-2009qn}
%
A.S. Josan, A.M. Green and K.A. Malik,  \emph{Generalised constraints on the
  curvature perturbation from primordial black holes},
\href{http://dx.doi.org/10.1103/PhysRevD.79.103520}{\emph{Phys.\ Rev.} {\bf D 79} (2009) 103520} [\arXivid{0903.3184}] [\inspire{EPRINT+arXiv:0903.3184}].

\bibitem{Cheung-ml-2007st}
%
C.~Cheung, P.~Creminelli, A.L. Fitzpatrick, J.~Kaplan and L.~Senatore,
   \emph{The effective field theory of inflation},
\jhep{03}{2008}{014} [\arXivid{0709.0293}] [\inspire{EPRINT+arXiv:0709.0293}].

\bibitem{Kobayashi-ml-2010cm}
%
T.~Kobayashi, M.~Yamaguchi and J.~Yokoyama,  \emph{G-inflation: inflation
  driven by the galileon field},
\href{http://dx.doi.org/10.1103/PhysRevLett.105.231302}{\emph{Phys.\ Rev.\
  Lett.} {\bf 105} (2010) 231302} [\arXivid{1008.0603}] [\inspire{EPRINT+arXiv:1008.0603}].

\bibitem{Kobayashi-ml-2011nu}
%
T.~Kobayashi, M.~Yamaguchi and J.~Yokoyama,  \emph{Generalized G-inflation:
  inflation with the most general second-order field equations},
\href{http://dx.doi.org/10.1143/PTP.126.511}{\emph{Prog.\ Theor.\ Phys.} {\bf
  126} (2011) 511} [\arXivid{1105.5723}] [\inspire{EPRINT+arXiv:1105.5723}].

\bibitem{Wang-ml-2014kqa}
%
Y.~Wang and W.~Xue,  \emph{Inflation and alternatives with blue tensor
  spectra},
\jcap{10}{2014}{075} [\arXivid{1403.5817}] [\inspire{EPRINT+arXiv:1403.5817}].

\bibitem{Golovnev-ml-2008cf}
%
A.~Golovnev, V.~Mukhanov and V.~Vanchurin,  \emph{Vector inflation},
\jcap{06}{2008}{009} [\arXivid{0802.2068}] [\inspire{EPRINT+arXiv:0802.2068}].

\bibitem{Himmetoglu-ml-2008zp}
%
B.~Himmetoglu, C.R. Contaldi and M.~Peloso,  \emph{Instability of anisotropic
  cosmological solutions supported by vector fields},
\href{http://dx.doi.org/10.1103/PhysRevLett.102.111301}{\emph{Phys.\ Rev.\
  Lett.} {\bf 102} (2009) 111301} [\arXivid{0809.2779}] [\inspire{EPRINT+arXiv:0809.2779}].

\bibitem{Maleknejad-ml-2011jw}
%
A.~Maleknejad and M.M. Sheikh-Jabbari,  \emph{Gauge-flation: inflation from
  non-abelian gauge fields},
\href{http://dx.doi.org/10.1016/j.physletb.2013.05.001}{\emph{Phys.\ Lett.}
  {\bf B 723} (2013) 224} [\arXivid{1102.1513}] [\inspire{EPRINT+arXiv:1102.1513}].

\bibitem{Adshead-ml-2012kp}
%
P.~Adshead and M.~Wyman,  \emph{Chromo-natural inflation: natural inflation on
  a steep potential with classical non-Abelian gauge fields},
\href{http://dx.doi.org/10.1103/PhysRevLett.108.261302}{\emph{Phys.\ Rev.\
  Lett.} {\bf 108} (2012) 261302} [\arXivid{1202.2366}] [\inspire{EPRINT+arXiv:1202.2366}].

\bibitem{Dimastrogiovanni-ml-2016fuu}
%
E.~Dimastrogiovanni, M.~Fasiello and T.~Fujita,  \emph{Primordial gravitational
  waves from axion-gauge fields dynamics},
\arXivid{1608.04216}
[\inspire{EPRINT+arXiv:1608.04216}].

\bibitem{Endlich-ml-2013jia}
%
S.~Endlich, B.~Horn, A.~Nicolis and J.~Wang,  \emph{Squeezed limit of the solid
  inflation three-point function},
\href{http://dx.doi.org/10.1103/PhysRevD.90.063506}{\emph{Phys.\ Rev.} {\bf D 90} (2014) 063506} [\arXivid{1307.8114}] [\inspire{EPRINT+arXiv:1307.8114}].

\bibitem{Koh-ml-2013msa}
%
S.~Koh, S.~Kouwn, O.-K. Kwon and P.~Oh,  \emph{Cosmological perturbations of a
  quartet of scalar fields with a spatially constant gradient},
\href{http://dx.doi.org/10.1103/PhysRevD.88.043523}{\emph{Phys.\ Rev.} {\bf D 88} (2013) 043523} [\arXivid{1304.7924}] [\inspire{EPRINT+arXiv:1304.7924}].

\bibitem{Cannone-ml-2014uqa}
%
D.~Cannone, G.~Tasinato and D.~Wands,  \emph{Generalised tensor fluctuations
  and inflation},
\jcap{01}{2015}{029} [\arXivid{1409.6568}] [\inspire{EPRINT+arXiv:1409.6568}].

\bibitem{Cannone-ml-2015rra}
%
D.~Cannone, J.-O. Gong and G.~Tasinato,  \emph{Breaking discrete symmetries in
  the effective field theory of inflation},
\jcap{08}{2015}{003} [\arXivid{1505.05773}] [\inspire{EPRINT+arXiv:1505.05773}].

\bibitem{Bartolo-ml-2015qvr}
%
N.~Bartolo, D.~Cannone, A.~Ricciardone and G.~Tasinato,  \emph{Distinctive
  signatures of space-time diffeomorphism breaking in EFT of inflation},
\jcap{03}{2016}{044} [\arXivid{1511.07414}] [\inspire{EPRINT+arXiv:1511.07414}].

\bibitem{Ricciardone-ml-2016aa}
%
A.~Ricciardone and G.~Tasinato,  \emph{Primordial gravitational waves in
  supersolid inflation},
\arXivid{1611.04516}
[\inspire{EPRINT+arXiv:1611.04516}].

\bibitem{Akhshik-ml-2014gja}
%
M.~Akhshik, R.~Emami, H.~Firouzjahi and Y.~Wang,  \emph{Statistical
  anisotropies in gravitational waves in solid inflation},
\jcap{09}{2014}{012} [\arXivid{1405.4179}] [\inspire{EPRINT+arXiv:1405.4179}].

\bibitem{Akhshik-ml-2014bla}
%
M.~Akhshik,  \emph{Clustering fossils in solid inflation},
\jcap{05}{2015}{043} [\arXivid{1409.3004}] [\inspire{EPRINT+arXiv:1409.3004}].

\bibitem{Horndeski-ml-1974wa}
%
G.W. Horndeski,  \emph{Second-order scalar-tensor field equations in a
  four-dimensional space},
\href{http://dx.doi.org/10.1007/BF01807638}{\emph{Int.\ J.\ Theor.\ Phys.}
  {\bf 10} (1974) 363}
[\inspire{J+\%22Int.J.Theor.Phys.,10,363\%22}].

\bibitem{Creminelli-ml-2014wna}
%
P.~Creminelli, J.~Gleyzes, J.~Nore\~na and F.~Vernizzi,  \emph{Resilience of the
  standard predictions for primordial tensor modes},
\href{http://dx.doi.org/10.1103/PhysRevLett.113.231301}{\emph{Phys.\ Rev.\
  Lett.} {\bf 113} (2014) 231301} [\arXivid{1407.8439}] [\inspire{EPRINT+arXiv:1407.8439}].

\bibitem{Higuchi-ml-1986py}
%
A.~Higuchi,  \emph{Forbidden mass range for spin-2 field theory in de~Sitter
  space-time},
\href{http://dx.doi.org/10.1016/0550-3213(87)90691-2}{\emph{Nucl.\ Phys.}
  {\bf B 282} (1987) 397}
[\inspire{J+\%22Nucl.Phys.,B282,397\%22}].

\bibitem{Arkani-Hamed-ml-2015bza}
%
N.~Arkani-Hamed and J.~Maldacena,  \emph{Cosmological collider physics},
\arXivid{1503.08043}
[\inspire{EPRINT+arXiv:1503.08043}].

\bibitem{Horava-ml-2009uw}
%
P.~Ho\v{r}ava,  \emph{Quantum gravity at a Lifshitz point},
\href{http://dx.doi.org/10.1103/PhysRevD.79.084008}{\emph{Phys.\ Rev.} {\bf D 79} (2009) 084008} [\arXivid{0901.3775}] [\inspire{EPRINT+arXiv:0901.3775}].

\bibitem{Blas-ml-2009my}
%
D.~Blas, D.~Comelli, F.~Nesti and L.~Pilo,  \emph{Lorentz breaking massive
  gravity in curved space},
\href{http://dx.doi.org/10.1103/PhysRevD.80.044025}{\emph{Phys.\ Rev.} {\bf D 80} (2009) 044025} [\arXivid{0905.1699}] [\inspire{EPRINT+arXiv:0905.1699}].

\bibitem{Cai-ml-2015yza}
%
Y.~Cai, Y.-T. Wang and Y.-S. Piao,  \emph{Is there an effect of a nontrivial
  $c_T$ during inflation?},
\href{http://dx.doi.org/10.1103/PhysRevD.93.063005}{\emph{Phys.\ Rev.} {\bf D 93} (2016) 063005} [\arXivid{1510.08716}] [\inspire{EPRINT+arXiv:1510.08716}].

\bibitem{Cai-ml-2016ldn}
%
Y.~Cai, Y.-T. Wang and Y.-S. Piao,  \emph{Propagating speed of primordial
  gravitational waves and inflation},
\href{http://dx.doi.org/10.1103/PhysRevD.94.043002}{\emph{Phys.\ Rev.} {\bf D 94} (2016) 043002} [\arXivid{1602.05431}] [\inspire{EPRINT+arXiv:1602.05431}].

\bibitem{Bartolo-ml-2013msa}
%
N.~Bartolo, S.~Matarrese, M.~Peloso and A.~Ricciardone,  \emph{Anisotropy in
  solid inflation},
\jcap{08}{2013}{022} [\arXivid{1306.4160}] [\inspire{EPRINT+arXiv:1306.4160}].

\bibitem{Bartolo-ml-2014xfa}
%
N.~Bartolo, M.~Peloso, A.~Ricciardone and C.~Unal,  \emph{The expected
  anisotropy in solid inflation},
\jcap{11}{2014}{009} [\arXivid{1407.8053}] [\inspire{EPRINT+arXiv:1407.8053}].

\bibitem{Bordin-ml-2016ruc}
%
L.~Bordin, P.~Creminelli, M.~Mirbabayi and J.~Nore\~na,  \emph{Tensor squeezed
  limits and the Higuchi bound},
\jcap{09}{2016}{041} [\arXivid{1605.08424}] [\inspire{EPRINT+arXiv:1605.08424}].

\bibitem{Ade-ml-2013nlj}
%
{\scshape Planck} collaboration, P.A.R. Ade et~al.,  \emph{Planck 2013
  results. XXIII. Isotropy and statistics of the CMB},
\href{http://dx.doi.org/10.1051/0004-6361/201321534}{\emph{Astron.\
  Astrophys.} {\bf 571} (2014) A23} [\arXivid{1303.5083}] [\inspire{EPRINT+arXiv:1303.5083}].

\bibitem{Ade-ml-2015hxq}
%
{\scshape Planck} collaboration, P.A.R. Ade et~al.,  \emph{Planck 2015
  results. XVI. Isotropy and statistics of the CMB},
\href{http://dx.doi.org/10.1051/0004-6361/201526681}{\emph{Astron.\
  Astrophys.} {\bf 594} (2016) A16} [\arXivid{1506.07135}] [\inspire{EPRINT+arXiv:1506.07135}].

\bibitem{Endlich-ml-2012pz}
%
S.~Endlich, A.~Nicolis and J.~Wang,  \emph{Solid inflation},
\jcap{10}{2013}{011} [\arXivid{1210.0569}] [\inspire{EPRINT+arXiv:1210.0569}].

\bibitem{Ade-ml-2013ydc}
%
{\scshape Planck} collaboration, P.A.R. Ade et~al.,  \emph{Planck 2013
  Results. XXIV. Constraints on primordial non-Gaussianity},
\href{http://dx.doi.org/10.1051/0004-6361/201321554}{\emph{Astron.\
  Astrophys.} {\bf 571} (2014) A24} [\arXivid{1303.5084}] [\inspire{EPRINT+arXiv:1303.5084}].

\bibitem{Schmidt-ml-2015xka}
%
F.~Schmidt, N.E. Chisari and C.~Dvorkin,  \emph{Imprint of inflation on galaxy
  shape correlations},
\jcap{10}{2015}{032} [\arXivid{1506.02671}] [\inspire{EPRINT+arXiv:1506.02671}].

\bibitem{Chisari-ml-2016xki}
%
N.E. Chisari, C.~Dvorkin, F.~Schmidt and D.~Spergel,  \emph{Multitracing
  anisotropic non-gaussianity with galaxy shapes},
\arXivid{1607.05232}
[\inspire{EPRINT+arXiv:1607.05232}].

\bibitem{Meerburg-ml-2016ecv}
%
P.D. Meerburg, J.~Meyers, A.~van Engelen and Y.~Ali-Ha\"{\i}moud,  \emph{CMB B-mode non-Gaussianity},
\href{http://dx.doi.org/10.1103/PhysRevD.93.123511}{\emph{Phys.\ Rev.} {\bf D 93} (2016) 123511} [\arXivid{1603.02243}] [\inspire{EPRINT+arXiv:1603.02243}].

\bibitem{GarciaBellido-ml-1996qt}
%
J.~Garc\'ia-Bellido, A.D. Linde and D.~Wands,  \emph{Density perturbations and
  black hole formation in hybrid inflation},
\href{http://dx.doi.org/10.1103/PhysRevD.54.6040}{\emph{Phys.\ Rev.} {\bf D 54} (1996) 6040} [\astroph{9605094}] [\inspire{EPRINT+astro-ph/9605094}].

\bibitem{Nakamura-ml-1997sm}
%
T.~Nakamura, M.~Sasaki, T.~Tanaka and K.S. Thorne,  \emph{Gravitational waves
  from coalescing black hole MACHO binaries},
\href{http://dx.doi.org/10.1086/310886}{\emph{Astrophys.\ J.} {\bf 487}
  (1997) L139} [\astroph{9708060}] [\inspire{EPRINT+astro-ph/9708060}].

\bibitem{Clesse-ml-2015wea}
%
S.~Clesse and J.~Garc\'{\i}a-Bellido,  \emph{Massive primordial black holes from
  hybrid inflation as dark matter and the seeds of galaxies},
\href{http://dx.doi.org/10.1103/PhysRevD.92.023524}{\emph{Phys.\ Rev.} {\bf D 92} (2015) 023524} [\arXivid{1501.07565}] [\inspire{EPRINT+arXiv:1501.07565}].

\bibitem{Clesse-ml-2016ajp}
%
S.~Clesse and J.~Garc\'{\i}a-Bellido,  \emph{Detecting the gravitational wave
  background from primordial black hole dark matter},
\arXivid{1610.08479}
[\inspire{EPRINT+arXiv:1610.08479}].

\bibitem{Bird-ml-2016dcv}
%
S.~Bird et~al.,  \emph{Did LIGO detect dark matter?},
\href{http://dx.doi.org/10.1103/PhysRevLett.116.201301}{\emph{Phys.\ Rev.\
  Lett.} {\bf 116} (2016) 201301} [\arXivid{1603.00464}] [\inspire{EPRINT+arXiv:1603.00464}].

\bibitem{Clesse-ml-2016vqa}
%
S.~Clesse and J.~Garc\'{\i}a-Bellido,  \emph{The clustering of massive primordial
  black holes as dark matter: measuring their mass distribution with advanced
  LIGO},
\href{http://dx.doi.org/10.1016/j.dark.2016.10.002}{\emph{Phys.\ Dark
  Univ.} {\bf 10} (2016) 002} [\arXivid{1603.05234}] [\inspire{EPRINT+arXiv:1603.05234}].

\bibitem{Sasaki-ml-2016jop}
%
M.~Sasaki, T.~Suyama, T.~Tanaka and S.~Yokoyama,  \emph{Primordial black hole
  scenario for the gravitational-wave event GW150914},
\href{http://dx.doi.org/10.1103/PhysRevLett.117.061101}{\emph{Phys.\ Rev.\
  Lett.} {\bf 117} (2016) 061101} [\arXivid{1603.08338}] [\inspire{EPRINT+arXiv:1603.08338}].

\bibitem{Cholis-ml-2016kqi}
%
I.~Cholis et~al.,  \emph{Orbital eccentricities in primordial black hole
  binaries},
\href{http://dx.doi.org/10.1103/PhysRevD.94.084013}{\emph{Phys.\
  Rev.} {\bf D 94} (2016) 084013} [\arXivid{1606.07437}] [\inspire{EPRINT+arXiv:1606.07437}].

\bibitem{Renaux-Petel-ml-2015mga}
%
S.~Renaux-Petel and K.~Turzy\'nski,  \emph{Geometrical destabilization of
  inflation},
\href{http://dx.doi.org/10.1103/PhysRevLett.117.141301}{\emph{Phys.\ Rev.\
  Lett.} {\bf 117} (2016) 141301} [\arXivid{1510.01281}] [\inspire{EPRINT+arXiv:1510.01281}].

\bibitem{GarciaBellido-ml-1997wm}
%
J.~Garc\'ia-Bellido and A.D. Linde,  \emph{Preheating in hybrid inflation},
\href{http://dx.doi.org/10.1103/PhysRevD.57.6075}{\emph{Phys.\ Rev.} {\bf D 57} (1998) 6075} [\hepph{9711360}] [\inspire{EPRINT+hep-ph/9711360}].

\bibitem{Clesse-ml-2010iz}
%
S.~Clesse,  \emph{Hybrid inflation along waterfall trajectories},
\href{http://dx.doi.org/10.1103/PhysRevD.83.063518}{\emph{Phys.\ Rev.} {\bf D 83} (2011) 063518} [\arXivid{1006.4522}] [\inspire{EPRINT+arXiv:1006.4522}].

\bibitem{Clesse-ml-2013jra}
%
S.~Clesse, B.~Garbrecht and Y.~Zhu,  \emph{Non-gaussianities and curvature
  perturbations from hybrid inflation},
\href{http://dx.doi.org/10.1103/PhysRevD.89.063519}{\emph{Phys.\ Rev.} {\bf D 89} (2014) 063519} [\arXivid{1304.7042}] [\inspire{EPRINT+arXiv:1304.7042}].

\bibitem{Phinney-ml-2001di}
%
%
E.S. Phinney,  \emph{A practical theorem on gravitational wave backgrounds},
\astroph{0108028}
[\inspire{EPRINT+astro-ph/0108028}].

\end{thebibliography}
\end{document}